\documentclass[longauth]{aa} % for the long lists of affiliations
\usepackage{graphicx}
\usepackage{txfonts}
\usepackage{xcolor}
\usepackage{multirow}
\usepackage[]{hyperref}
\usepackage{caption}

\newcommand{\tess}{\textit{TESS}\xspace}
\newcommand{\cheops}{\textit{CHEOPS}\xspace}

\newcommand{\harps}{\textit{HARPS}\xspace}
\newcommand{\Prot}{P_{\mathrm{rot},\star}}
\newcommand{\ttt}{\texttt{t}}
\newcommand{\ttxy}{(\texttt{xy})}
\newcommand{\ttdx}{\texttt{pc1}}
\newcommand{\ttdy}{\texttt{pc2}}
\newcommand{\pyde}[0]{\textsc{PyDE}}
\newcommand{\emcee}[0]{\textsc{emcee}}
\newcommand{\trades}[0]{\textsc{TRADES}}

\newcommand{\up}[2]{\ensuremath{\mathcal{U}(#1, #2)}}
\newcommand{\hgp}[2]{\ensuremath{\mathcal{H_{G}}(#1, #2)}}

\usepackage{chemformula}
\newcommand{\micron}{$\mu$m}
\newcommand{\microns}{\micron}

\usepackage[normalem]{ulem}

\renewcommand{\arraystretch}{1.25}

\begin{document}

\title{Radii, masses, and transit-timing variations of the three-planet system orbiting the naked-eye star TOI-396\thanks{Based on observations performed with the 3.6\,m telescope at the European Southern Observatory (La Silla, Chile) under programmes 1102.C-0923, 1102.C-0249, 0102.C-0584, and 60.A-9700.}}

\author{A.~Bonfanti\inst{1} $^{\href{https://orcid.org/0000-0002-1916-5935}{\includegraphics[scale=0.5]{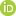}}}$
\and I.~Amateis\inst{2,1,3}
\and D.~Gandolfi\inst{2} $^{\href{https://orcid.org/0000-0001-8627-9628}{\includegraphics[scale=0.5]{figures/orcid.jpg}}}$
\and L.~Borsato\inst{4} $^{\href{https://orcid.org/0000-0003-0066-9268}{\includegraphics[scale=0.5]{figures/orcid.jpg}}}$
\and J.~A.~Egger\inst{5} $^{\href{https://orcid.org/0000-0003-1628-4231}{\includegraphics[scale=0.5]{figures/orcid.jpg}}}$
\and P.~E.~Cubillos\inst{1,6} $^{\href{https://orcid.org/0000-0002-1347-2600}{\includegraphics[scale=0.5]{figures/orcid.jpg}}}$
\and D.~Armstrong\inst{7,8} $^{\href{https://orcid.org/0000-0002-5080-4117}{\includegraphics[scale=0.5]{figures/orcid.jpg}}}$
\and I.~C.~Leão\inst{9} $^{\href{https://orcid.org/0000-0001-5845-947X}{\includegraphics[scale=0.5]{figures/orcid.jpg}}}$
\and M.~Fridlund\inst{10,11} $^{\href{https://orcid.org/0000-0002-0855-8426}{\includegraphics[scale=0.5]{figures/orcid.jpg}}}$
\and B.~L.~Canto~Martins\inst{9,12} $^{\href{https://orcid.org/0000-0001-5578-7400}{\includegraphics[scale=0.5]{figures/orcid.jpg}}}$
\and S.~G.~Sousa\inst{13} $^{\href{https://orcid.org/0000-0001-9047-2965}{\includegraphics[scale=0.5]{figures/orcid.jpg}}}$
\and J.~R.~De~Medeiros\inst{9} $^{\href{https://orcid.org/0000-0001-8218-1586}{\includegraphics[scale=0.5]{figures/orcid.jpg}}}$
\and L.~Fossati\inst{1} $^{\href{https://orcid.org/0000-0003-4426-9530}{\includegraphics[scale=0.5]{figures/orcid.jpg}}}$
\and V.~Adibekyan\inst{13} $^{\href{https://orcid.org/0000-0002-0601-6199}{\includegraphics[scale=0.5]{figures/orcid.jpg}}}$
\and A.~Collier~Cameron\inst{14} $^{\href{https://orcid.org/0000-0002-8863-7828}{\includegraphics[scale=0.5]{figures/orcid.jpg}}}$
\and S.~Grziwa\inst{15} $^{\href{https://orcid.org/0000-0003-3370-4058}{\includegraphics[scale=0.5]{figures/orcid.jpg}}}$
\and K.~W.~F.~Lam\inst{16} $^{\href{https://orcid.org/0000-0002-9910-6088}{\includegraphics[scale=0.5]{figures/orcid.jpg}}}$
\and E.~Goffo\inst{17} $^{\href{https://orcid.org/0000-0001-9670-961X}{\includegraphics[scale=0.5]{figures/orcid.jpg}}}$
\and L.~D.~Nielsen\inst{18} $^{\href{https://orcid.org/0000-0002-5254-2499}{\includegraphics[scale=0.5]{figures/orcid.jpg}}}$
\and F.~Rodler\inst{19} 
\and J.~Alarcon\inst{20}
\and J.~Lillo-Box\inst{21} $^{\href{https://orcid.org/0000-0003-3742-1987}{\includegraphics[scale=0.5]{figures/orcid.jpg}}}$
\and W.~D.~Cochran\inst{22,23} $^{\href{https://orcid.org/0000-0001-9662-3496}{\includegraphics[scale=0.5]{figures/orcid.jpg}}}$
\and R.~Luque\inst{24} $^{\href{https://orcid.org/0000-0002-4671-2957}{\includegraphics[scale=0.5]{figures/orcid.jpg}}}$
\and S.~Redfield\inst{25} $^{\href{https://orcid.org/0000-0003-3786-3486}{\includegraphics[scale=0.5]{figures/orcid.jpg}}}$
\and N.~C.~Santos\inst{13,26} $^{\href{https://orcid.org/0000-0003-4422-2919}{\includegraphics[scale=0.5]{figures/orcid.jpg}}}$
\and S.~C.~C.~Barros\inst{13,26} $^{\href{https://orcid.org/0000-0003-2434-3625}{\includegraphics[scale=0.5]{figures/orcid.jpg}}}$
\and D.~Bayliss\inst{7,8} $^{\href{https://orcid.org/0000-0001-6023-1335}{\includegraphics[scale=0.5]{figures/orcid.jpg}}}$
\and X.~Dumusque\inst{27} $^{\href{https://orcid.org/0000-0002-9332-2011}{\includegraphics[scale=0.5]{figures/orcid.jpg}}}$
\and M.~A.~F.~Keniger\inst{7,8} 
\and J.~Livingston\inst{28,29,30} $^{\href{https://orcid.org/0000-0002-4881-3620}{\includegraphics[scale=0.5]{figures/orcid.jpg}}}$
\and F.~Murgas\inst{31,32}~$^{\href{https://orcid.org/0000-0001-9087-1245}{\includegraphics[scale=0.5]{figures/orcid.jpg}}}$
\and G.~Nowak\inst{33} $^{\href{https://orcid.org/0000-0002-7031-7754}{\includegraphics[scale=0.5]{figures/orcid.jpg}}}$
\and A.~Osborn\inst{34} ${\href{https://orcid.org/0000-0002-5899-7750}{\includegraphics[scale=0.5]{figures/orcid.jpg}}}$
\and H.~P.~Osborn\inst{5,35} $^{\href{https://orcid.org/0000-0002-4047-4724}{\includegraphics[scale=0.5]{figures/orcid.jpg}}}$
\and E.~Pallé\inst{31,32} $^{\href{https://orcid.org/0000-0003-0987-1593}{\includegraphics[scale=0.5]{figures/orcid.jpg}}}$
\and C.~M.~Persson\inst{11} $^{\href{https://orcid.org/0000-0003-1257-5146}{\includegraphics[scale=0.5]{figures/orcid.jpg}}}$
\and L.~M.~Serrano\inst{2} $^{\href{https://orcid.org/0000-0001-9211-3691}{\includegraphics[scale=0.5]{figures/orcid.jpg}}}$
\and P.~A.~Strøm\inst{7,8}
\and S.~Udry\inst{27} $^{\href{https://orcid.org/0000-0001-7576-6236}{\includegraphics[scale=0.5]{figures/orcid.jpg}}}$
\and P.~J.~Wheatley\inst{7,8} $^{\href{https://orcid.org/0000-0003-1452-2240}{\includegraphics[scale=0.5]{figures/orcid.jpg}}}$
}

\institute{Austrian Academy of Sciences, Schmiedlstrasse 6, A-8042 Graz, Austria\\
          \email{andrea.bonfanti@oeaw.ac.at}
\and
Dipartimento di Fisica, Università degli Studi di Torino, via Pietro Giuria 1, I-10125, Torino, Italy
\and
Department of Physics and Astronomy, Uppsala University, Box 516, Uppsala SE-75120, Sweden
\and
INAF, Osservatorio Astronomico di Padova, Vicolo dell'Osservatorio 5, 35122 Padova, Italy
\and
Weltraumforschung und Planetologie, Physikalisches Institut, University of Bern, Gesellschaftsstrasse 6, 3012 Bern, Switzerland
\and
INAF, Osservatorio Astrofisico di Torino, Via Osservatorio, 20, I-10025 Pino Torinese To, Italy 
\and
Department of Physics, University of Warwick, Coventry CV4 7AL, UK
\and
Centre for Exoplanets and Habitability, University of Warwick, Gibbet Hill Road, Coventry CV4 7AL, UK
\and
Departamento de Física Teórica e Experimental, Universidade Federal do Rio Grande do Norte, Campus Universitário, Natal, RN, 59072-970, Brazil
\and
Leiden Observatory, University of Leiden, PO Box 9513, 2300 RA Leiden, The Netherlands
\and
Chalmers University of Technology, Department of Space, Earth and Environment, Onsala Space Observatory, SE-439 92 Onsala, Sweden
\and
INAF-Osservatorio Astrofisico di Arcetri, Largo E. Fermi 5 Florence, Italy
\and
Instituto de Astrof\'isica e Ci\^encias do Espa\c{c}o, Universidade do Porto, CAUP, Rua das Estrelas, 4150-762 Porto, Portugal
\and
School of Physics and Astronomy, Physical Science Building, North Haugh, St Andrews, United Kingdom
\and
Rhenish Institute for Environmental Research, Dep. of Planetary Research, University of Cologne, Aachener Str. 209, 50931 Cologne, Germany
\and
Institute of Planetary Research, German Aerospace Center (DLR), Rutherfordstrasse 2, D-12489 Berlin, Germany
\and
Thüringer Landessternwarte Tautenburg, Sternwarte 5, D-07778 Tautenburg, Germany
\and
University Observatory Munich, Ludwig-Maximilians-Universit\"{a}t, Scheinerstr. 1, 81679 Munich, Germany
\and
European Southern Observatory, Alonso de Cordova 3107, Vitacura, Casilla, 19001, Santiago, Chile
\and
European Southern Observatory, Alonso de Cordova 3107, Vitacura, Santiago de Chile, Chile
\and
Centro de Astrobiolog\'ia (CAB), CSIC-INTA, Camino Bajo del Castillo s/n, 28692, Villanueva de la Ca\~nada (Madrid), Spain
\and
McDonald Observatory, The University of Texas, Austin Texas USA
\and
Center for Planetary Systems Habitability, The University of Texas, Austin Texas
\and
Department of Astronomy \& Astrophysics, University of Chicago, Chicago, IL 60637, USA
\and
Astronomy Department and Van Vleck Observatory, Wesleyan University, Middletown, CT 06459, USA
\and
Departamento de F\'isica e Astronomia, Faculdade de Ci\^encias, Universidade do Porto, Rua do Campo Alegre, 4169-007 Porto, Portugal
\and
Department of Astronomy of the University of Geneva, chemin Pegasi 51, 1290 Versoix, Switzerland
\and
Astrobiology Center, NINS, 2-21-1 Osawa, Mitaka, Tokyo 181-8588, Japan
\and
National Astronomical Observatory of Japan, NINS, 2-21-1 Osawa, Mitaka, Tokyo 181-8588, Japan
\and
Astronomical Science Program, Graduate University for Advanced Studies, SOKENDAI, 2-21-1, Osawa, Mitaka, Tokyo, 181-8588, Japan
\and
Instituto de Astrof\'isica de Canarias (IAC), calle V\'ia L\'actea s/n, 38205 La Laguna, Tenerife, Spain
\and
Departamento de Astrof\'isica, Universidad de La Laguna (ULL), 38206 La Laguna, Tenerife, Spain
\and
Institute of Astronomy, Faculty of Physics, Astronomy and Informatics, Nicolaus Copernicus University, Grudzi\c{a}dzka 5, 87-100 Toru\'n, Poland
\and
Department of Physics and Astronomy, McMaster University, 1280 Main St W, Hamilton, ON, L8S 4L8, Canada
\and
Department of Physics, ETH Zurich, Wolfgang-Pauli-Strasse 2, CH-8093 Zurich, Switzerland}
\date{}

  \abstract
   {TOI-396 is an F6\,V bright naked-eye star ($V$\,$\approx$\,6.4) orbited by three small ($R_p$\,$\approx$\,2~$R_\oplus$) transiting planets discovered thanks to space-based photometry from two \tess sectors. The orbital periods of the two innermost planets, namely TOI-396\,b and c, are close to the 5:3 commensurability ($P_b$\,$\sim$\,3.6\,d and $P_c$\,$\sim$\,6.0\,d), suggesting that the planets might be trapped in a mean motion resonance (MMR).}
   {To measure the masses of the three planets, refine their radii, and investigate whether planets b and c are in MMR, we carried out \harps radial velocity (RV) observations of TOI-396 and retrieved archival high-precision transit photometry from four \tess\ sectors.} 
   {We extracted the RVs via a skew-normal fit onto the \harps cross-correlation functions and performed a Markov chain Monte Carlo joint analysis of the Doppler measurements and transit photometry while employing the breakpoint method to remove stellar activity from the RV time series. We also performed a transit timing variation (TTV) dynamical analysis of the system and simulated the temporal evolution of the TTV amplitudes of the three planets following an N-body numerical integration.}
   {Our analysis confirms that the three planets have similar sizes ($R_b=2.004_{-0.047}^{+0.045}\,R_{\oplus}$; $R_c=1.979_{-0.051}^{+0.054}\,R_{\oplus}$; $R_d=2.001_{-0.064}^{+0.063}\,R_{\oplus}$) and is thus in agreement with previous findings. However, our measurements are $\sim$\,1.4 times more precise thanks to the use of two additional \tess sectors. For the first time, we have determined the RV masses for TOI-396\,b and d, finding them to be $M_b=3.55_{-0.96}^{+0.94}\,M_{\oplus}$ and $M_d=7.1\pm1.6\,M_{\oplus}$, which implies bulk densities of $\rho_b=2.44_{-0.68}^{+0.69}$ g\,cm$^{-3}$ and $\rho_d=4.9_{-1.1}^{+1.2}$ g\,cm$^{-3}$, respectively. Our results suggest a quite unusual system architecture, with the outermost planet being the densest. Based on a frequency analysis of the \harps\ activity indicators and \tess light curves, we find the rotation period of the star to be $\Prot=6.7\pm1.3$~d, in agreement with the value predicted from $\log{R'_{\mathrm{HK}}}$-based empirical relations. The Doppler reflex motion induced by TOI-396\,c remains undetected in our RV time series, likely due to the proximity of the planet's orbital period to the star's rotation period. We also discovered that TOI-396 b and c display significant TTVs. While the TTV dynamical analysis returns a formally precise mass for TOI-396\,c of $M_{c,\mathrm{dyn}}=2.24^{+0.13}_{-0.67}\,M_{\oplus}$, the result might not be accurate, owing to the poor sampling of the TTV phase. We also conclude that TOI-396\,b and c are close to but out of the 5:3 MMR.}
   {A TTV dynamical analysis of additional transit photometry evenly covering the TTV phase and super-period is likely the most effective approach for precisely and accurately determining the mass of TOI-396\,c. Our numerical simulation suggests TTV semi-amplitudes of up to 5 hours over a temporal baseline of $\sim$\,5.2 years, which should be duly taken into account when scheduling future observations of TOI-396.}

   \keywords{planets and satellites: fundamental parameters --
             stars: fundamental parameters --
             techniques: photometric --
             techniques: radial velocities
             }

   \maketitle
%
%________________________________________________________________

\section{Introduction}
Multi-planet systems enable us to place significantly stronger constraints on formation and evolution mechanisms compared to single-planet systems \citep[e.g.][]{lissauer2011,fabrycky2014,winn2015,mishra2023}. As a matter of fact, the temporal evolution of the gas content in the proto-planetary disc influences planet migration \citep[e.g.][]{lin1979,goldreich1980,tanaka2002,dangelo2008,alexander2009}, which shapes the orbital architecture of a planetary system. The latter is further expected to correlate with the planet composition \citep[e.g.][]{thiabaud2014,thiabaud2015,walsh2015,bergner2020,li2020} that can be inferred once the physical parameters of the planets are known \citep[e.g.][]{dorn2017,otegi2020internalStructure,leleu2021,haldemann2024}.

If planets are found in mean motion resonance \citep[MMR; e.g.][]{lee2002,correia2018}, systems can also shed light on the migration mechanisms during formation, as well as on the impact of tidal effects occurring later on \citep[e.g.][]{delisle2012,izidoro2017}. In addition, planets in, or close to, MMR likely exhibit transit timing variations \citep[TTVs; see e.g.][]{agol2005,AgolFabrycky2018,leleu2021} that enable one to infer planetary masses without necessarily relying on radial velocity (RV) measurements, which are not always possible \citep[e.g.][]{hatzes2016}.

Multi-planet systems also give insights into correlations between physical and orbital parameters of exoplanets. For example, \citet{weiss2018} noticed that planets in adjacent orbits usually show similar radii, hence the ``peas in a pod'' label to describe this scenario. \citet{weiss2018} further noticed that the outermost planet is the largest in the majority of cases, which agrees with the observed bulk planet density ($\rho_p$) trend highlighted by \citet{mishra2023}, where $\rho_p$ decreases with the distance from the stellar host as outer planets are expected to be richer in volatiles (thus bigger and less dense).

The object TOI-396 represents an interesting laboratory to test these theories as it is the brightest star known so far to host three transiting planets \citep{Vanderburg2019} after $\nu^2$\,Lupi \citep{delrez2021}. After analysing two sectors from the Transiting Exoplanet Survey Satellite \citep[\tess;][]{ricker2015}, \citet{Vanderburg2019} found that the three planets have radii of $\sim$\,2\,$R_{\oplus}$ and orbital periods of $\sim$\,3.6, $\sim$\,6.0, and $\sim$\,11.2 d, with TOI-396\,c and b showing a period commensurability close to the 5:3 ratio. Following the notation introduced in \citet{mishra2023}, the three planets are `similar' in terms of radii, and one may wonder whether this architecture class is kept also in the mass-period diagram. \citet{mishra2023} found a positive and strong correlation of the coefficients of similarity between radii and masses, though exceptions are possible \citep[e.g.][]{weiss2014,otegi2020MR,otegi2022}.

In this work, we complement the photometric analysis of new \tess light curves (LCs) with RV observations taken with the High Accuracy Radial Velocity Planet Searcher \citep[\harps;][]{mayor2003} spectrograph to refine the planet radii and constrain for the first time the planetary masses. Considering the possible 5:3 MMR between TOI-396\,c and b, we also simulate the evolution in time of the TTV amplitudes.
This paper is organised as follows: Section~\ref{sec:star} presents the stellar properties, and Sect.~\ref{sec:data} describes the photometric and RV data. After outlining the method to jointly analyse the \tess LCs and the \harps RV time series in Sect.~\ref{sec:methods}, we present the corresponding results in Sect.~\ref{sec:results}. 
We attempt to dynamically model TTV and RV data simultaneously and track the temporal evolution of the TTV signals in Sect.~\ref{sec:dynamicTTV}, and we study the planets' internal structure in Sect.~\ref{sec:internalStructure} and explore the prospects for characterising the system with the James Webb Space Telescope \citep[JWST;][]{gardner2006} in Sect.~\ref{sec:JWST}. Finally Sect.~\ref{sec:conclusions} gathers the conclusions.

\section{Host star characterisation}\label{sec:star}
TOI-396 is an F6\,V \citep{gray2006} bright naked-eye star with an apparent visual magnitude of $V$\,$\approx$\,6.4 \citep{perryman1997}. It is located $\sim$\,31.7 pc away and is visible in the constellation of Fornax in the southern hemisphere. It is member of a visual binary system and its companion HR\,858\,B is a faint M-dwarf ($G$\,$\sim$\,16 mag), about 8.4\arcsec away from the main component.

We co-added 78 \harps spectra (see Sect.~\ref{sec:HARPS_RV} for further details) and then modelled it with   
Spectroscopy Made Easy\footnote{\url{http://www.stsci.edu/~valenti/sme.html}.} 
\citep[\texttt{SME};][]{Piskunov2017} version 5.2.2. \texttt{SME} computes synthetic spectra from a grid of well established stellar atmosphere models and adjusts a chosen free parameter based on comparison with the observed spectrum. Here we used the stellar atmosphere grid \textsc{Atlas12} \citep{Kurucz2013} together with atomic line lists from the \textsc{Vald} database \citep{Piskunov95} in order to produce the synthetic spectra. We modelled one parameter at a time utilising spectral features sensitive to different photospheric parameters iterating until convergence of all free parameters. Throughout the modelling, we held the macro- and micro-turbulent velocities, $v_{\rm mac}$ and $v_{\rm mic}$, fixed to 6~km\,s$^{-1}$ \citep{Doyle2014} and 1.34~km\,s$^{-1}$ \citep{bruntt2010}, respectively.
A description of the modelling procedure is detailed in \citet{persson2018}. 
Finally, we obtained $T_{\mathrm{eff}}=6354\pm70$ K, $\mathrm{[Fe/H]}=0.025\pm0.050$, $\log{g}=4.30\pm0.06$, and $v\sin{i_{\star}}=7.5\pm0.2$~km\,s$^{-1}$.

To double-check the derived spectroscopy parameters we performed an additional analysis employing ARES+MOOG \citep[][]{Sousa-21, Sousa-14, Santos-13}. In detail, we used the latest version of ARES \footnote{The last version, ARES v2, can be downloaded at \url{https://github.com/sousasag/ARES}.} \citep{Sousa-07, Sousa-15} to consistently measure the equivalent widths (EW) for the list of iron lines presented in \citet[][]{Sousa-08}. Following a minimisation process, we then find the ionisation and excitation equilibria to converge for the best set of spectroscopic parameters. This process uses a grid of Kurucz model atmospheres \citep{Kurucz-93} and the radiative transfer code MOOG \citep{Sneden-73}. We obtained $T_{\mathrm{eff}}=6389\pm67$ K, $\mathrm{[Fe/H]}=-0.014\pm0.045$ dex, $\log{g}=4.58\pm0.11$, and $v_\mathrm{mic}= 1.54\pm0.04$ km\,s$^{-1}$. In this process we also derived a more accurate trigonometric surface gravity ($\log{g_{\mathrm{trig}}}=4.34\pm0.02$) using recent GAIA data following the same procedure as described in \citet[][]{Sousa-21}. In the end ARES+MOOG provides consistent values when compared with the ones derived with \texttt{SME}.

Using the \texttt{SME} stellar atmospheric parameters, we determined the abundances of Mg and Si following the classical curve-of-growth analysis method described in \citet{Adibekyan-12, Adibekyan-15}. Similar to the stellar parameter determination, we used ARES to measure the EWs of the spectral lines of these elements and a grid of Kurucz model atmospheres \citep{Kurucz-93} along with the radiative transfer code MOOG to convert the EWs into abundances, assuming local thermodynamic equilibrium.

The stellar radius $R_{\star}$, mass $M_{\star}$, and age $t_{\star}$ were derived homogeneously using the isochrone placement algorithm \citep{bonfanti2015,bonfanti2016} and its capability of interpolating a flexible set of input parameters within pre-computed grids of PARSEC\footnote{\url{http://stev.oapd.inaf.it/cgi-bin/cmd}.} v1.2S \citep{marigo2017} isochrones and tracks. For each magnitude listed in Table~\ref{tab:star}, we performed an isochrone placement run by inserting the spectroscopic parameters derived above, the Gaia parallax $\pi$ \citep[][offset-corrected following \citealt{lindegren2021}]{GaiaDR3-2023}, and the magnitude value to obtain an estimate for the stellar radius, mass, and age along with their uncertainties. From these results, we built the corresponding Gaussian probability density functions (PDFs) and then we merged (i.e. we summed) the PDFs derived from the different runs to obtain robust estimates for $R_{\star}$, $M_{\star}$, and $t_{\star}$. The radius $R_{\star}$ derives essentially from $T_{\mathrm{eff}}$, $\pi$, and the stellar magnitude, while $M_{\star}$ and $t_{\star}$ are much more model-dependent; therefore we conservatively inflated their internal uncertainties by $4\%$ and $20\%$, respectively, following \citet{bonfanti2021}. Our adopted stellar parameters are listed in Table~\ref{tab:star}.

\begin{table}
\caption{Stellar properties of TOI-396.}
\label{tab:star}
\centering
\begin{tabular}{l c c}
\hline\hline   
\noalign{\smallskip}
\multicolumn{1}{l}{\multirow{6}*{Star names}} & \multicolumn{2}{l}{TOI-396} \\
\multicolumn{1}{l}{}               & \multicolumn{2}{l}{TIC 178155732} \\
\multicolumn{1}{l}{}               & \multicolumn{2}{l}{HR 858} \\
\multicolumn{1}{l}{}               & \multicolumn{2}{l}{HD 17926} \\
\multicolumn{1}{l}{}               & \multicolumn{2}{l}{HIP 13363} \\
\multicolumn{1}{l}{}               & \multicolumn{2}{l}{Gaia DR3 5064574724769475968} \\
\noalign{\smallskip}
\hline
\noalign{\smallskip}
Parameter & Value & Source \\
\noalign{\smallskip}
\hline
\noalign{\smallskip}
  RA\; [h:min:s] & 02:51:56.25 & Gaia DR3 \\
  Dec\; [$^\circ$:$\arcmin$:$\arcsec$] & $-$30:48:52.26 & Gaia DR3 \\
  $B$ & $6.862\pm0.015$ & \citet{perryman1997} \\
  $V$ & $6.382\pm0.010$ & \citet{perryman1997} \\
  $B_T$ & $6.956\pm0.015$ & \citet{hog2000} \\
  $V_T$ & $6.438\pm0.010$ & \citet{hog2000} \\
  $J$ & $5.473\pm0.027$ & \citet{cutri2003} \\
  $H$ & $5.225\pm0.034$ & \citet{cutri2003} \\
  $K$ & $5.149\pm0.020$ & \citet{cutri2003} \\
  $G$ & $6.2650\pm0.0028$ & Gaia DR3 \\
  $G_{\mathrm{RP}}$ & $6.5100\pm0.0029$ & Gaia DR3 \\
  $G_{\mathrm{BP}}$ & $5.8555\pm0.0038$ & Gaia DR3 \\
  $\pi$\; [mas] & $31.564\pm0.035$ & Gaia DR3\tablefootmark{(a)} \\
\noalign{\smallskip}
\hline
\noalign{\smallskip}
  $T_{\mathrm{eff}}$\; [K] & $6354\pm70$ & Spectroscopy \\\relax
  [Fe/H]                   & $0.025\pm0.050$ & Spectroscopy \\\relax
  [Mg/H]                   & $-0.01\pm0.07$ & Spectroscopy \\\relax
  [Si/H]                   & $-0.03\pm0.05$ & Spectroscopy \\
  $\log{g}$\; [cgs]        & $4.30\pm0.06$ & Spectroscopy \\
  $v\sin{i_{\star}}$\; [km\,s$^{-1}$] & $7.5\pm0.2$ & Spectroscopy \\
  log\,R$^\prime_\mathrm{HK}$ & $-4.926\pm0.014$ & Spectroscopy \\
  $\Prot$ \;[d] & $6.7\pm1.3$ & log\,R$^\prime_\mathrm{HK}$ \\
  $R_{\star}$\; [$R_{\odot}$] & $1.258\pm0.019$ & $\pi$ \& photometry\tablefootmark{(b)} \\
  $M_{\star}$\; [$M_{\odot}$] & $1.204\pm0.052$ & Isochrones \\
  $t_{\star}$\; [Gyr] & $2.0\pm0.6$ & Isochrones \\
  $L_{\star}$\; [$L_{\odot}$] & $2.31\pm0.12$ & From $T_{\mathrm{eff}}$ \& $R_{\star}$ \\
  $\rho_{\star}$\; [$\rho_{\odot}$] & $0.605\pm0.038$ & From $M_{\star}$ \& $R_{\star}$ \\
\noalign{\smallskip}
\hline                        
\end{tabular}
\tablefoot{RA \& DEC are reported as in J2000 reference frame. Values in the bottom half of the table have been derived as part of this paper. \\ \tablefoottext{a}{Zero-point correction from \citet{lindegren2021} applied.} \\ \tablefoottext{b}{$R_{\star}$ is computed within the isochrone placement interpolation scheme as PARSEC isochrones convert stellar luminosity into absolute magnitudes via a series of bolometric correction tables built as explained in \citet{marigo2017} and references therein.}}
\end{table}

\section{Observational data}\label{sec:data}

\subsection{\tess\ photometry}

As presented in \citet{Vanderburg2019}, TOI-396 was photometrically monitored by \tess during the first year of its nominal mission in Sector 3 from 20 September to 17 October 2018 (UT) in CCD~2 of Camera~2, and in Sector 4 from 19 October to 14 November 2018 (UT) in CCD~2 of Camera~1. TOI-396 was later re-observed by \tess during the first year of its extended mission in Sector 30 from 23 September to 20 October 2020 (UT) in CCD~2 of Camera~2, and in Sector 31 from 22 October to 16 November 2020 (UT) in CCD~2 of Camera~1. All data were collected at a 2-minute cadence, except for Sector 3 for which \tess only sent down data at 30-minute cadence.

We analyse all available \tess time series, including the Sector 3 and 4 data already presented in \citep{Vanderburg2019}. For Sector 3 we used the \tess Asteroseismic Science Operation Center (TASOC) photometry \citep{handberg2021,lund2021}, while for the other sectors we analysed the Presearch Data Conditioning Simple Aperture Photometry (PDCSAP) LCs generated by the \tess Science Processing Operation Center (SPOC) pipeline \cite{Jenkins2016}, which removes the majority of instrumental artefacts and systematic trends \citep{Smith2012,Stumpe2012,Stumpe2014}. 
We rejected those data marked with a bad quality flag and we performed a 5 median-absolute-deviation (MAD) clipping of flux data points. After that, we built our custom LCs by splitting the \tess sectors in temporal windows centred around the transit events keeping $\sim$\,4 h of out-of-transit data points both before and after the transit for de-trending purposes. We ended up with 41 \tess LCs reporting the epoch of observation (\texttt{t}), the flux and its error, and further ancillary vectors available from the \tess science data product, that is \textsc{mom\_centr1}, \textsc{mom\_centr2} (hereafter denoted with \texttt{x} and \texttt{y}, respectively), \textsc{pos\_corr1}, and \textsc{pos\_corr2} (hereafter denoted with \texttt{pc1} and \texttt{pc2}, respectively)\footnote{Please refer to \url{https://tasoc.dk/docs/EXP-TESS-ARC-ICD-TM-0014-Rev-F.pdf} for further details about the output of the \tess science data products.}.

\subsection{\harps\ high-resolution spectroscopy}\label{sec:HARPS_RV}

We performed the radial velocity (RV) follow-up of TOI-396 with the \harps spectrograph mounted at the ESO-3.6\,m telescope at La Silla Observatory (Chile). We acquired 77 high resolution ($R$\,$\approx$\,115\,000) spectra between 31 January and 27 July 2019 (UT), covering a baseline of $\sim$\,177 days, as part of the follow-up programs of \tess\ systems carried out with the \harps\ spectrograph (IDs: 1102.C-0923, 1102.C-0249, 0102.C-0584; PIs: Gandolfi, Armstrong, De Medeiros). One additional spectrum was acquired during a technical night (ID: 60.A-9700) in February 2019. Following \citet{Dumusque2011b}, we averaged out p-mode stellar pulsations by setting the exposure time to 900\,s, which led to a median signal-to-noise (S/N) ratio of $\sim$320 per pixel at 550\,nm. We used the second fibre of the \harps\ spectrograph to simultaneously observe a Fabry-Perot lamp and trace possible instrumental drift down to the sub-metre per second level \citep{wildi2010}. We reduced the data using the dedicated \harps\ data reduction software \citep[\texttt{DRS};][]{Pepe2002,Lovis2007} and computed the cross-correlation function (CCF) for each spectrum using a G2 numerical mask \citep{Baranne1996}.

We then performed a skew-normal (SN) fit on each CCF \citep{simola2019} in orderto extract the stellar radial velocity along with its error, the full width at half maximum (FWHM$_{\mathrm{SN}}$), the contrast ($A$), and the skewness parameter ($\gamma$) of the CCF. The advantages of using an SN-fit rather than a Normal fit are thoroughly discussed in \citet{simola2019}, while the SN-fitting details are outlined in, for example, \citet{bonfanti2023,luque2023,fridlund2024}. After the SN-based extraction, we ended up with an RV time series, whose ancillary vectors (FWHM$_{\mathrm{SN}}$, $A$, $\gamma$) are the activity indicators used to constrain the polynomial basis to model and de-trend the RV component of the stellar activity (see Table~\ref{tab:RVdata}).

As stellar activity is not stationary, the correlations between the RV observations and the activity indicators are expected to change over time as discussed in \citet{Simola2022}, who proposed to apply the breakpoint (\textit{bp}) technique \citep{BaiPerron2003} to check whether these correlation changes are statistically significant. If so, the \textit{bp} algorithm finds the optimal locations of correlation changes, which defines the segmentation characterised by the lowest Bayesian Information Criterion \citep[BIC;][]{schwarz1978}. Indeed, we found one breakpoint at observation 48 (BJD = 2458667.940719; $\Delta\mathrm{BIC}\approx-14$ with respect to the zero-breakpoint solution). Thus, we split our RV time series in two piecewise stationary segments and de-trended it on a chunk-wise base, rather than performing a global de-trending over the whole time series, similarly to what has already been done in \citet{bonfanti2023,luque2023}. 

\section{Methods}\label{sec:methods}
We jointly analysed the \tess LCs and the RV time series within a Markov chain Monte Carlo (MCMC) framework using the MCMCI code \citep{bonfanti2020}. When fitting the \tess LCs extracted from Sector 3 that has a cadence of 30 min, we generated the transit model using a cadence of 2 min (the same as the other \tess sectors) and then rebinned it to 30 min following \citet{kipping2010}, who warns that long-cadence photometry may lead to retrieve erroneous system parameters.
We imposed Normal priors on the input stellar parameters (i.e. $T_{\mathrm{eff}}$, [Fe/H], $R_{\star}$, and $M_{\star}$), as derived in Sect.~\ref{sec:star} with a twofold aim: (i) the induced prior on the mean stellar density $\rho_{\star}$ (via $M_{\star}$ and $R_{\star}$) helps the convergence of the transit model; (ii) limb darkening (LD) coefficients for the \tess filter may be retrieved following interpolation within \textsc{Atlas9}-based\footnote{\url{http://kurucz.harvard.edu/grids.html}.} grids that were pre-computed using the \texttt{get\_lds.py} code\footnote{\url{https://github.com/nespinoza/limb-darkening}} by \citet{espinoza2015}. We then set Normal priors on the quadratic LD coefficients using the values coming from the grid interpolation (i.e. $u_{1,\tess}=0.2318\pm0.0065$ and $u_{2,\tess}=0.3085\pm0.0028$) after re-parameterising them following \citet{holman2006}.
On the planetary side, we adopted unbounded (except for the physical limits) uniform priors on the transit depth d$F$, the impact parameter $b$, the orbital period $P$, the transit timing $T_0$, and the RV semi-amplitude $K$, while we set the eccentricity $e=0$ and the argument of periastron $\omega=90^{\circ}$ for all planets. We come back to the assumption on the eccentricity below. The specific parameterisations of the jump parameters (aka step parameters) are outlined in \citet{bonfanti2020}.

The LCs and the RV time series were de-trended simultaneously during the MCMC analysis using polynomials. To assess the polynomial orders to be associated with the different de-trending parameters for each time series, we first launch several MCMC preliminary runs made of 10\,000 steps where we varied only one polynomial order at a time. We then selected the best de-trending baseline (see Table~\ref{tab:polyDetrending}) as the one having the lowest BIC. After that, we performed a preliminary MCMCI run comprising 200\,000 steps to evaluate the contribution of both the white and red noise in the LCs following \citet{pont2006,bonfanti2020}, so to properly rescale the photometric errors and get reliable uncertainties on the output parameters. Finally, three independent MCMCI runs comprising 200\,000 steps each (burn-in length equal to 20\%) were performed to build the posterior distributions of the output parameters after checking their convergence via the Gelman-Rubin statistic \citep[$\hat{R}$;][]{gelman1992}. 

We also tested the possibility of eccentric orbits by imposing uniform priors on ($\sqrt{e}\cos{\omega}$, $\sqrt{e}\cos{\omega}$) either bounded to imply $e\lesssim0.3$ or completely unbounded (except for the physical limits). The wider the eccentricity range to be explored by the MCMC scheme, the poorer the parameter convergence, which suggests that the available data are not enough to constrain the planetary eccentricities well. Moreover, the MCMCI runs with $e\neq0$ are disfavoured by the $\Delta$BIC criterion \citep[e.g.][]{kass95,trotta2007} as well, in fact we obtained $\Delta\mathrm{BIC}=\mathrm{BIC}_{e\neq0}-\mathrm{BIC}_{e=0}\gtrsim+100$. This is also in agreement with the simulations performed by \citet{Vanderburg2019}, who suggested that the periods' commensurability state of planets b and c is more likely maintained if the system is characterised by low eccentricities. Therefore, we adopted the circular solution as the reference one.

\section{LC and RV data analysis results}\label{sec:results}
\subsection{Joint LC and RV analysis with linear ephemerides}\label{sec:linearEphem}
As mentioned in Sect.~\ref{sec:methods}, we set $P$ and $T_0$ as free parameters under the control of a uniform prior, which implies assuming linear ephemerides. With this setup, we improved the transit depth precision of all three planets by a factor $\sim$\,1.4 if compared with the results of \citet{Vanderburg2019}. This improvement level is consistent with having twice the number of data points with respect to the LC analysis performed by \citet{Vanderburg2019}, as well as with low TTV amplitudes (see Sect.~\ref{sec:TTV}).

By combining the SN-fit onto the \harps CCFs along with the \textit{bp} method, we were able to estimate the masses of TOI-396\,b and TOI-396\,d to $M_b=3.56_{-0.94}^{+0.92}\,M_{\oplus}$ and $M_d=7.2\pm1.6\,M_{\oplus}$ (detections at the 3.8 and 4.5$\sigma$-level, respectively). Instead, we did not detect any significant Keplerian signal at the orbital period of TOI-396\,c within the RV time series. In detail, we obtained a median $K_c=0.28_{-0.20}^{+0.29}$ m\,s$^{-1}$ (3$\sigma$ upper limit $K_c^{\mathrm{up}}=1.2$ m\,s$^{-1}$), which implies $M_c=0.92_{-0.63}^{+0.94}\,M_{\oplus}$ ($M_c^{\mathrm{up}}=4.0\,M_{\oplus}$).

When combining the mass and radius values of the three planets, we obtain the following median estimates for the bulk planetary densities: $\rho_b=2.56_{-0.70}^{+0.71}$, $\rho_c=0.67_{-0.46}^{+0.69}$ ($\rho_c^{\mathrm{up}}=3.1$), and $\rho_d=5.1_{-1.2}^{+1.3}$ g\,cm$^{-3}$. We note that the RV-undetected TOI-396\,c would be the least dense planet, while the densest planet is the outermost one (i.e. TOI-396\,d), which constitutes a quite atypical architecture within the observed exoplanet population \citep[e.g.][]{ciardi2013,weiss2018,mishra2023}.
However, this conclusion is just tentative, given the uncertainties on the mean planetary densities. Moreover, the detection level of the RV-inferred parameters of TOI-396\,c is not statistically significant. We further tested whether including the MINERVA-Australis \citep{addison2019} RV data \citep[30 measurements as taken from][]{Vanderburg2019} can help detect the elusive planet. However, it turned out that their precision level ($\sim$\,6\,m\,s$^{-1}$) is not high enough to improve the characterisation of the system. In other words, the RV semi-amplitudes we obtained are consistent and indistinguishable within the statistical fluctuation with what was derived from the more precise \harps data set. This led us to further check that TOI-396\,c indeed belongs to this system (Sect.~\ref{sec:TTV}) and to investigate the reason for its RV non-detection (Sect.~\ref{sec:M_planet_c?}).

\subsection{Joint LC and RV analysis accounting for TTVs}\label{sec:TTV}
As planet c is undetected in the RV time series, one may wonder whether TOI-396\,c is a false positive. However, by using the \texttt{VESPA} tool \citep{morton2012,morton2015} that accounts for the constraints from the \tess LCs, imaging, and spectroscopy, \citet{Vanderburg2019} already computed that the false-positive probabilities (FPPs) are lower than $10^{-3}$ for all three planets.

\begin{figure}
    \centering
    \includegraphics[width=\columnwidth]{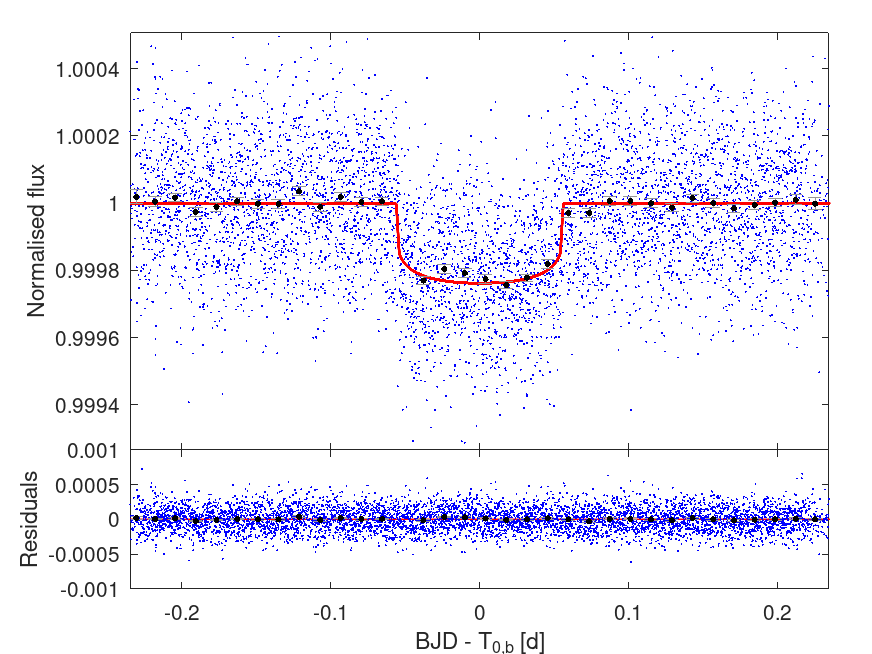} \\
    \includegraphics[width=\columnwidth]{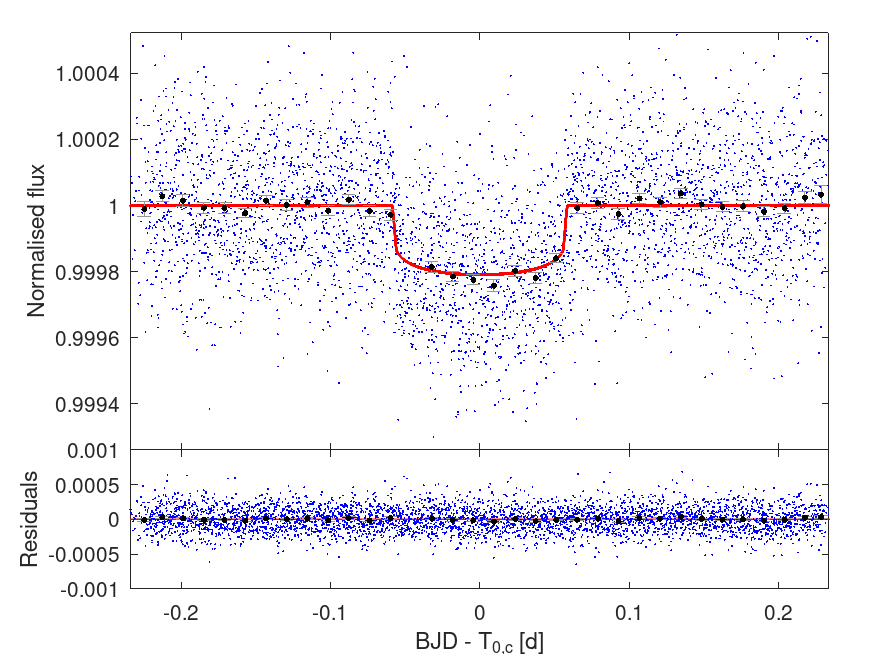} \\
    \includegraphics[width=\columnwidth]{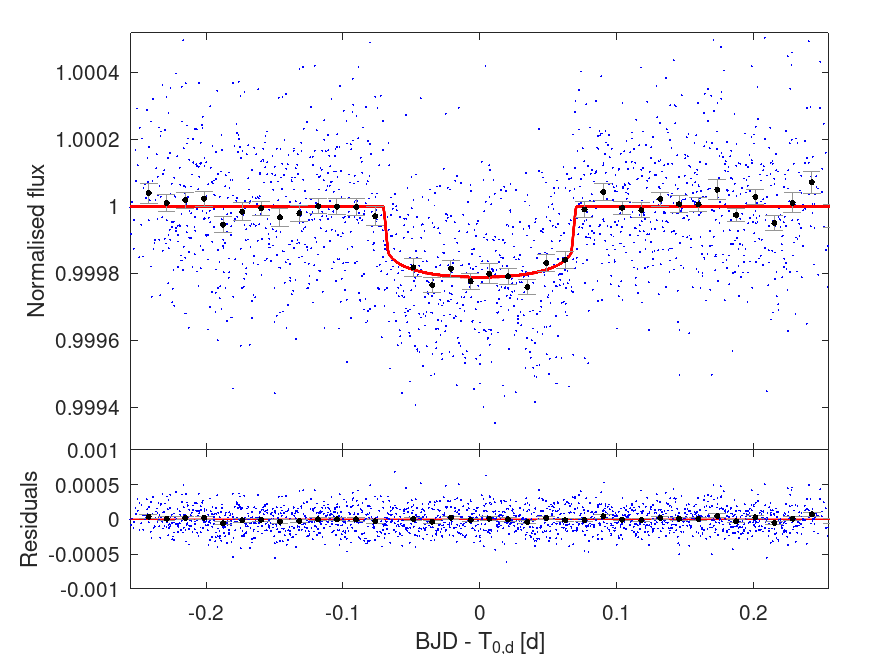}
    \caption{\tess detrended and phase-folded LCs (blue dots) of TOI-396\,b (\textit{Top panel}), TOI-396\,c (\textit{Middle panel}), and TOI-396\,d (\textit{Bottom panel}) with the transit model superimposed in red. The black markers are the binned data points (binning 20 min).}
    \label{fig:LCs}
\end{figure}

\begin{figure}[h!]
    \centering
    \includegraphics[width=\columnwidth]{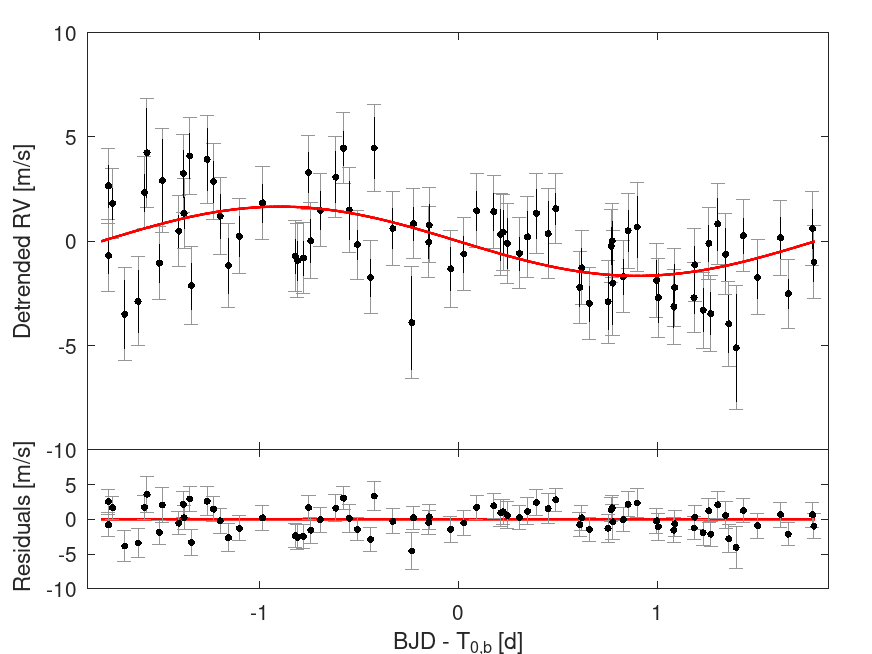} \\
    \includegraphics[width=\columnwidth]{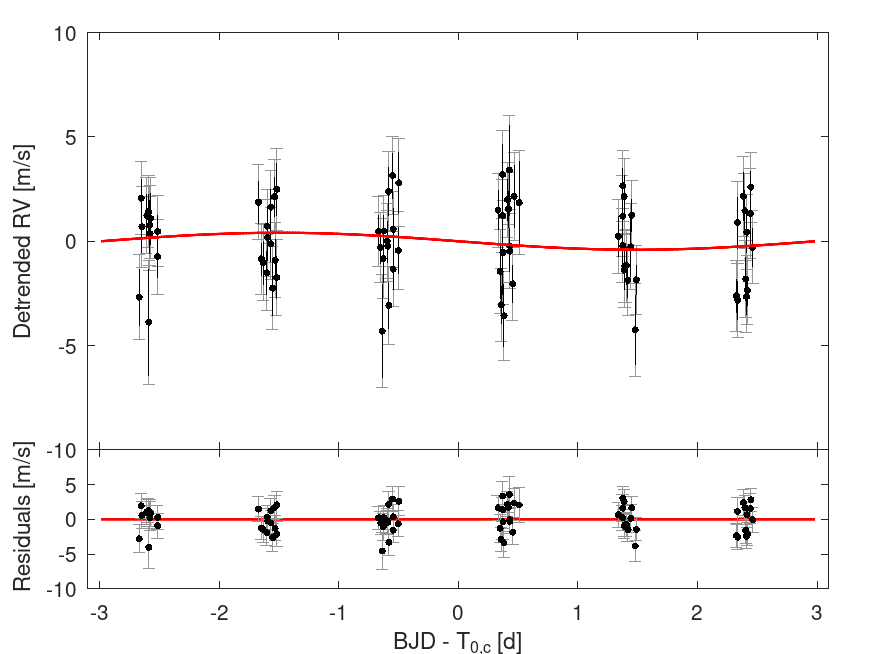} \\
    \includegraphics[width=\columnwidth]{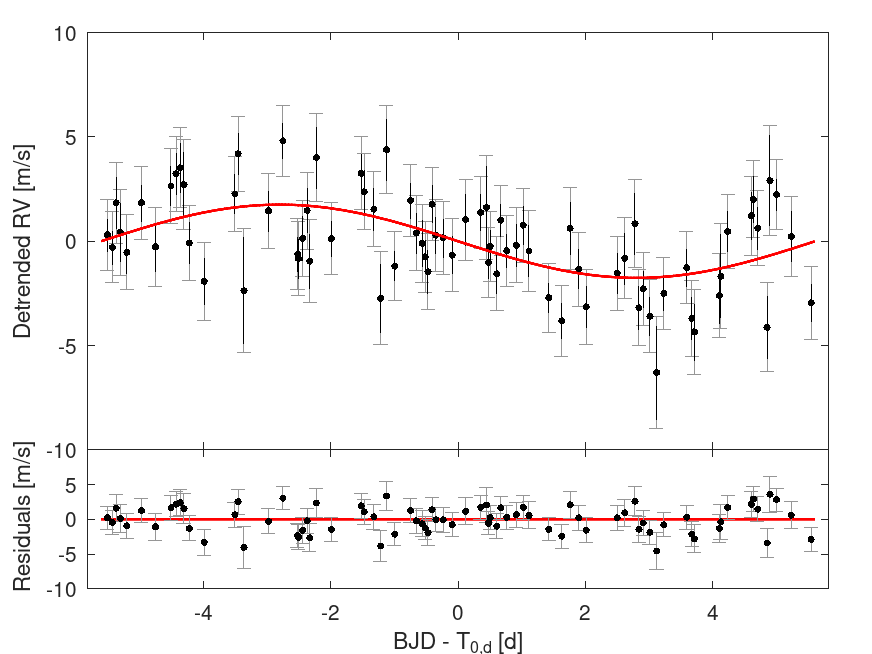}
    \caption{\harps detrended and phase-folded RV time series of TOI-396\,b (\textit{Top panel}), TOI-396\,c (\textit{Middle panel}), and TOI-396\,d (\textit{Bottom panel}) with the Keplerian model superimposed in red. For each planet, the time series were obtained after subtracting the RV contribution of the other planets. The error bars also account for the jitter contribution (displayed in grey).}
    \label{fig:RVs}
\end{figure}

In line with \citet{Vanderburg2019}, we also confirmed that the $P_c/P_b$ period ratio is commensurable and differs from the 5:3 ratio by less than 0.027\%. As planets with orbits in, or close to, resonances are likely to show TTVs, we decided to repeat the same MCMC analysis outlined above, but enabling the transit timings of each transit event to vary to then compute the TTV amplitude with respect to the linear ephemerides model derived in Sect.~\ref{sec:linearEphem}.
All jump parameters converged ($\hat{R}\lesssim1.01$). The medians of the posterior distributions of the most relevant system parameters along with the 68.3\% confidence intervals are listed in Table~\ref{tab:planets}. The phase-folded LCs of the three planets are shown in Figure~\ref{fig:LCs}, while the phase-folded RV time series are displayed in Figure~\ref{fig:RVs}. In particular, the middle panel of Fig.~\ref{fig:RVs} shows that, after subtracting the RV signals of both planets b and d, the RV time series looks flat consistently with the non-detection of TOI-396\,c.

\begin{table*}
\caption{System parameters as derived from the joint LC and RV analysis accounting for TTVs.}
\label{tab:planets}
\centering
\begin{tabular}{lllll}
\hline\hline
Parameter & Symbol & TOI-396\,b & TOI-396\,c\,\tablefootmark{(c)} & TOI-396\,d \\
\hline
\noalign{\smallskip}
Orbital period & $P$\,\tablefootmark{(a)} \;[d] & $3.585287_{-0.000012}^{+0.000009}$ & $5.973865_{-0.000016}^{+0.000015}$ & $11.230511_{-0.000045}^{+0.000043}$ \\
\noalign{\smallskip}
Transit timing & $T_0$ \tablefootmark{(a)} \;[BJD] & $8409.1900_{-0.0012}^{+0.0011}$ & $8415.63344_{-0.00093}^{+0.00099}$ & $8409.7324_{-0.0025}^{+0.0028}$ \\
\noalign{\smallskip}
Transit depth & d$F$ \;[ppm] & $213.7_{-6.4}^{+6.5}$ & $208.7_{-8.3}^{+8.5}$ & $213_{-11}^{+12}$ \\
\noalign{\smallskip}
Impact parameter & $b$ \; & $0.587_{-0.028}^{+0.025}$ & $0.701_{-0.019}^{+0.017}$ & $0.712_{-0.023}^{+0.026}$ \\
\noalign{\smallskip}
RV semi-amplitude & $K$ \;[m\,s$^{-1}$] & $1.30_{-0.35}^{+0.34}$ & $0.28_{-0.19}^{+0.29}$  & $1.78\pm0.40$ \\
\noalign{\smallskip}
Transit duration & $W$ \;[h] & $2.705_{-0.034}^{+0.038}$ & $2.843_{-0.043}^{+0.046}$ & $3.47_{-0.13}^{+0.08}$ \\
\noalign{\smallskip}
Semi-major axis & $a$ \;[AU] & $0.04888\pm0.00066$ & $0.06870\pm0.00092$ & $0.1046\pm0.0014$ \\
\noalign{\smallskip}
Orbital inclination & $i$ \;[$^{\circ}$] & $85.98_{-0.25}^{+0.26}$ & $86.59_{-0.14}^{+0.15}$ & $87.72_{-0.11}^{+0.10}$ \\
\noalign{\smallskip}
Scaled semi-major axis & $a/R_{\star}$ \; & $8.37_{-0.16}^{+0.17}$ & $11.77_{-0.22}^{+0.24}$ & $17.93_{-0.34}^{+0.37}$ \\
\noalign{\smallskip}
Equilibrium temperature & $T_{\mathrm{eq}}$\,\tablefootmark{(b)} \;[K] & $1552_{-23}^{+22}$ & $1309\pm19$ & $1061_{-16}^{+15}$ \\
\noalign{\smallskip}
Eccentricity & $e$ \; & $0$ (fixed) & $0$ (fixed) & $0$ (fixed) \\
\noalign{\smallskip}
Argument of pericentre & $\omega$ \;[$^{\circ}$] & $90$ (fixed) & $90$ (fixed) & $90$ (fixed) \\
\noalign{\smallskip}
Planet radius & $R_p$ \;[$R_{\oplus}$] & $2.004_{-0.047}^{+0.045}$ & $1.979_{-0.051}^{+0.054}$ & $2.001_{-0.064}^{+0.063}$ \\
\noalign{\smallskip}
Planet mass & $M_p$ \;[$M_{\oplus}$] & $3.55_{-0.96}^{+0.94}$ & $0.90_{-0.63}^{+0.94}$ & $7.1\pm1.6$ \\
\noalign{\smallskip}
Planet density & $\rho_p$ \;[g\,cm$^{-3}$] & $2.44_{-0.68}^{+0.69}$ & $0.65_{-0.45}^{+0.67}$ & $4.9_{-1.1}^{+1.2}$ \\
\noalign{\smallskip}
\hline
\noalign{\smallskip}
LD coefficient 1 & $u_{1,\mathrm{TESS}}$\; & $0.2318_{-0.0064}^{+0.0066}$ & \\
\noalign{\smallskip}
LD coefficient 2 & $u_{2,\mathrm{TESS}}$\; & $0.3084_{-0.0029}^{+0.0030}$ & \\
\noalign{\smallskip}
RV jitter & $\sigma_{\harps}$\;[m\,s$^{-1}$] & $1.490_{-0.022}^{+0.041}$ & \\
\noalign{\smallskip}
\hline
\end{tabular}
\tablefoot{All jump parameters, but the LD coefficients, were subject to unbounded uniform priors following the parameterisations specified in \citet{bonfanti2020}; see text for further details. \\
\tablefoottext{a}{Uncertainties from the run assuming linear ephemerides (no TTVs). $T_0$ values are shifted by $-2\,450\,000$.} \;
\tablefoottext{b}{Assuming zero albedo and full recirculation.} \;
\tablefoottext{c}{The 3$\sigma$ upper limits on the RV semi-amplitude, the planet mass, and the mean planet density of TOI-396\,c are $K_c^{\mathrm{up}}=1.2$ m\,s$^{-1}$, $M_c^{\mathrm{up}}=3.8\,M_{\oplus}$, and $\rho_c^{\mathrm{up}}=2.9$ g\,cm$^{-3}$, respectively.}
}
\end{table*}

This analysis gives $R_b=2.004_{-0.047}^{+0.045}\,R_{\oplus}$, $R_c=1.979_{-0.051}^{+0.054}\,R_{\oplus}$, and $R_d=2.001_{-0.064}^{+0.063}\,R_{\oplus}$ and confirms the RV detection of TOI-396\,b and TOI-396\,d (at the 3.8 and 4.5$\sigma$-level, respectively) with $M_b=3.55_{-0.96}^{+0.94}\,M_{\oplus}$ and $M_d=7.1\pm1.6\,M_{\oplus}$. These outcomes are consistent with the results obtained from the analysis that assumes linear ephemerides (Sect.~\ref{sec:linearEphem}).

\begin{figure}
    \centering
    \includegraphics[width=1.1\columnwidth]{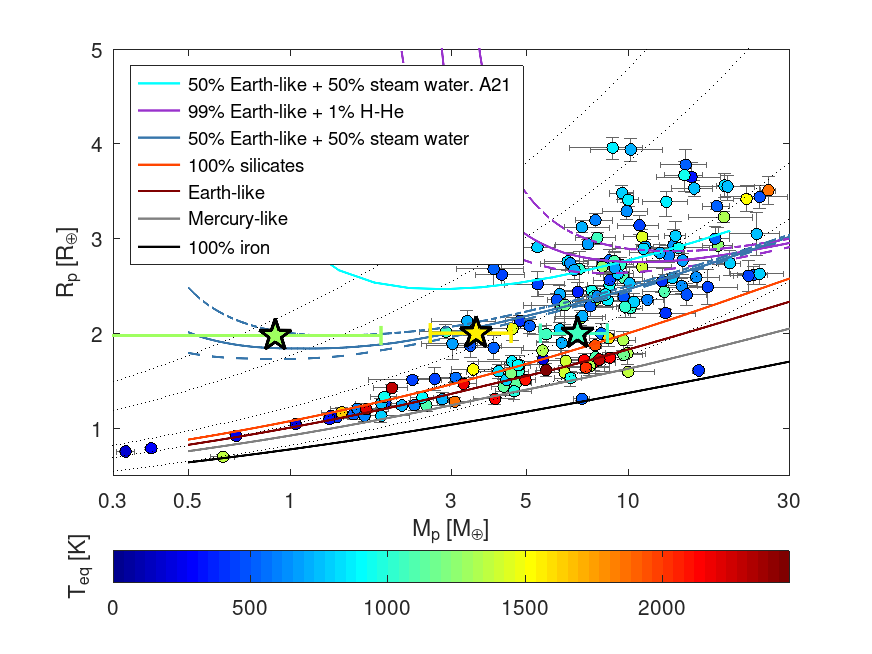}
    \caption{Mass-radius diagram showing the three planets orbiting TOI-396 (star symbol) along with the exoplanets whose density is more significant than the 3$\sigma$ level (circle). All the markers are colour-coded according to the equilibrium temperature ($T_{\mathrm{eq}}$) of the planets. The thick lines are the theoretical MR BICEPS models as detailed in the legend, except for the light-blue line denoted with A21, which is taken from \citet{aguichine2021}. Earth-like means 32.5\% iron + 67.5\% silicates, while Mercury-like means 70\% iron + 30\% silicates. Models were computed for $T_{\mathrm{eq}}=T_{\mathrm{eq},b}=1552$ K (dashed-dotted lines), $T_{\mathrm{eq}}=T_{\mathrm{eq},c}=1309$ K (solid lines), and $T_{\mathrm{eq}}=T_{\mathrm{eq},d}=1061$ K (dashed lines). The difference between the A21 model and its BICEPS counterpart is due to different assumptions for e.g. pressure-temperature profiles and opacities. The dotted black lines are the iso-density loci of points corresponding to 0.5, 1, 3, 5, 10 g\,cm$^{-3}$ (going from top to bottom). We recall that the mass estimate of TOI-396 c (the leftmost star symbol) is not statistically significant and its 3$\sigma$ upper limit is $M_c^{\mathrm{up}}$\,$\sim$\,4$\,M_{\oplus}$.}
    \label{fig:MRdiagram}
\end{figure}

Fig.~\ref{fig:MRdiagram} shows the three planets of the system (star symbol) along with other exoplanets whose density is more significant than the 3$\sigma$ level\footnote{The planet properties were downloaded from the Nasa Exoplanet Archive as of 5 June 2024: \url{https://exoplanetarchive.ipac.caltech.edu/}} (circular marker) in the mass-radius (MR) diagram. The superimposed theoretical MR models as taken from \citet[A21]{aguichine2021} and \citet{haldemann2024} help guide the eye; however, we warn the reader of possible degeneracies occurring in the MR plane. A thorough internal structure analysis of the well characterised TOI-396\,b and d is presented in Sect.~\ref{sec:internalStructure}.

\begin{figure}
    \centering
    \includegraphics[width=\columnwidth]{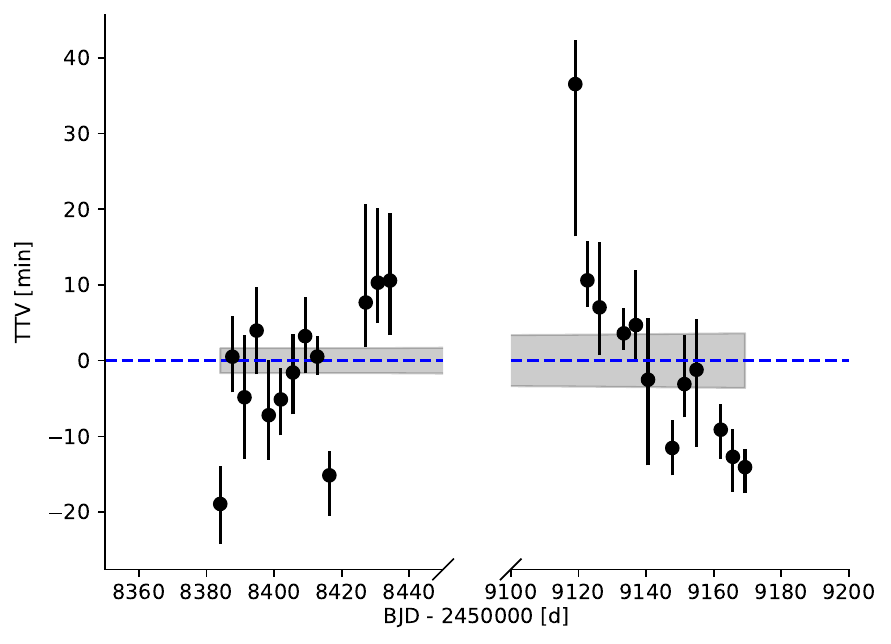} \\
    \includegraphics[width=\columnwidth]{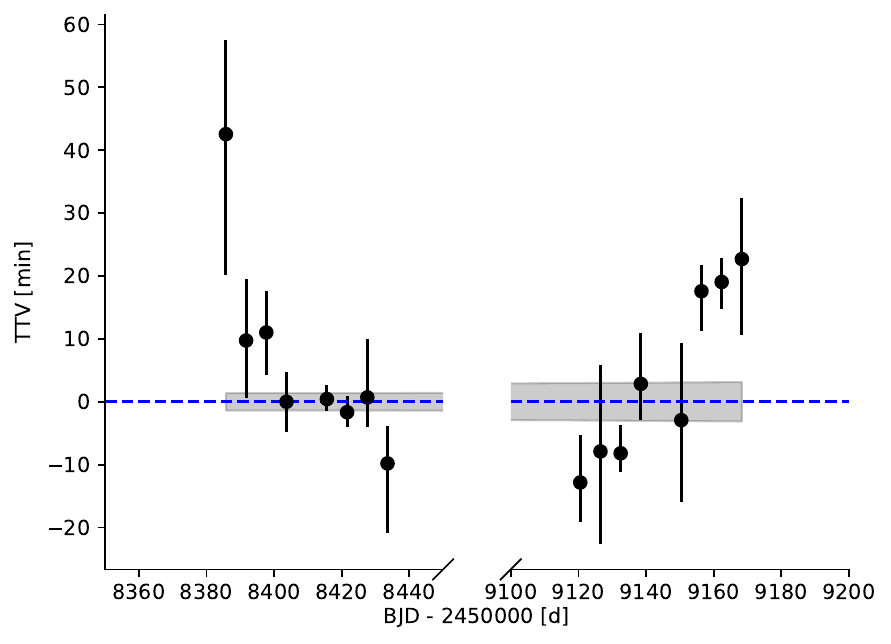} \\
    \includegraphics[width=\columnwidth]{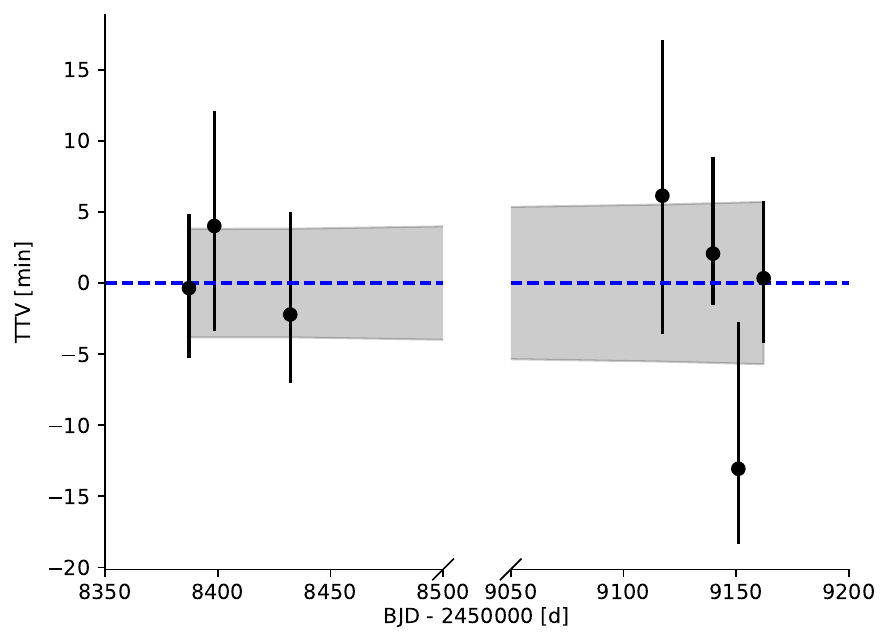} \\
    \caption{TTV amplitudes obtained for TOI-396\,b (\textit{Top panel}), TOI-396\,c (\textit{Middle panel}), and TOI-396\,d (\textit{Bottom panel}). The grey shaded region highlights the 1$\sigma$ uncertainty region as derived from error propagation of the linear ephemerides.}
    \label{fig:TTVall}
\end{figure}

We found significant TTV signals for planet b and c (see Figure~\ref{fig:TTVall}, Top and Middle panels) as expected from the commensurability of their periods.
The TTV statistical significance of TOI-396\,b and c is evident even by eye when comparing the data points location with the shaded regions that represent the 1$\sigma$ uncertainties of the linear ephemerides. For each planet, we further quantified the reduced-$\chi^2$ ($\hat{\chi}^2$) characterising the TTV amplitudes via
\begin{equation}
    \hat{\chi}^2_j = \frac{1}{N_{\mathrm{tr},j}-2}\displaystyle\sum_{i=1}^{N_{\mathrm{tr},j}} \left(\frac{\mathrm{TTV}_{j,i}}{\sigma_{j,i}}\right)^2
    \label{eq:chi2TTV}
\end{equation}
where $N_{\mathrm{tr},j}$ is the number of transit event of the $j$-th planet. We obtained $\hat{\chi}^2_b=4.4$, $\hat{\chi}^2_c=4.1$, and $\hat{\chi}^2_d=0.7$ for planets b, c, and d, respectively, which confirms the significance of the TTV amplitudes for TOI-396\,b and c.

Furthermore, the TTV amplitudes of TOI-396\,b and c exhibit a clear anti-correlation pattern that we highlight by superimposing the TTV measurements in Figure~\ref{fig:TTVbc}. This is a typical signature of gravitational interaction between the two planets, which confirms that TOI-396\,c belongs to the system despite its elusiveness in the RV time series. For each planet, the timing of each transit event and the corresponding TTV amplitude computed with respect to the linear ephemerides are listed in Table~\ref{tab:TTV}.

\subsection{Discussion regarding why TOI-396 c is not detected in the RV time series}\label{sec:M_planet_c?}

Magnetic activity combined with stellar rotation induces RV variations that can hide, affect, or even mimic planetary signals \citep[e.g.][]{Queloz2001,Hatzes2010,Dumusque2011a,Haywood2014,suarezMascareno2017,Gandolfi2017}. A possible explanation for the non-detection of the Doppler reflex motion induced by TOI-396\,c is that stellar activity destructively interferes with the Keplerian signal of the planet. This may happen if the star has a rotation period comparable to the orbital period of the planet \citep[e.g.][]{vanderburg2016}. Disentangling the planetary signal from stellar activity and retrieving the Doppler motion induced by the orbiting planet would then be challenging \citep[e.g.][]{dragomir2012,kossakowski2022}. 

\begin{figure*}
    \centering
    \includegraphics[width=0.80\textwidth]{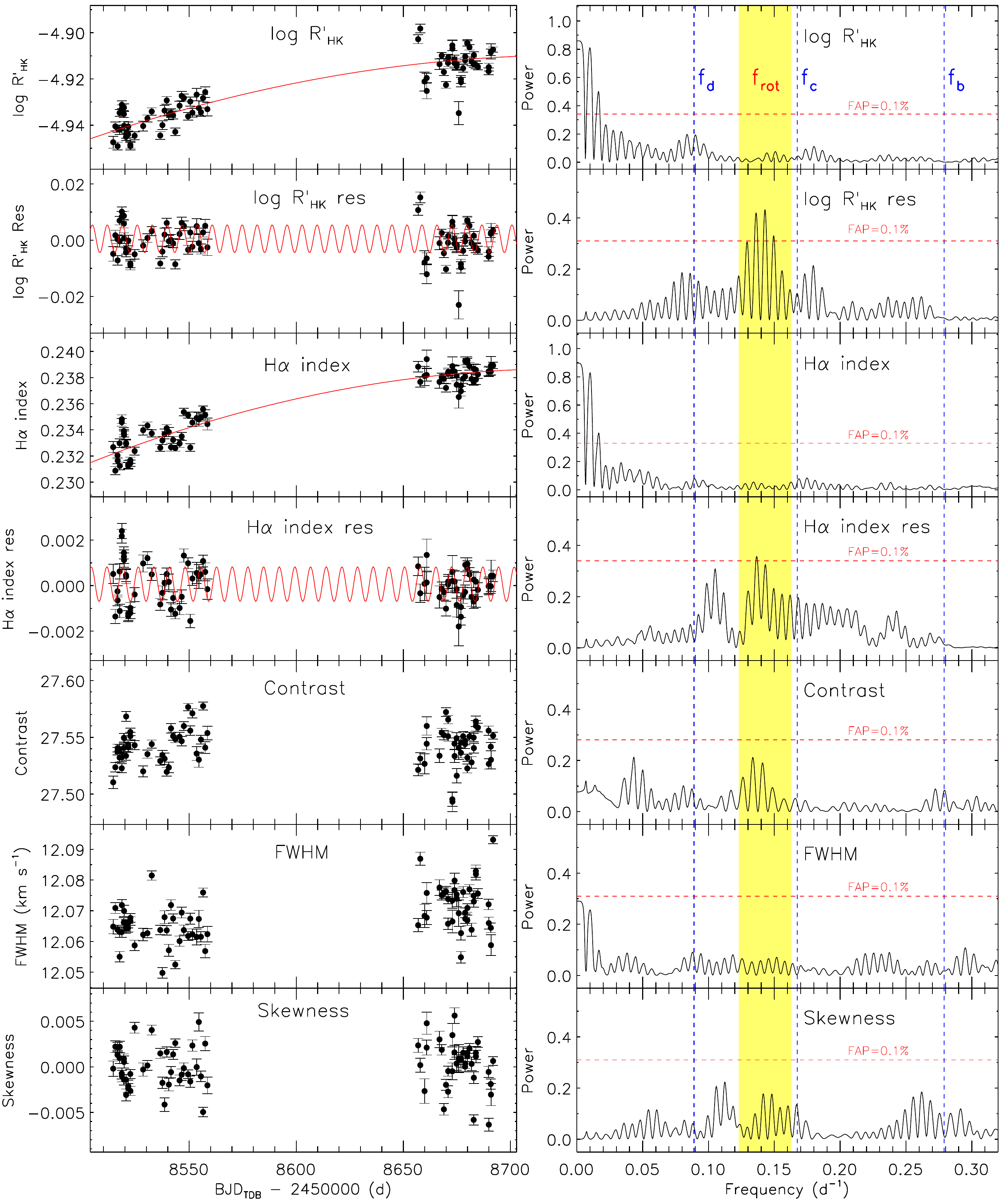}
    \caption{Time series (left panels) and GLS periodograms (right panels) of the line profile variation diagnostics and activity indicators extracted from TOI-396's HARPS spectra. In the left panels, the red curves in the first four panels mark the quadratic trends and sine functions as obtained from the best fit to the most significant peaks (false alarm probability FAP\,$<$\,0.1\%) identified in the corresponding GLS periodograms. In the right panels, the vertical dashed blue lines mark the orbital frequencies of the tree transiting planets. The yellow area encompasses the peaks likely due to stellar rotation. The horizontal dashed red lines mark the 0.1\% false alarm probability.}
    \label{fig:gls-activity-SN}
\end{figure*}

In order to investigate this hypothesis, we performed a frequency analysis of the line profile variation diagnostics (FWHM, contrast, and skewness) and activity indicators (log\,R$^\prime_\mathrm{HK}$ and H$\alpha$ indexes).
Figure~\ref{fig:gls-activity-SN} displays the time series (left column), along with the respective generalised Lomb-Scargle \citep[GLS,][]{Zechmeister2009} periodograms (right column). We assessed the significance of the peaks detected in the power spectra by estimating their false alarm probability (FAP), that is the probability that noise could produce a peak with power higher than the one we found in the time series. To account for the possible presence of non-Gaussian noise in the data, we estimated the FAP using the bootstrap randomisation method \citep[see, e.g.,][]{Murdoch1993,Kuerster1997,Hatzes2019}. Briefly, we computed the GLS periodogram of 10$^5$ mock time series obtained by randomly shuffling the data points along with their error bars, while keeping the timestamps fixed. We defined the FAP as the fraction of those mock periodograms whose highest power exceeds the power of the real data at any frequency. In the present work, we considered a peak to be significant if its false alarm probability is FAP\,$<$\,0.1\,\%.

We found that the time series of the log\,R$^\prime_\mathrm{HK}$ and H$\alpha$ indexes display long-term trends likely due to the magnetic cycles of the star (Fig.~\ref{fig:gls-activity-SN}, left column, first and third panels). In the Fourier domain, these trends translate into a significant (FAP\,$<$\,0.1\,\%) excess of power at frequencies lower than the spectral resolution\footnote{The spectral resolution is defined as the inverse of the time baseline. Our observations cover a baseline of about 177.4\,d, which yields a spectral resolution of $\sim$0.0056\,d$^{-1}$.} of our HARPS observations (Fig.~\ref{fig:gls-activity-SN}, right column, first and third panels). We modelled these long-term signals as quadratic trends and subtracted the best-fitting parabolas from the respective time-series. The GLS periodograms of the residuals of the activity indicators display significant peaks between $\sim$6 and 8\,d (i.e, between $\sim$0.0125 and 0.167 d$^{-1}$; Fig.~\ref{fig:gls-activity-SN}, yellow area). The peaks are equally spaced by $\sim$0.0068\,d$^{-1}$, which coincides with a peak found in the periodogram of the window function. Given the current data at our disposal, we are not able to distinguish between true frequencies and aliases. Although not significant, the power spectra of the contrast and skewness show peaks in the same frequency range, suggesting that the rotation period of the star might be $\sim$6--8\,d. 

We note that the periodogram of the FWHM also displays an excess of power at low frequencies. However, the corresponding peak remains below our 0.1\,\%-FAP significance threshold (see Fig.~\ref{fig:gls-activity-SN}, second to last panel). Yet, if we apply the same procedure described above and remove this signal by fitting a quadratic trend to the FWHM time series, we find no significant peak in the residuals.

The projected equatorial velocity of the star ($v\sin{i_{\star}}\,=7.5\,\pm0\,.2$\,km\,s$^{-1}$), along with its radius ($R_\star\,=1.258\,\pm\,0.019$\,$R_\odot$), yields an upper limit for the rotation period of $P_\mathrm{rot}^\mathrm{up}=8.5\,\pm\,0.3$\,d. Using the mean value\footnote{Rather than assigning to the mean an uncertainty equal to the standard error, we adopted a more conservative uncertainty equal to the standard deviation of our 78 measurements.} of log\,R$^\prime_\mathrm{HK}$\,=\,$-4.926\pm0.014$, we inferred a stellar rotation period of 6.7\,$\pm$\,1.3\,d and 6.9\,$\pm$\,1.3\,d from the empirical equations of \citet{Noyes1984} and \citet{Mamajek2008}, respectively. In addition, by inputting the isochronal age into the gyrochronological relation from \citet{barnes10}, we computed a stellar rotation period of $7.1_{-1.1}^{+1.0}$\,d. These results corroborate our interpretations that the peaks between 6 and 8~d significantly detected in the power spectra of the activity indicators originate from stellar rotation.

To check whether a quasi-periodic signal compatible with $\sim$7~d is also present in the photometric data, for each \tess sector we extracted custom LCs from pixel data using \texttt{lightkurve} \citep{lightkurve2018}. In detail, we adopted the default quality bitmask and set the aperture to `all', which corresponds to an aperture larger than the one used by the official SAP pipeline. In fact, larger apertures mitigate the effect of slow image drifts that could interfere with slow flux changes, such as the 6--8\,d rotation period signals we aim to detect. After removing the temporal windows containing the transit events, we computed the GLS periodograms of these \texttt{lightkurve}-based LCs. The FAP was computed following the same bootstrap technique outlined above for the RV activity indicators.
The four periodograms (Fig.~\ref{fig:gls-photometry}, first column) exhibit very significant peaks at $\sim$\,7.7, 7.9, 7.5, and 6.8 days for \tess Sectors 3, 4, 30, and 31, respectively. Except for Sector 30, they are not the most prominent peaks; however, they persist even after removing the most significant signals (Fig.~\ref{fig:gls-photometry}, second column). Whereas we acknowledge that the likely rotation period of the star is close to the first harmonic of the orbital period of \tess around the Earth ($\Prot$\,$\sim$\,$\frac{1}{2}P_{\mathrm{TESS}}$\,$\sim$\,14 days), the photometric signal at 6--8~d is significant in all the four \tess sectors and persists after pre-whitening the data. This signal is consistent with the rotation period detected in the HARPS activity indicators and inferred from log\,R$^\prime_\mathrm{HK}$, $v\sin{i_{\star}}$, and gyrochronology, suggesting it is astrophysical in nature and due to the presence of active regions carried around by stellar rotation.
Assuming the log\,R$^\prime_\mathrm{HK}$-based $\Prot=6.7\pm1.3$~d as our reference estimate, the orbital period of TOI-396\,c ($P_c\sim6$ d) is close to $\Prot$, which may explain the non-detection of planet c within the RV time series.

If some kind of destructive interference between the Keplerian signal of planet c and the stellar activity has occurred, any artificial Keplerian signals with period $P=P_c$ added to the observed RV time series should in principle be retrieved. To test this hypothesis, we considered Keplerian signals of the following form
\begin{equation}
RV_{\mathrm{art}}=- K_{\mathrm{art}} \sin{\left[\frac{2 \pi}{P} (t-T_{0})\right]}
\label{eq:artificial-rv-signal}
\end{equation}
and generated four different RV time series by separately adding to the \harps time series synthetic RV signals following Equation~(\ref{eq:artificial-rv-signal}) with $P=P_c$, $T_0=T_{0,c}$, and $K_{\mathrm{art}}=K_{\mathrm{in}}$, where $K_{\mathrm{in}}$ are the four different amplitude values listed in the first column of Table~\ref{tab:artificial-rv-testC}.
For each RV time series, we then performed an MCMCI analysis to retrieve the RV semi-amplitude of the artificial signal ($K_{\mathrm{out}}$; see Table~\ref{tab:artificial-rv-testC}). 

\begin{table}
\caption{Radial velocity semi-amplitudes $K_{\mathrm{out}}$ as retrieved from MCMCI analyses of the RV time series obtained by adding an artificial Keplerian signal with period $P=P_c$ and semi-amplitudes $K_{\mathrm{in}}$ to the original \harps time series.}
\label{tab:artificial-rv-testC}
\centering
\begin{tabular}{c c c c c}
\hline
\hline
   $K_{\mathrm{in}}$ & $\rho_p$ & $M_p$ & $K_{\mathrm{out}}$ & Detection  \\\relax
   [m\,s$^{-1}$] & [g\,cm$^{-3}$] & $[M_\oplus]$ & [m\,s$^{-1}$] & level \\
\hline
      0.51 & $1.0$ & $1.6$ & $0.68 \pm 0.36$ & $\sim 2\sigma$ \\
      0.77 & $1.5$ & $2.3$ & $0.93 \pm 0.42$ & $\sim 2\sigma$ \\
       1.0 & $1.9$ & $3.0$ & $1.15 \pm 0.38$ &  $\sim 3\sigma$ \\
$K_d=1.79$ & $3.5$ & $5.4$ & $1.94 \pm 0.38$ &  $\sim 5\sigma$ \\        
\hline
\end{tabular}
\tablefoot{Columns $\rho_p$ and $M_p$ translate the injected synthetic signals into the corresponding physical parameters of the planet, while the last column quantifies the $K_{\mathrm{out}}$ detection level in terms of $\sigma$.}
\end{table}

We note that the resulting $K_{\mathrm{out}}\approx K_{\mathrm{in}} + K_c$, where $K_c = 0.28_{-0.19}^{+0.29}$ m\,s$^{-1}$ is the RV semi-amplitude of planet c as derived from the analysis on the original RV time series. As we essentially retrieved what we inserted in the \harps time series, we may conclude that the destructive interference between the RV signals induced by the star and by planet c has already occurred and any further RV signal added to the RV time series is detected.

As $\Prot$ is not exactly equal to $P_c$, we then repeated the test outlined above, but this time we injected into the original \harps time series artificial Keplerian signals with $P=\Prot$.
The $K_{\mathrm{out}}$ values obtained by the MCMC analyses depending on the different $K_{\mathrm{in}}$ values are reported in Table~\ref{tab:artificial-rv-test-Prot}. The $K_{\mathrm{out}}$ values are systematically and significantly smaller than the corresponding $K_{\mathrm{in}}$ values, which a posteriori supports the conclusion that the stellar rotation period is around 6--8\,d. Furthermore, a planet with this orbital period would be firmly detected if its RV semi-amplitude were greater than $K_d$, which is the largest RV semi-amplitude detected for this system. Instead, by injecting a Keplerian signal with $K_{\mathrm{in}}=K_d$ and $P=\Prot$, the MCMCI analysis is able to barely detect (at $\sim$\,2$\sigma$) a planetary signal whose amplitude is half of that expected. The detection level increases when $K_{\mathrm{in}}$ increases; however, we still underestimate $K_{\mathrm{out}}$. These tests prove that it is difficult to retrieve planetary signals with $P$\,$\sim$\,$\Prot$. 

In summary, we conclude that stellar activity is responsible for generating spurious RV signals whose harmonics also include the stellar rotation period. As a consequence, it is hard to reliably detect planets with orbital periods comparable to $\Prot$ via the RV technique and Table~\ref{tab:artificial-rv-test-Prot} quantifies the magnitude of this effect. We note that the $K_c$ we obtained from the MCMCI analysis in Sect.~\ref{sec:TTV} is comparable to the $K_{\mathrm{out}}$ retrieved when inserting an artificial signal having $K_{\mathrm{in}}=1.0$ m\,s$^{-1}$, which let us to speculate that TOI-396\,c might have $M_c$\,$\sim$\,3.0\,$M_{\oplus}$ and $\rho_c$\,$\sim$\,2.0 g\,cm$^{-3}$.

\begin{table}
\caption{Same as Table~\ref{tab:artificial-rv-testC}, but this time the Keplerian signals with $K=K_{\mathrm{in}}$ have period $P=\Prot$.}
\label{tab:artificial-rv-test-Prot}
\centering
\scriptsize
\begin{tabular}{c c c c c c c}
\hline
\hline
   $K_{\mathrm{in}}$ & $\rho_p$ & $M_p$ & $K_{\mathrm{out}}$ & Detection & $\frac{\Delta K}{K}$ &  $\Delta K$ \\\relax
   [m\,s$^{-1}$] & [g\,cm$^{-3}$] & [$M_\oplus$] & [m\,s$^{-1}$] & level &  & [$\sigma$] \\
\hline
       1.0 & $2.0$ & $3.1$   & $0.29 \pm 0.28$ & $\sim 1\sigma$ & $-71$\% & $-2.5$ \\
$K_d=1.79$ & $3.6$ & $5.6$   & $0.80 \pm 0.39$ & $\sim 2\sigma$ & $-55$\%  & $-2.5$ \\
       3.0 & $6.1$ & $9.4$   & $1.90 \pm 0.40$ & $\sim 5\sigma$ &  $-37$\% & $-2.8$ \\
    $10.0$ & $20.2$ & $31.4$ & $9.02 \pm 0.36$ & $\sim 25\sigma$ & $-10$\% & $-2.7$\\        
\hline
\end{tabular}
\tablefoot{Column $\frac{\Delta K}{K}$ gives the relative difference (in percentage) between the obtained RV semi-amplitude ($K_{\mathrm{out}}$) and the expected one ($K_{\mathrm{in}}$), while the last column $\Delta K \equiv K_{\mathrm{out}}-K_{\mathrm{in}}$ lists the semi-amplitude difference in terms of the 1$\sigma$ uncertainty of $K_{\mathrm{out}}$.}
\end{table}

\section{Joint RV and TTV dynamical analysis}\label{sec:dynamicTTV}

As mentioned, the anti-correlation pattern of the observed TTVs (see Fig.~\ref{fig:TTVbc}) is a typical signal of the dynamical interaction between TOI-396\,b and c.
However, the data in our hands are not enough to currently derive meaningful planetary masses from the TTVs. Indeed, the photometric observations are clustered in two groups (about two years apart), which results in a partial coverage of the curvature of the TTVs.
This prevents us from accurately mapping the full phases and super-periods of the TTV signals. Hence a dynamical fit onto the transit times would lead to a high fraction of low-amplitude TTVs with a short super-period or to a low fraction of high-amplitude long-period TTV signals.

Despite this, we attempted to run a dynamical joint fit of the TTV and RV data set with
\trades{}\footnote{\url{https://github.com/lucaborsato/trades}}
\citep{Borsato2014A&A...571A..38B,Borsato2019MNRAS.484.3233B,Borsato2021MNRAS.506.3810B},
a \texttt{Fortran-python} code developed to model TTVs and RVs simultaneously along with N-body integration.
We have taken and fixed the stellar mass and radius values from Table~\ref{tab:star}.
We further fixed the orbital inclinations $i$ to the values in Table~\ref{tab:planets} and the longitude of the ascending nodes to $\Omega=180^{\circ}$\footnote{
We assumed the same reference system of \citet{Winn2010} and \citet{Borsato2014A&A...571A..38B} and later used in \citet{Borsato2019MNRAS.484.3233B,Borsato2021MNRAS.506.3810B,Borsato2024A&A...689A..52B,Nascimbeni2023A&A...673A..42N,nascimbeni2024}.
}
for all the planets.
We fitted the planetary masses scaled by the stellar mass ($M_\mathrm{b,c,d}/M_\star$), the orbital periods ($P$),
the eccentricities ($e$) and the arguments of the pericentre ($\omega$) in the form ($\sqrt{e}\cos\omega$, $\sqrt{e}\sin\omega$),
and the mean longitudes ($\lambda$\footnote{The mean longitude is defined as $\lambda =\omega + \Omega + \mathcal{M}$,
where $\omega$ is the argument of the pericentre, $\Omega$ is the longitude of the ascending node, and $\mathcal{M}$ is the mean anomaly.}).
We also fitted for an RV offset ($\gamma_\mathrm{RV}$) and for
an RV jitter term ($\sigma_\mathrm{jitter}$) by adopting $\log_{2}\sigma_\mathrm{jitter}$ as step parameter, although we used the de-trended RV data set as derived from Sect.~\ref{sec:TTV}, where a jitter term was already included in the RV errors.
All fitting parameters were subject to uniform priors that account for their respective physical boundaries (see Table~\ref{tab:trades_parameters});
for the eccentricities we applied a log-penalty ($\log\mathrm{p}_e$) based on the half-Gaussian ($e=0, \sigma_e = 0.083$)
from \citet{VanEylen2019AJ....157...61V}.

The reference time for the dynamical integration of the orbital parameters was set at $T_\mathrm{ref} = 2\,458\,379\,\mathrm{BJD_{TDB}}$, that is before all available observations.
We combined the quasi-global differential evolution \citep{Storn1997}
optimisation algorithm implemented in \pyde{} \citep{Parviainen2016} with the Affine Invariant MCMC Ensemble sampler \citep{AffineInvariantGoodmanWeare2010} 
\emcee{} implemented by \citet{DFM2019JOSS....4.1864F,Foreman2013}.
We first ran \pyde{} and evolved 68 different initial configurations of parameter sets for $50\,000$ generations 
(number of steps for which each parameter is evolved).
To perform the dynamical analysis in an MCMC fashion, we then ran \emcee{}, assuming as starting point the outcome obtained with \pyde. We set up 68 chains for $600\,000$ steps each.
To sample the parameter space efficiently, we mixed the \texttt{DEMove()} and \texttt{DESnookerMove()} differential evolution moves\footnote{See \emcee{} documentation: 
\url{https://emcee.readthedocs.io/en/stable/tutorials/moves/}}
in the proportion 80\%--20\% \citep{DENelson2014ApJS..210...11N,terBraak2008}.
We repeated the sequence \pyde+\emcee{} twice, 
with different seeds for the random number generator.
The chains reached convergence according  to visual inspection
and statistical indicators, such as the Gelman-Rubin $\hat{R}$, the
Geweke's statistic \citep{geweke1991},
and the auto-correlation function\footnote{See e.g. \citet{AffineInvariantGoodmanWeare2010}
and the \emcee{} documentation at \url{https://emcee.readthedocs.io/en/stable/tutorials/autocorr/\#autocorr}}.
For both runs, we applied a conservative thinning factor of 100 and
discarded the first $50\%$ of the chains (burn-in).
For each run, we derived the reference outcome (hereinafter also referred to as best-fit), as the maximum-a-posteriori (MAP) parameters' set,
that is the set of parameters that maximises the log-probability\footnote{
The log-probability is given by $\log\mathcal{P} = \log\mathcal{L} + \log\mathrm{p}_e$,
where $\log\mathcal{L}$ is the log-likelihood and
$\log\mathrm{p}_e$ is the log-penalty of the eccentricities.
}. The uncertainty of each parameter is quantified by the high-density interval (HDI) at 68.27\%\footnote{
An HDI at 68.27\% is the equivalent of the $15.87^\mathrm{th}-84.14^\mathrm{th}$ percentile range of a Normal distribution that provides the 1$\sigma$ standard confidence interval.} of its posterior distribution.
For each parameter we computed a Z-score defined as
$\text{Z-score} = |\mathrm{MAP}_1 - \mathrm{MAP}_2| / \sqrt{\max|\mathrm{ERR}_1|^{2} + \max|\mathrm{ERR}_2|^{2}}$,
where the subscripts denote the two different runs and $\mathrm{ERR} = \mathrm{HDI} - \mathrm{MAP}$.
It turned out that $\text{Z-score}<1$ for each parameter, which allows us to merge the posterior distributions deriving from the two runs to finally compute the MAP and the respective HDI from the merged posterior distributions.
We further checked the Hill stability of the system \citep{sundman1913} when assuming the entire merged posterior distributions by calculating the angular momentum deficit \citep[AMD,][]{Laskar1997A&A...317L..75L,Laskar2000PhRvL..84.3240L,LaskarPetit2017A&A...605A..72L} criterion
\citep[Eq. 26 from][]{Petit2018A&A...617A..93P}.
Table~\ref{tab:trades_parameters} lists the parameters returned by \trades{} (MAP and HDI) along with their respective priors.

We note that the dynamical integration with TRADES allowed for a significant detection (>3$\sigma$) of the mass of TOI-396\,c, that is $M_{c,\mathrm{dyn}}=2.24_{-0.67}^{+0.13}\,M_{\oplus}$. However, as explained above, \tess data do not allow us to fully map the TTV pattern.
Therefore, the present-day estimate of $M_{c,\mathrm{dyn}}$ only provides an indication of the possible mass of the planet that might not be accurate, despite its formal precision. Indeed, to accurately and reliably determine planetary masses via TTVs, it is necessary to monitor the TTV signal by benefiting of a full sampling coverage, as demonstrated for example by the cases of Kepler-9 \citep{holman2010} and K2-24 \citep{petigura2016}, whose orbital parameters and masses were comprehensively revisited by \citet{Borsato2014A&A...571A..38B} and \citet{petigura2018,nascimbeni2024}, respectively. In detail, \citet{holman2010} had reported the masses of Kepler-9\,b and c with a precision better than 8\%, by performing a TTV analysis on the first three quarters of the Kepler data. Later on, by benefiting of twelve sectors of Kepler data that enabled the full mapping of the TTV phase, \citet{Borsato2014A&A...571A..38B} obtained TTV-based masses differing by a factor $\sim$\,2 from the estimate of \citet{holman2010}. Similarly, by accounting on more photometric data, \citet{petigura2018,nascimbeni2024} found that the mass of K2-24\,c is lower by almost 2\,$\sigma$ than the estimate of \citet{petigura2016} who had claimed a detection at the $\sim$\,4\,$\sigma$ level.

\begin{figure*}
\includegraphics[width=0.49\textwidth]{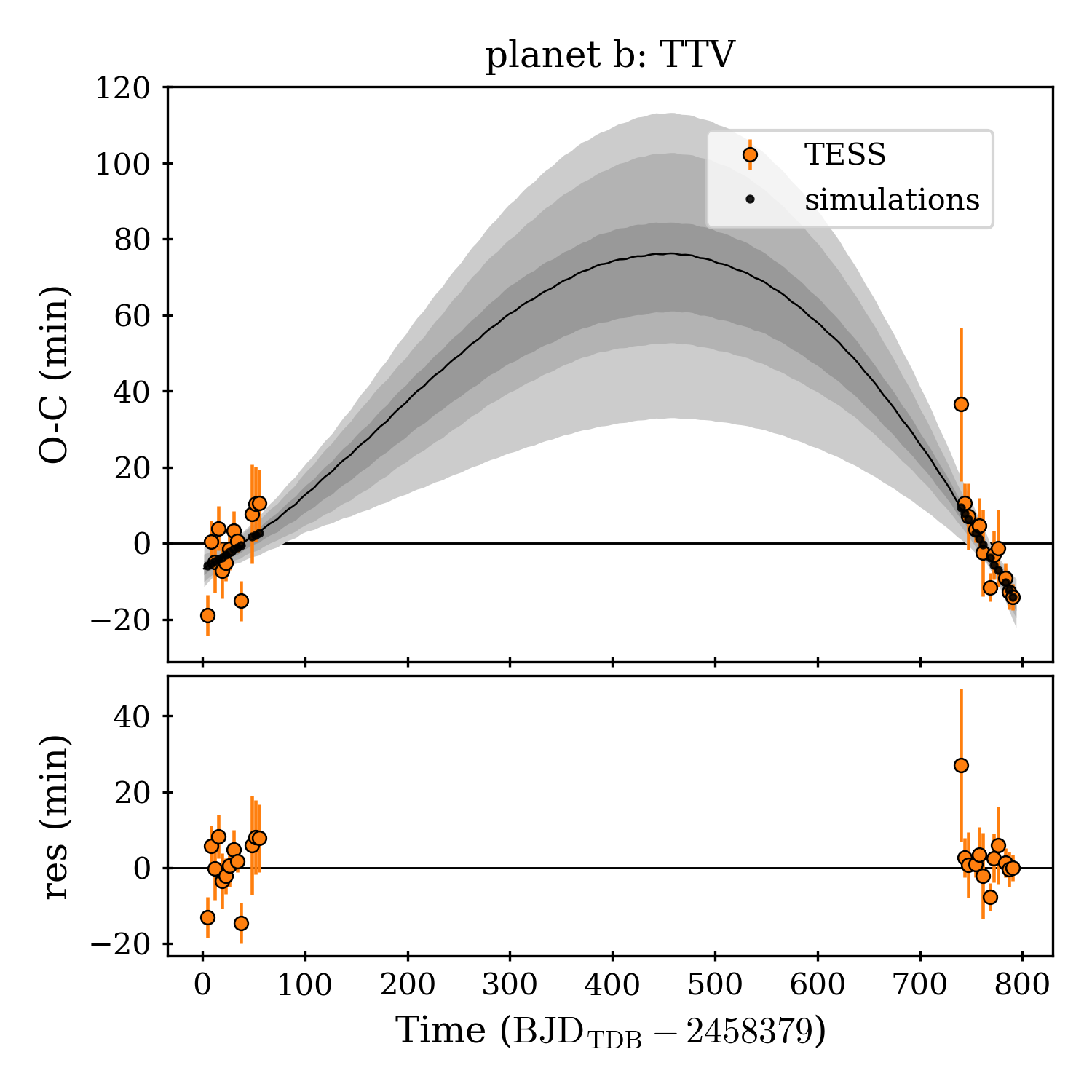}
\includegraphics[width=0.49\textwidth]{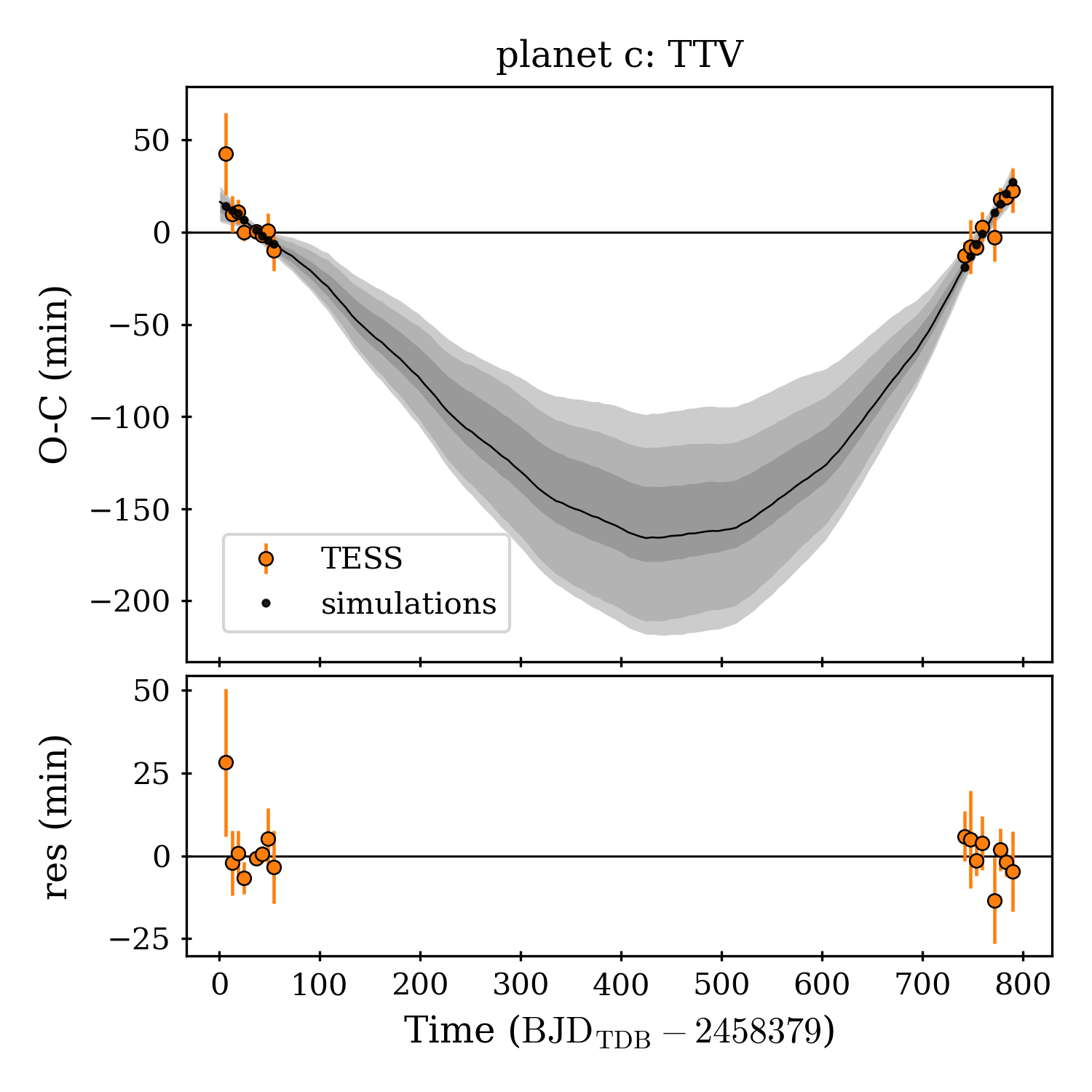}
  \caption{Observed minus calculated synthetic diagrams derived from the joint RV and TTV dynamical analysis with \trades{}
  for planet b (\textit{left panel}) and c (\textit{right panel}).
  The $O-C$ for the best-fit (MAP) model is plotted with a black line, while the observed data points are the orange circles. The shaded grey regions displays the confidence intervals at 1, 2, and 3$\sigma$, as inferred from the 200 samples randomly drawn from the merged posterior distributions.
  Residuals are shown in the lower panels.
  }
  \label{fig:oc_trades_fit_1}
\end{figure*}

\begin{figure*}
\includegraphics[width=0.465\textwidth]{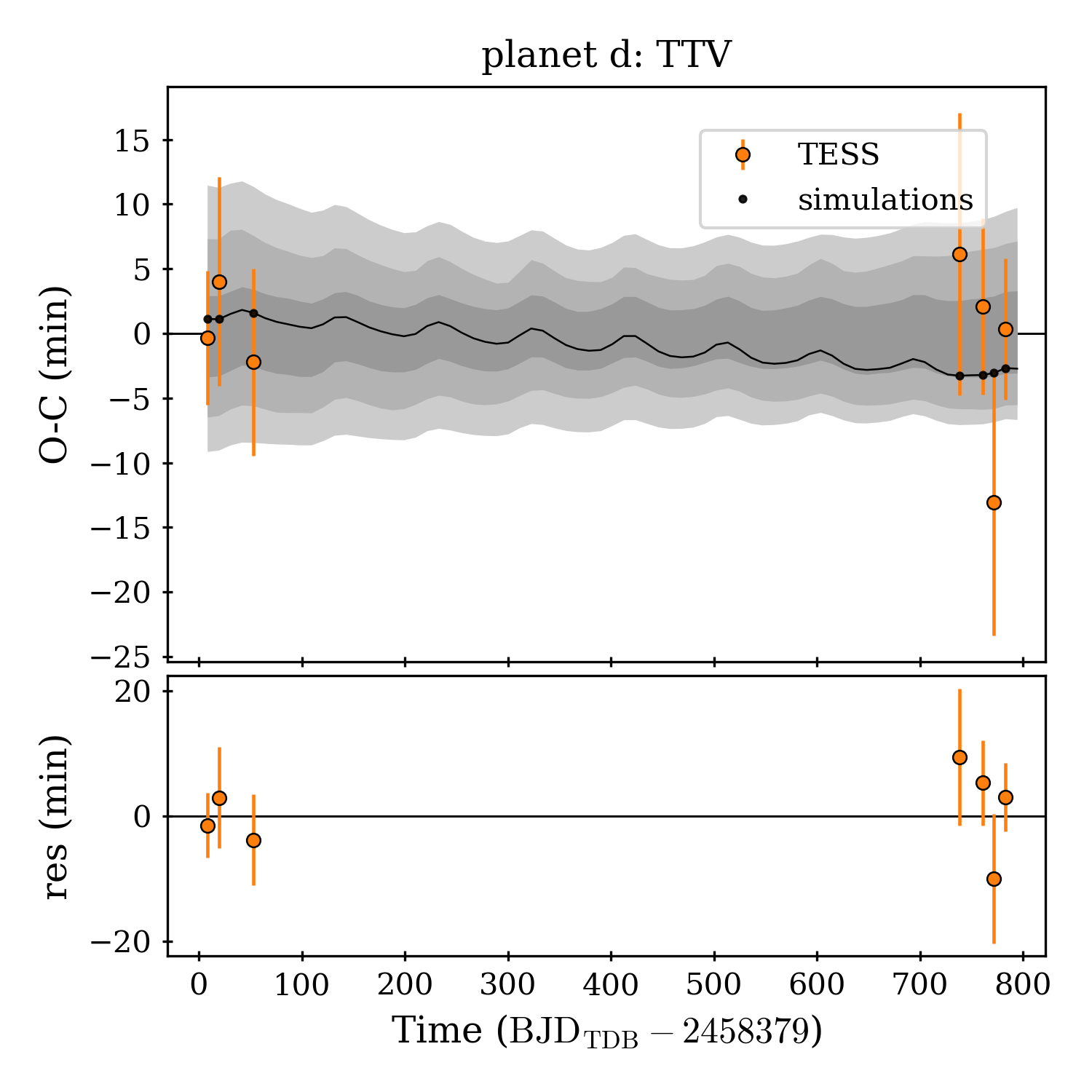}
\includegraphics[width=0.49\textwidth]{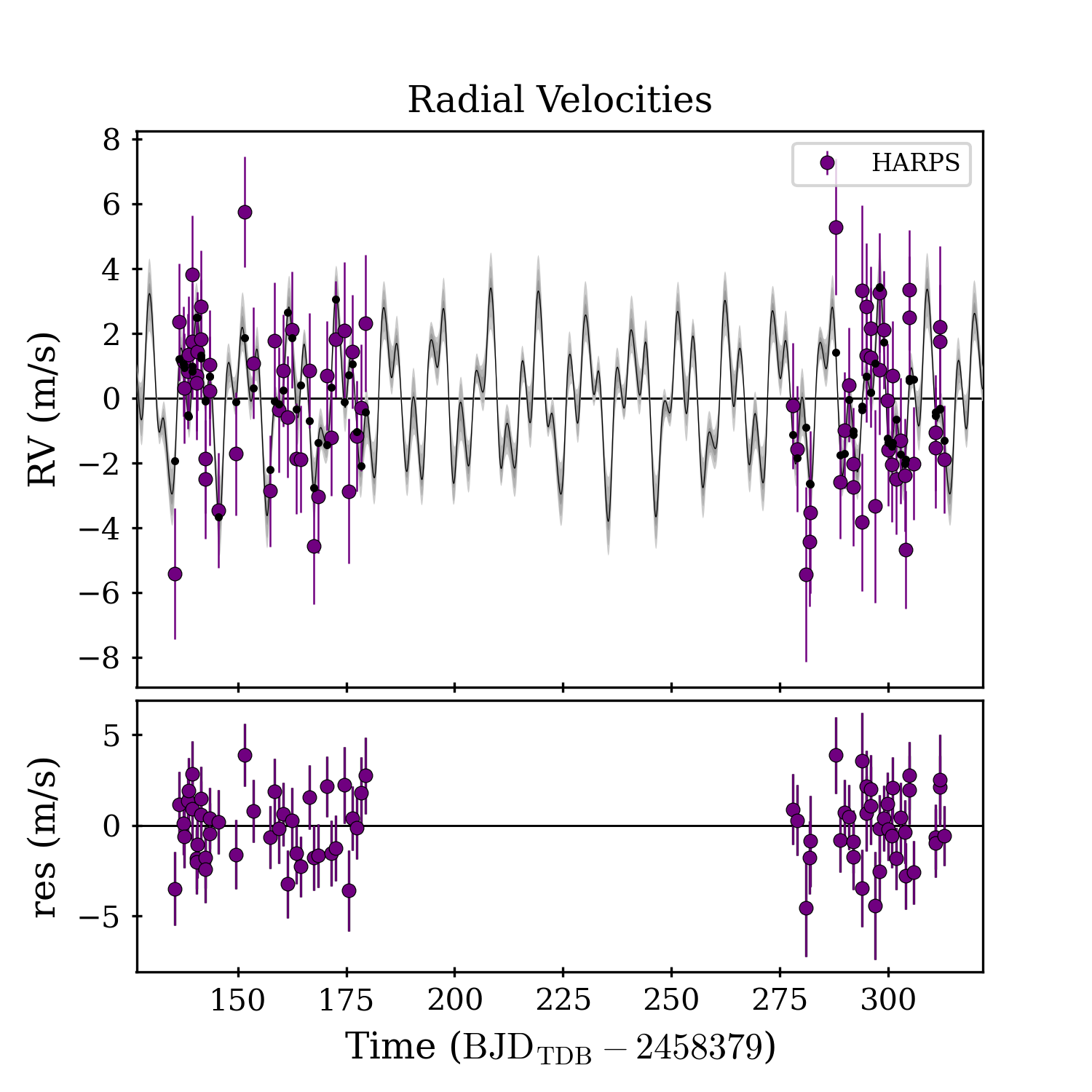}
  \caption{\textit{Left panel}: Same as Fig.~\ref{fig:oc_trades_fit_1}, but for TOI-396\,d. 
  \textit{Right panel}: Combined RV model of the three planets (black line) superimposed to the de-trended \harps observations (purple circles). The grey shaded area is determined from the 200 sets of system parameters randomly drawn from the merged posterior distributions as obtained from the joint dynamical analysis with \trades{}.
  }
  \label{fig:oc_trades_fit_2}
\end{figure*}

After extracting all the synthetic transit timings $T_{\mathrm{tr}}$ (i.e. the `observed': O) from \trades{}' analysis, we computed the `calculated' (C) counterpart according to the linear ephemerides model based on what listed in
% on $T_{\mathrm{tr}}$,
Table~\ref{tab:planets}. We plotted the $O-C$ as a function of time as well as the RV best-fit model in Figures~\ref{fig:oc_trades_fit_1} and \ref{fig:oc_trades_fit_2}.

\begin{figure}
\includegraphics[width=0.49\textwidth]{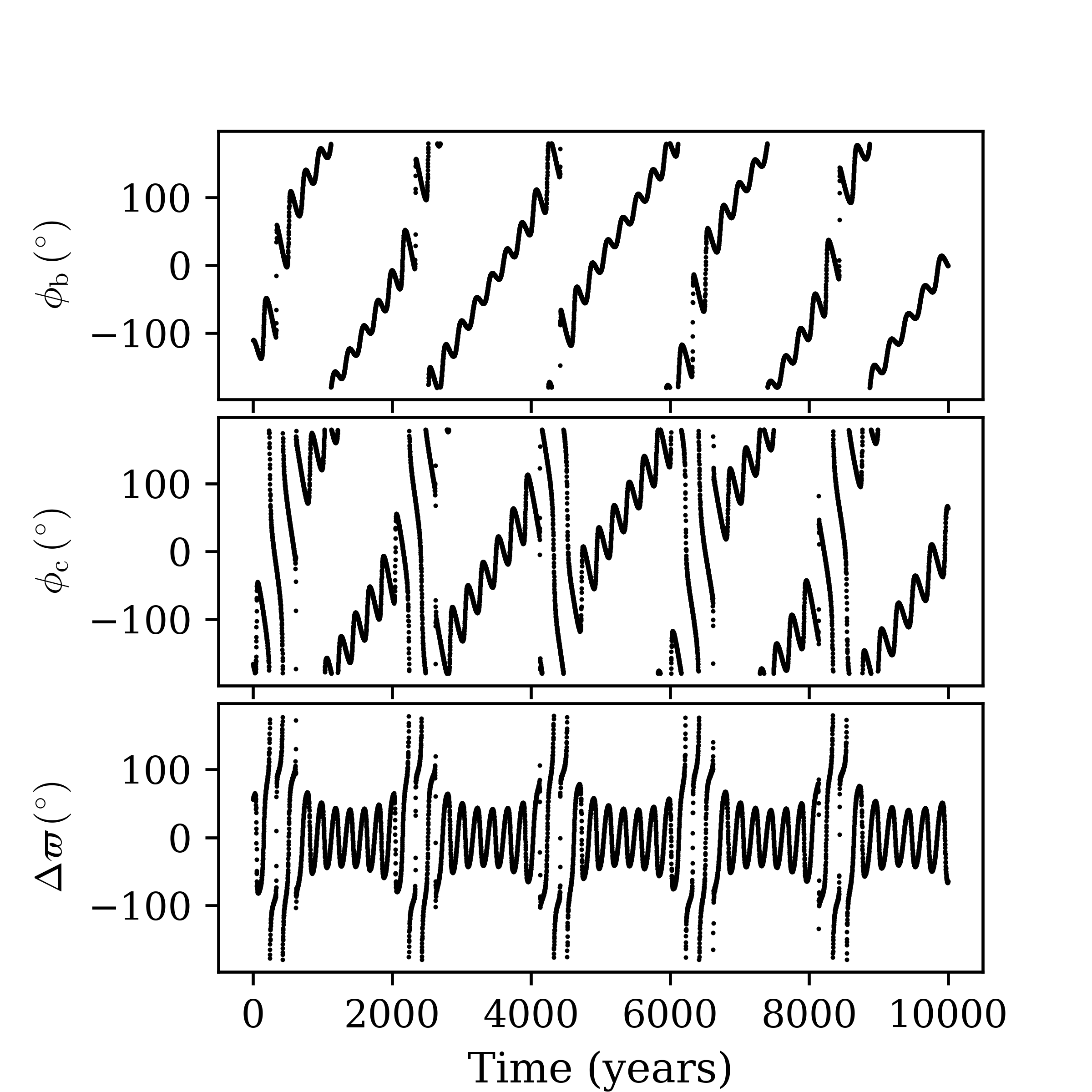}
  \caption{Temporal evolution of the critical resonance angles $\phi_{\mathrm{b}}$ (\textit{top panel}) and $\phi_{\mathrm{c}}$ (\textit{middle panel}) as well as of $\Delta\varpi = (\phi_\mathrm{b} - \phi_\mathrm{c}) / q$ (\textit{bottom panel}) as inferred when assuming the MAP parameters derived by the \trades{} dynamical analysis and integrated with \textsc{rebound+whfast} for 10\,000 years.
  }
  \label{fig:evolution_mmr}
\end{figure}

Additionally, we investigated if the best-fit configuration is in or close to an MMR.
We integrated the MAP parameters with the N-body code \textsc{rebound} \citep{rebound}
and the symplectic Wisdom-Holman integrator \textsc{whfast} \citep{reboundwhfast,wh}
for 10\,000 years.
We computed the evolution of the critical resonance angles of TOI-396\,b and TOI-396\,c
\begin{equation}
\begin{split}
\phi_\mathrm{b} &= p \lambda_\mathrm{b} - (p+q) \lambda_\mathrm{c} + q \varpi_\mathrm{b} \\
\phi_\mathrm{c} &= p \lambda_\mathrm{b} - (p+q) \lambda_\mathrm{c} + q \varpi_\mathrm{c},  \end{split}
\end{equation}
where $p=3$ and $q=2$ (for a second order 5:3 MMR), while $\varpi\equiv\omega+\Omega$ is the longitude of the pericentre.
We also computed the evolution of $\Delta\varpi\equiv\varpi_b-\varpi_c = (\phi_\mathrm{b} - \phi_\mathrm{c}) / q$.
In case of MMR, we expect that both $\phi_b$ and $\phi_c$ librate (i.e. oscillate) around a fixed value for the entire orbital integration. Instead, if these angles circulate, that is they span the full 0$^{\circ}$--360$^{\circ}$ range (or equivalently the $-180^{\circ}$--$180^{\circ}$ range), then the planet pair is not in resonance.
We found that both $\phi_\mathrm{b}$ and $\phi_\mathrm{c}$ circulate (see the two upper panels of Fig.~\ref{fig:evolution_mmr}), 
which indicates that the system is not in an exact 5:3 MMR.
Even if $\Delta\varpi$ seems to oscillate around $0^{\circ}$, it circulates every $\sim 2\,000$ years (bottom panel of Fig.~\ref{fig:evolution_mmr}), which further confirms that the system is not trapped in an MMR state.
We also found the same behaviour for the resonant angles of 200 random samples drawn from the posterior distribution (not shown here).
Our conclusions are consistent with the simulations performed by \citet{Vanderburg2019}, who found that most realisations of the system are not in resonance.

\begin{figure}
\includegraphics[width=0.49\textwidth]{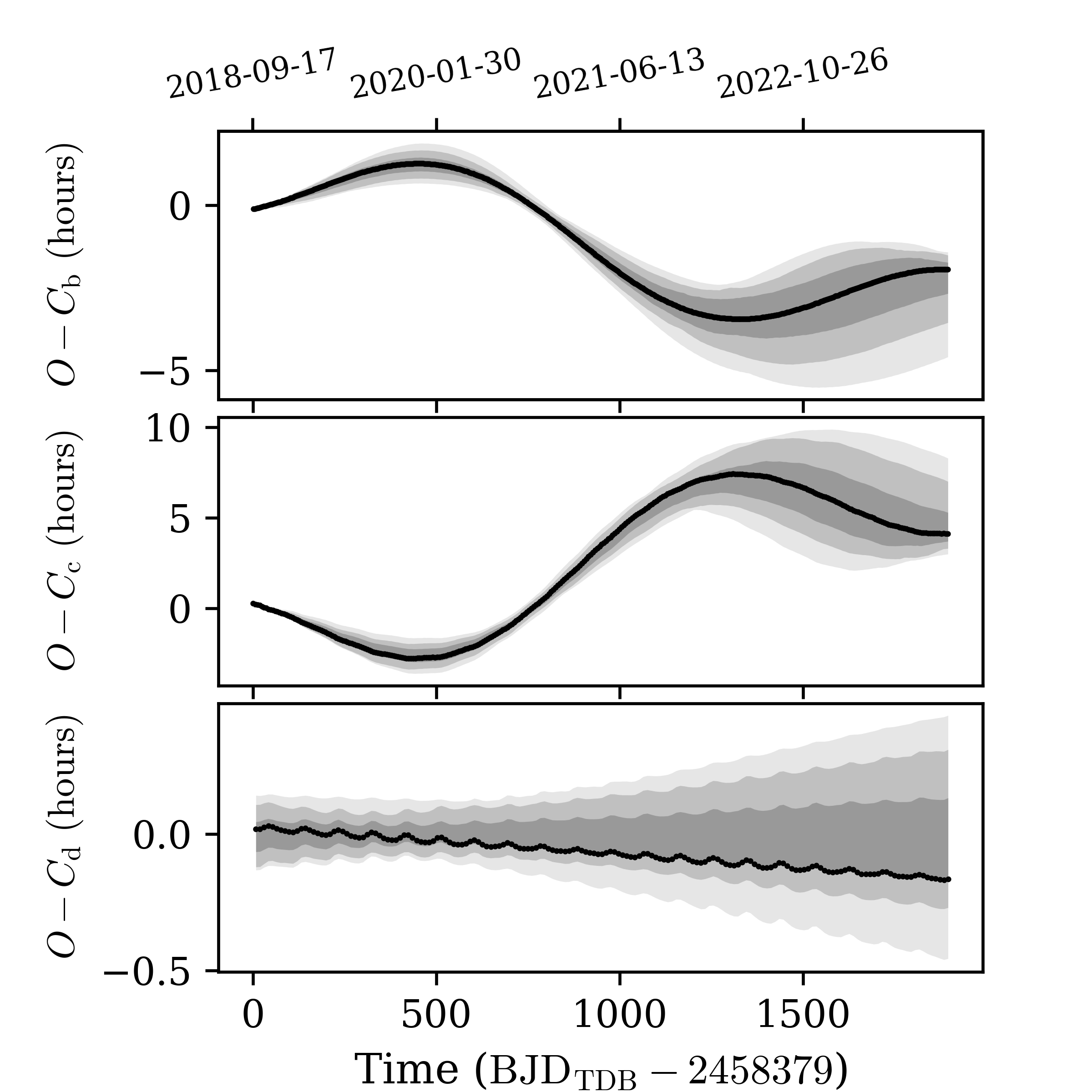}
  \caption{Synthetic $O-C$ diagrams obtained after performing forward numerical N-body simulations with \trades{} (integration
  of 5.2 years). The $C$ represents the timings calculated from the linear ephemerides in Table~\ref{tab:planets}.
  The MAP model is plotted with a black line, while the confidence intervals at 1, 2, and 3$\sigma$ are marked as shaded grey regions and come from 200 random samples drawn from the merged posterior distributions derived from the joint RV and TTV dynamical analysis.
  }
  \label{fig:oc_trades_predicted}
\end{figure}

As shown in Fig.~\ref{fig:oc_trades_fit_1}, the MAP parameter set
predicts that the TTV super-period is longer than the time spanning the
two clustered TESS observations.
We decided to track the potential evolution of the TTV signals over a temporal baseline of $\sim$\,5.2 years.
To this end, we ran forward numerical N-body simulations with \trades{}, setting the initial conditions of the orbital parameters to be integrated at $t=T_{\mathrm{ref}}$.
We ran a first simulation taking all the parameters from the MAP solution.
Then, we further ran 200 simulations, where the sets of the system's parameters were randomly drawn from the merged posterior distributions.
We computed the synthetic (i.e. observed: $O$) transit times $T_{\mathrm{tr}}$ and created a simulated $O-C$ plot against time, where the $C$ counterpart of $T_{\mathrm{tr}}$ was computed assuming the linear ephemerides of Table~\ref{tab:planets}. 
The results are displayed in Fig.~\ref{fig:oc_trades_predicted} that emphasises a progressive drift of the $O-C$ values inferred from the MAP parameters (black line) with respect to the zero value. As a consequence, the TTV amplitudes of planets b and c increase with time. In particular, the semi-amplitude of the $O-C$ black curves are about two and five hours for b and c, respectively.
The TTV super-period seems to be roughly equal to or larger than the integration time span, that is $\sim$\,5\,years.
The variance of the TTV amplitude (shaded area) is inferred from the results of the additional 200 simulations and reflects the widths of the posterior distributions from which the system's parameters were drawn. The remarkable $O-C$ drifts (i.e. the poorly constrained linear ephemerides) combined with the uncertainty on the TTV amplitudes make challenging to plan future observations of planets b and c, as the actual transit timings might differ from the linear ephemerides predictions by $\sim$\,5 and $\sim$\,10 hours for TOI-396\,b and TOI-396\,c, respectively.
\tess will not observe the target in the foreseeable future.\footnote{\url{https://heasarc.gsfc.nasa.gov/wsgi-scripts/TESS/TESS-point_Web_Tool/TESS-point_Web_Tool/wtv_v2.0.py/}}

\begin{table*}
\centering\small
\caption{\label{tab:trades_parameters}
    Best-fit parameters (MAP and HDI at 68.27\%) along with their respective priors as inferred from the dynamical joint modelling of RVs and TTVs with \trades{}.
}
    \begin{tabular}{l c c c c c c}
    \hline
    \hline
     & \multicolumn{2}{c}{TOI-396\,b} & \multicolumn{2}{c}{TOI-396\,c} & \multicolumn{2}{c}{TOI-396\,d} \\
     & MAP (HDI) & Prior & MAP (HDI) & Prior & MAP (HDI) & Prior \\
    \hline
    $M_\mathrm{p}/M_\star$ \; [$10^{-6}$] &
    $8.62_{-2.26}^{+0.76}$ & \up{0.02}{52.0} &
    $5.59_{-1.69}^{+0.25}$ & \up{0.02}{52.0} &
    $17_{-1}^{+3}$ & \up{0.02}{52.0} \\

    $P$ \;[d] &
    $3.585535_{-0.000171}^{+0.000092}$ & \up{3}{4} &
    $5.97291_{-0.00031}^{+0.00058}$ & \up{5}{7} &
    $11.230355_{-0.000016}^{+0.000173}$ & \up{10}{12} \\

    $\sqrt{e}\cos\omega$ &
    $-0.099_{-0.014}^{+0.155}$ & \up{-\sqrt{0.5}}{\sqrt{0.5}} &
    $-0.264_{-0.017}^{+0.060}$ & \up{-\sqrt{0.5}}{\sqrt{0.5}} &
    $-0.168_{-0.035}^{+0.069}$ & \up{-\sqrt{0.5}}{\sqrt{0.5}} \\
    $\sqrt{e}\sin\omega$ &
    $-0.268_{-0.031}^{+0.038}$ & \up{-\sqrt{0.5}}{\sqrt{0.5}} &
    $-0.066_{-0.042}^{+0.062}$ & \up{-\sqrt{0.5}}{\sqrt{0.5}} &
    $0.220_{-0.071}^{+0.062}$ & \up{-\sqrt{0.5}}{\sqrt{0.5}} \\

    $\lambda$ \;[$^{\circ}$] &
    $122.53_{-5.50}^{+0.21}$ & \up{0}{360} &
    $230.04_{-3.53}^{+0.61}$ & \up{0}{360} &
    $9.93_{-2.99}^{+0.55}$   & \up{0}{360} \\

    $M_\mathrm{p}$ \; [$M_\oplus$] & 
    $3.46_{-0.97}^{+0.27}$ & (derived) &
    $2.24_{-0.67}^{+0.13}$ & (derived) &
    $6.89_{-0.60}^{+1.41}$ & (derived) \\

    $e$ & 
    $0.082_{-0.020}^{+0.011}$ & \hgp{0}{0.083} &
    $0.0741_{-0.0275}^{+0.0071}$ & \hgp{0}{0.083} &
    $0.077_{-0.031}^{+0.013}$ & \hgp{0}{0.083} \\

    $\omega$ \; [$^{\circ}$] & 
    $250_{-5}^{+31}$ & (derived) &
    $194_{-13}^{+12}$ & (derived) &
    $127_{-18}^{+16}$ & (derived) \\

    $\mathcal{M}$ \; [$^{\circ}$] & 
    $53_{-37}^{+5}$ & (derived) &
    $216_{-14}^{+12}$ & (derived) &
    $63_{-15}^{+16}$ & (derived) \\

    $\rho_\mathrm{p}$ \; [$\rho_\oplus$] & 
    $0.426_{-0.136}^{+0.013}$ & (derived) &
    $0.288_{-0.096}^{+0.003}$ & (derived) &
    $0.842_{-0.123}^{+0.12}$ & (derived) \\

    $\rho_\mathrm{p}$ \; [g\,cm$^{-3}$] & 
    $2.339_{-0.750}^{+0.073}$ & (derived) &
    $1.582_{-0.528}^{+0.019}$ & (derived) &
    $4.624_{-0.676}^{+0.704}$ & (derived) \\

    \noalign{\smallskip}
    \hline
    \noalign{\smallskip}
      & MAP (HDI) & Prior & & & & \\
        
    $\gamma_\mathrm{RV}$ \; [m\,s$^{-1}$] &
    $0.059_{-0.234}^{+0.095}$ & \up{-5005}{+5005} & & & & \\
    
    $\log_{2}\sigma_\mathrm{jitter}$ &
    $-32_{-11}^{+19}$ & \up{-49.83}{6.65} & & & & \\ 
    
    $\sigma_\mathrm{jitter}$ \; [m\,s$^{-1}$] &
    $\ll 10^{-5}$ & (derived) & & & & \\
        
    \hline
    \end{tabular}
\tablefoot{All parameters have been defined at the reference time 
$T_\mathrm{ref} = \mathrm{BJD_{TDB}} - 2\,450\,000= 8379$.
\up{X}{Y} means uniform distribution between $X$ and $Y$ values;
\hgp{\mu}{\sigma} means half-normal distribution with mean $\mu$ and standard deviation $\sigma$.
}
\end{table*}

\section{Internal structure}\label{sec:internalStructure}
Using the masses and transit depths reported in Table~\ref{tab:planets}, we ran the neural network based internal structure modelling framework \texttt{plaNETic}\footnote{\url{https://github.com/joannegger/plaNETic}} \citep{Egger+2024} to infer the internal structure of TOI-396~b and d. \texttt{plaNETic} uses a full grid accept-reject sampling method in combination with a deep neural network (DNN) that was trained on the forward model of BICEPS \citep{haldemann2024} to infer the internal structure of observed planets. Each planet is modelled as a three-layered structure: an inner iron-dominated core, a silicate mantle and a fully mixed envelope made up of water and H/He. In the case of multi-planet systems, all planets are modelled simultaneously.

As modelling the internal structure of exoplanets is a highly degenerate problem, the resulting inferred structure is, at least to a certain extent, dependent on the chosen priors. To mitigate this effect, we ran a total of six models assuming six different combinations of priors. Most importantly, we use two different priors for the water content of the modelled planet, one motivated by a formation scenario outside the iceline (case~A, water-rich) and one compatible with a formation inside the iceline (case~B, water-poor). For both of these water priors, we choose three different options for the planetary Si/Mg/Fe ratios. In a first case, we assume that these match the stellar Si/Mg/Fe ratios exactly \cite{thiabaud2015}. Second, we assume that the planet is enriched in iron compared to its host star by using the fit of \cite{Adibekyan2021}. For option~3, we model the planet independent of the stellar Si/Mg/Fe ratios, but just sampling the planetary ratios uniformly from the simplex where the molar Si, Mg and Fe ratios add up to~1, with an upper bound of 0.75 for Fe. These priors are described in more detail in \cite{Egger+2024}.

\begin{figure*}
    \centering
    \includegraphics[width=\linewidth]{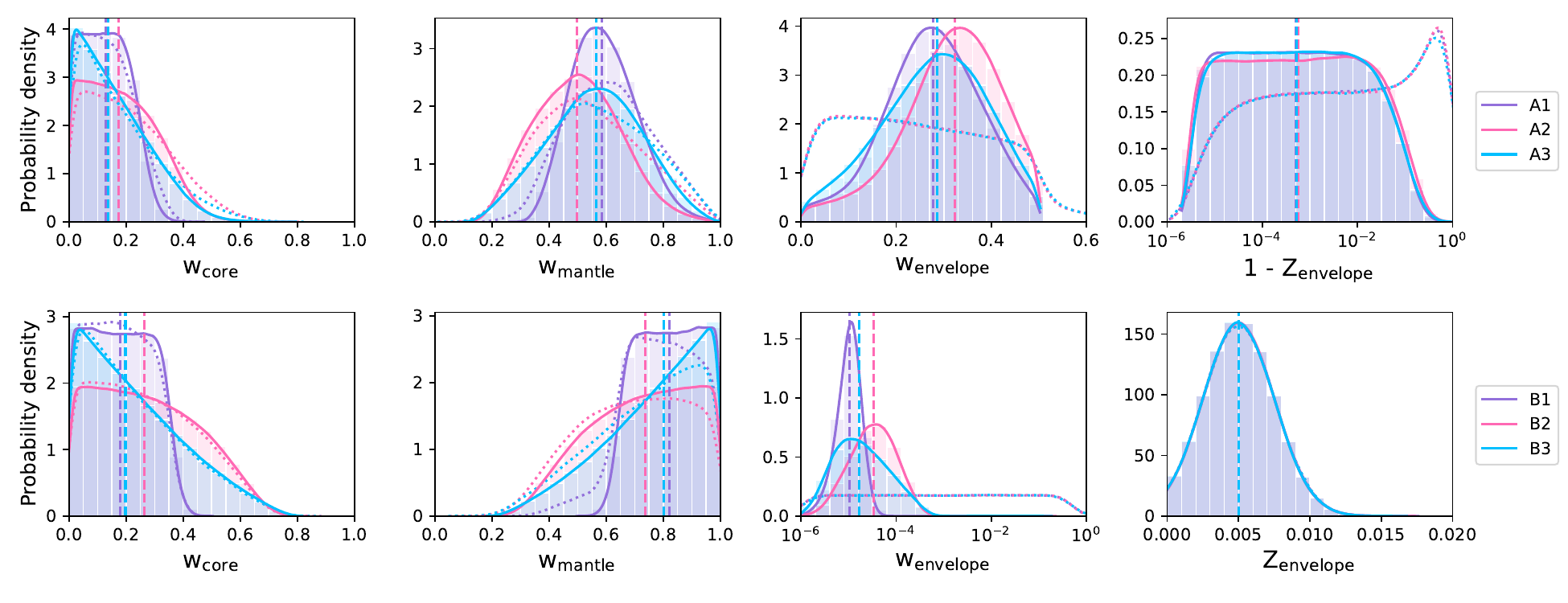}
    \caption{Inferred posteriors for the most important internal structure parameters of TOI-396 b. The depicted parameters are the mass fractions of the inner core (w$_\mathrm{core}$), mantle (w$_\mathrm{mantle}$) and envelope layers (w$_\mathrm{envelope}$) with respect to the total planet mass, and the mass fraction of water in the envelope (Z$_\mathrm{envelope}$). The top row shows the results when assuming a water prior motivated by a formation of the planet outside the iceline (case A), while the bottom row uses a water prior compatible with a formation inside the iceline (case B). At the same time, we run models with three different compositional priors for the planetary Si/Mg/Fe ratios: stellar (purple, option 1), iron-enriched compared to the star (pink, option 2) and sampled using a uniform prior (blue, option 3). The dotted lines show the prior distributions, while the dashed vertical lines show the median values of the posteriors.}
    \label{fig:int_struct_b}
\end{figure*}

\begin{figure*}
    \centering
    \includegraphics[width=\linewidth]{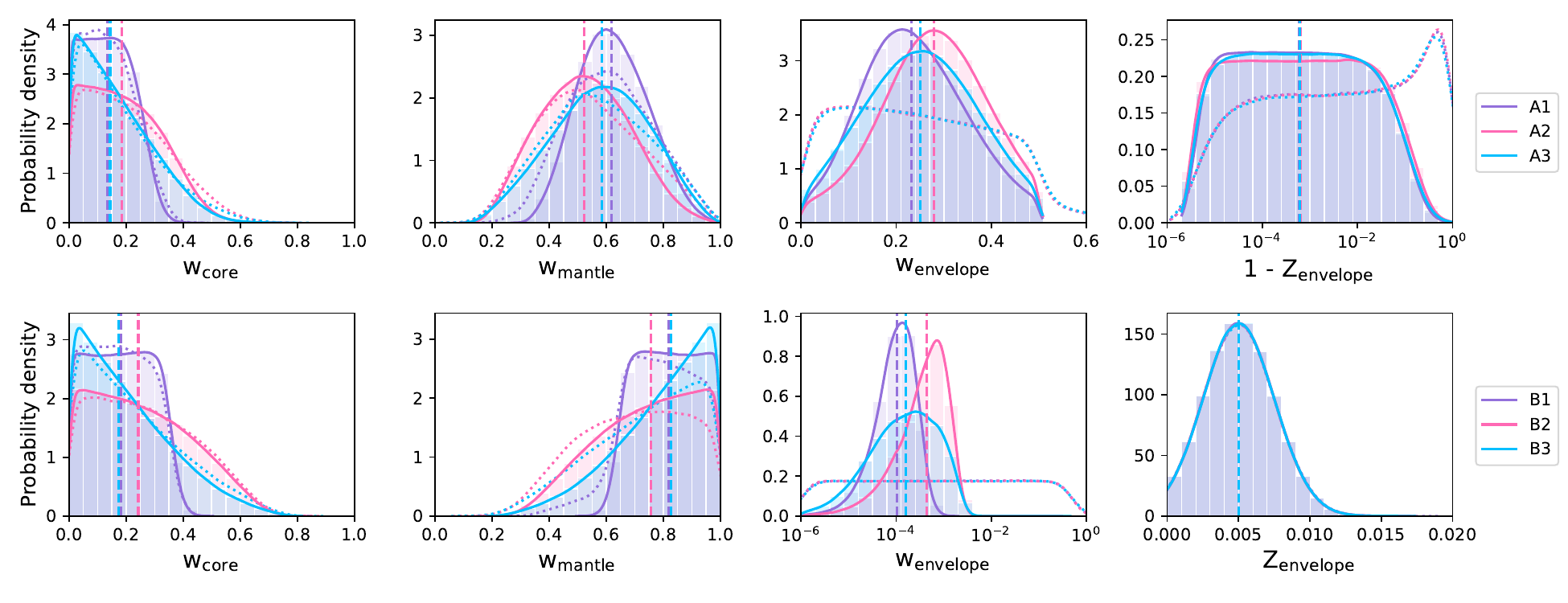}
    \caption{Same as Figure \ref{fig:int_struct_b} but for TOI-396 d.}
    \label{fig:int_struct_d}
\end{figure*}

Figures~\ref{fig:int_struct_b} and \ref{fig:int_struct_d} show the resulting posteriors of the most important interior structure parameters for TOI-396~b and d, respectively, in comparison with the chosen priors (dotted lines). Tables~\ref{tab:internal_structure_results_b} and \ref{tab:internal_structure_results_d} in the Appendix summarise the median and one sigma error intervals for the full set of internal structure parameters. For both planets, the posterior distributions for the core and mantle mass fractions largely agree with the chosen priors for each of the six models that we ran. 
Indeed, the only planetary structure parameters for which the observational data contributed to their characterisation were the envelope mass fractions.
If we assume that the planets formed outside the iceline, we find envelope mass fractions of $28\pm10$\%~(A1), $32^{+9}_{-11}$\%~(A2) and $30^{+11}_{-13}$\%~(A3) for planet~b and $23^{+12}_{-10}$\%~(A1), $28\pm11$\%~(A2) and $26^{+12}_{-13}$\%~(A3) for planet~d, with water mass fractions in the envelope of almost 100\%. Conversely, if the planets were to have formed inside the iceline, we infer envelope mass fractions of the order of $10^{-5}$ for planet~b and $10^{-4}$ for planet~d, almost entirely made up of H/He.

\section{JWST characterisation prospects}\label{sec:JWST}

\begin{figure}
\centering
\includegraphics[width=\linewidth]{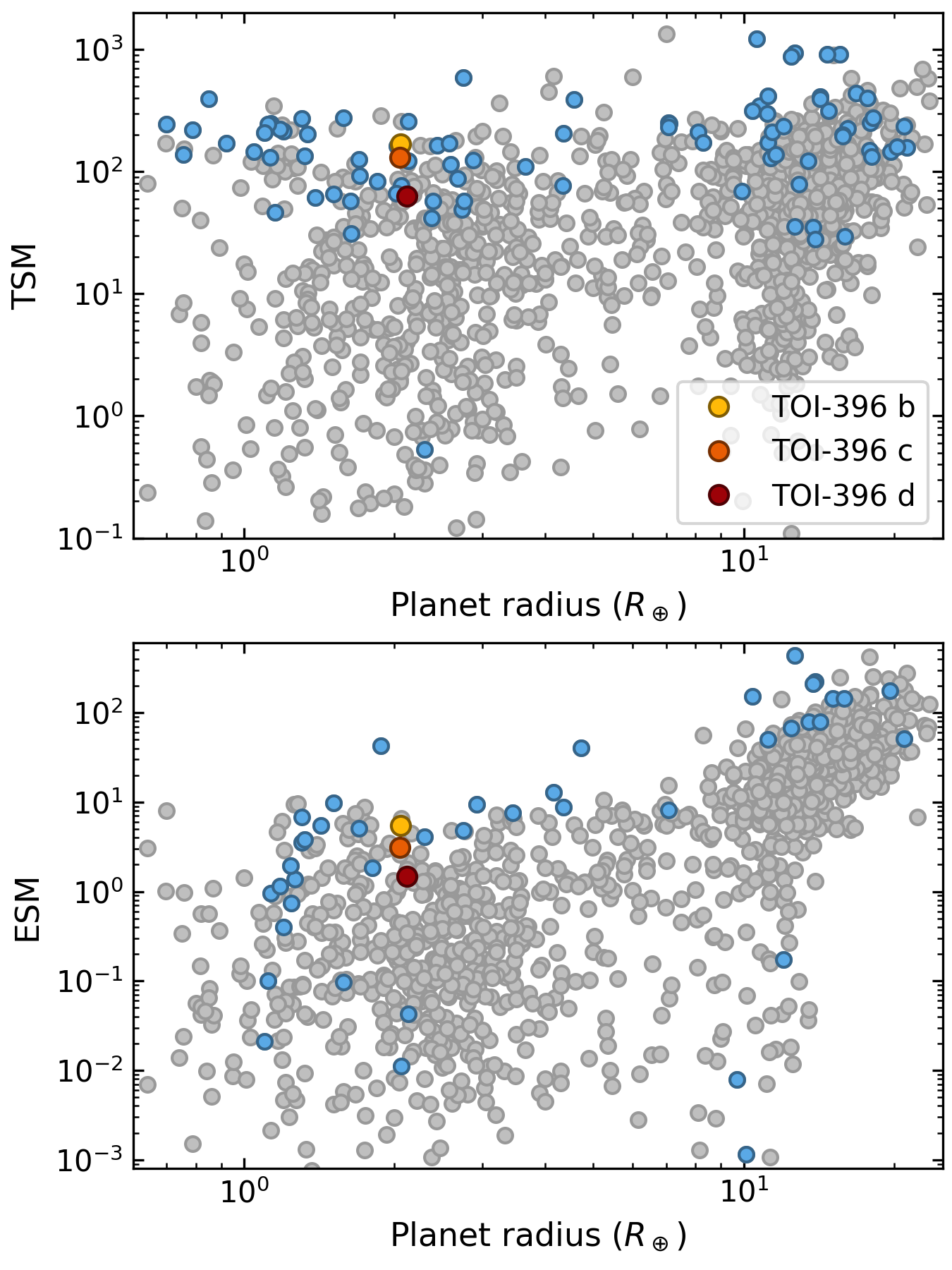}
\caption{Transit (\textit{top panel}) and eclipse (\textit{bottom panel}) spectroscopic metrics for the TOI-396 planets (see legend). The metrics were calculated using the 2MASS Ks-band magnitude. The grey markers show the metrics for the known sample of transiting exoplanets to date. The blue markers show the metrics for targets with approved JWST programmes.}
\label{fig:jwst_metrics}
\end{figure}

\begin{figure}
\centering
\includegraphics[width=\linewidth]{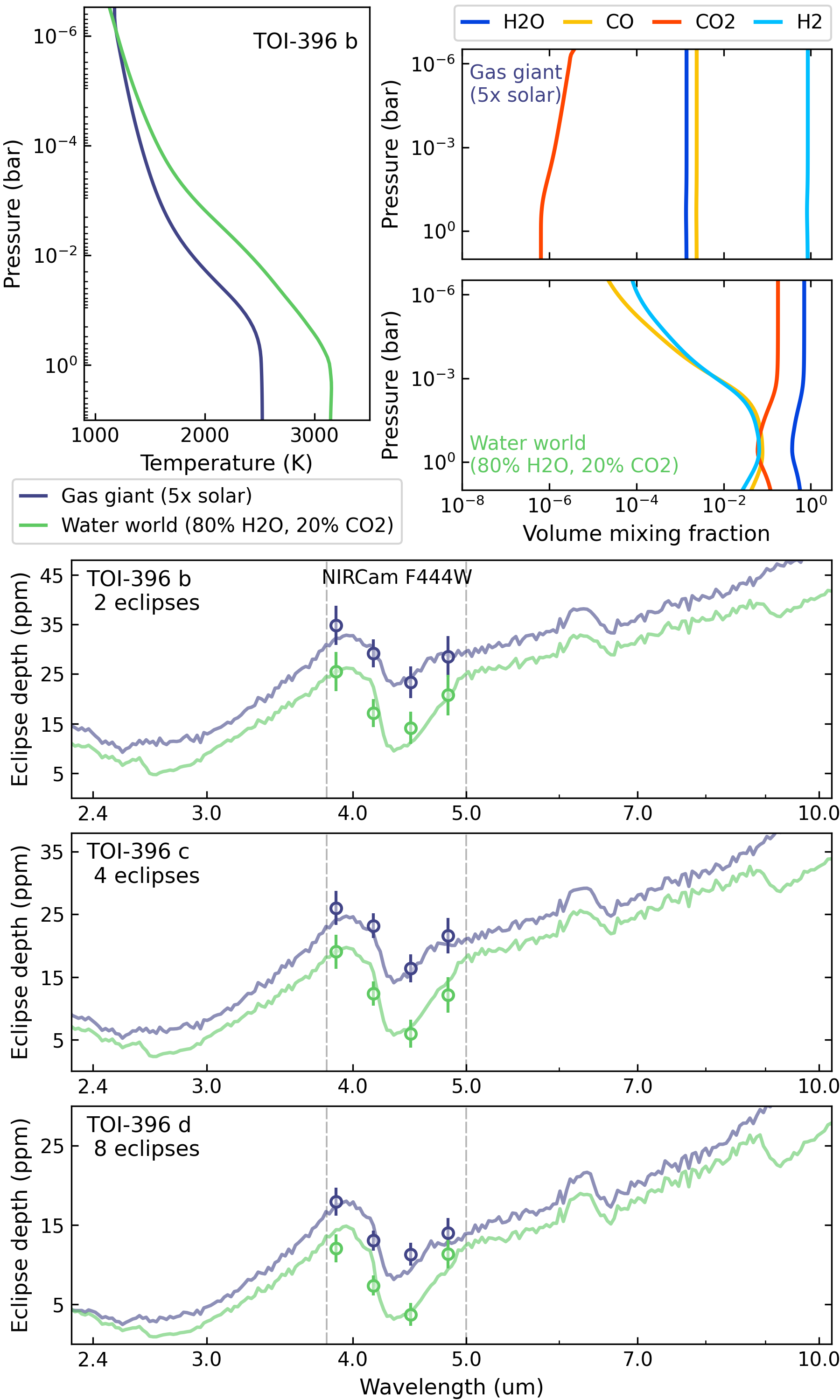}
\caption{Simulations of the atmospheric pressure profiles, eclipse spectra, and JWST observations.  \textit{Top-left panel}:
  Radiative-equilibrium thermal profile of TOI-396\,b assuming a
  gas-giant atmosphere (5$\times$ solar metallicity) and a secondary
  atmosphere (80\% \ch{H2O} plus 20\% \ch{CO2}). \textit{Top-right panels}: Volume mixing ratio of TOI-396\,b for the most relevant species shaping the infrared spectrum. 
  \textit{Bottom panels}:
  Synthetic secondary eclipse spectra of the three TOI-396 planets (solid curves). The round markers with error bars show a realisation of JWST observation with NIRCam/F444W (and their
  expected uncertainties) when accumulating 2, 4, and 8 observations
  for planets b, c, and d, respectively. The vertical dashed lines mark the spectral window covered by NIRCam/F444W.}
\label{fig:jwst_simulations}
\end{figure}

All three planets in the TOI-396 system share similar radii
($\sim$2~$R_\oplus$), but span a wide range of masses
(0.9--7.1~$M_\oplus$), which leaves open the question of whether they
have primary or secondary atmospheres.  Furthermore, the progression
of bulk densities with distance from the host star varies in ways that
cannot be described by simple formation and evolution models \citep[e.g.][]{weiss2018,mishra2023}.

Given the bright host star and combination of planetary masses, radii, and equilibrium temperatures, the three planets have favourable metrics for atmospheric characterisation in both transmission and emission among sub-Neptunes \citep[][see Fig.\ \ref{fig:jwst_metrics}]{KemptonEtal2018paspTransitSpectroscopicMetric}.
This makes the TOI-396 system a highly valuable laboratory to study
the formation and evolution of planetary systems.  Thus, we explored the prospects for characterisation with JWST.  We focused these simulations on emission observations, but we note that transmission and emission have their own advantages and disadvantages in terms of achievable science goals and challenges.

We employed the open-source \textsc{Pyrat Bay} modelling framework
\citep{CubillosBlecic2021mnrasPyratBay} to compute synthetic spectra
of the TOI-396 planets.
These models consist of 1D cloud-free atmospheres in radiative,
thermochemical, and hydrostatic equilibrium (Cubillos et al., in
prep.).
%%%
We varied the models' atmospheric elemental content to explore the
wide range of compositions that the planets span.  For this comparison
we settled on two models to represent a primary- and a
secondary-atmosphere scenario: the first is a gas giant with a
5$\times$ solar metallicity, the second is a water world with a 80\%
\ch{H2O} plus 20\% \ch{CO2} composition \citep[based on the C/O ratios
  seen in the solar system minor bodies, see, e.g.,][]{MummaCharnley2011araaCometsChemicalComposition,
  McKayEtal2019ajPanSTARRScometCO}.
%%%
For the thermochemical-equilibrium calculations we considered a set of
45 neutral and ionic species, which are the main actors determining
the thermal structure.  For the radiative-transfer calculation we
considered opacities from molecular species for CO, \ch{CO2},
\ch{CH4}, \ch{H2O}, HCN, \ch{NH3}, and \ch{C2H2} from \textsc{hitemp} and \textsc{ExoMol} \citep[][]{RothmanEtal2010jqsrtHITEMP,
  TennysonEtal2016jmsExomol}; Na and K resonant lines
\citep{BurrowsEtal2000apjBDspectra}; H, H$_2$, and He Rayleigh
\citep{Kurucz1970saorsAtlas}; and H$_2$--H$_2$ and H$_2$--He
collision-induced absorption \citep{BorysowEtal2001jqsrtH2H2highT,
  Borysow2002jqsrtH2H2lowT, RichardEtal2012jqsrtCIA}.  We
pre-processed the large \textsc{ExoMol} line lists with the \textsc{Repack}
algorithm \citep{Cubillos2017apjRepack} to extract the dominant
transitions.
%%%
Figure \ref{fig:jwst_simulations} (top panels) shows the resulting
thermal and composition structure for TOI-396\,b (planets c and d follow
a similar trend).

The infrared synthetic emission spectra
(Fig.\ \ref{fig:jwst_simulations}, bottom panels) are mainly shaped by
\ch{H2O}, \ch{CO2}, and \ch{CO} features.  At most wavelengths the
primary- and secondary-atmosphere scenarios roughly differ by an
offset, which would be hard to distinguish unless the energy budget of
the planets are known.  In contrast, the 4--5 {\micron} window shows
the most distinctive spectral features; here the strong \ch{CO2}
absorption band at 4.4~{\microns} mainly allows one to distinguish
primary from secondary atmospheres.  Thus, in the following we focus
on this region of the spectrum.

We simulated JWST observations using the \textsc{Pandeia} exposure time
calculator \citet{PontoppidanEtal2016spiePandeia}.  The
brightness of TOI-396 limits the instrument selection to NIRCam (F444W
filter) to avoid saturation.  We selected the fastest readout and
subarray modes, 5 groups per integration, to optimise the S/N.  We
generated a distribution of (noised up) realisations for each model to
estimate how many eclipses are required to distinguish between primary and
secondary atmospheres at the 3$\sigma$ level.  We found that 2, 4, and
8 eclipses (for planets b, c, and d, respectively) would be sufficient
to differentiate between these two models.  Figure
\ref{fig:jwst_simulations} shows one of those random realisations when
including the required number of eclipses.
The decreasing equilibrium
temperature of the planets as they are located further away from
TOI-396 plays the major role in the decreasing S/N for planets c and d.

\section{Conclusions}\label{sec:conclusions}
The object TOI-396 is an F6\,V bright naked-eye star orbited by three planets of almost equal size, and the two inner planets are close to but out of a 5:3 MMR.
A photometric analysis of the system was already performed by \citet{Vanderburg2019}, but by benefiting from two additional \tess sectors, we improved the precision on the planet radii by a factor of $\sim$\,1.4, obtaining $R_b=2.004_{-0.047}^{+0.045}\,R_{\oplus}$, $R_c=1.979_{-0.051}^{+0.054}\,R_{\oplus}$, and $R_d=2.001_{-0.064}^{+0.063}\,R_{\oplus}$.

We determined the masses of the planets by extracting the RV time series from \harps CCFs using an SN fit followed by a joint LC and RV MCMC analysis, where the RV de-trending uses the breakpoint method.
We obtained a firm detection of the RV signals of planets b and d, deriving $M_b=3.55_{-0.96}^{+0.94}\,M_{\oplus}$ and $M_d=7.1\pm1.6\,M_{\oplus}$, but we can provide only a 3$\sigma$ upper limit for the mass of TOI-396\,c of $M_c^{\mathrm{up}}=3.8\,M_{\oplus}$. This yields the following mean planet densities: $\rho_b=2.44_{-0.68}^{+0.69}$, $\rho_c^{\mathrm{up}}=2.9$, and $\rho_d=4.9_{-1.1}^{+1.2}$ g\,cm$^{-3}$, implying a quite unusual system architecture \citep{mishra2023} where the mid planet is the least dense and the outermost planet is the densest.

The reason for the RV non-detection of any Keplerian signal at $P=P_c$\,$\sim$\,6 d is likely to be ascribed to the vicinity of $P_c$ to the stellar rotation period. As a matter of fact, from the GLS periodograms of both the RV-related activity indices and the \tess raw LCs and from $\log{R'_{\mathrm{HK}}}$-based empirical relations, we consistently inferred $\Prot=6.7\pm1.3$ d. After injecting synthetic Keplerian signals at $P=\Prot$ and different semi-amplitudes ($K_{\mathrm{in}}$) into the RV time series, we empirically find that the RV semi-amplitudes output by the MCMC analyses ($K_{\mathrm{out}}$) are systematically lower than the input ones by almost 3$\sigma$, and they are statistically non-significant as far as $K_{\mathrm{in}}\lesssim K_d$. In addition, we find that $K_{\mathrm{out}}\approx K_c$ when considering a planet with $M_p$\,$\sim$\,3\,$M_{\oplus}$ (i.e. $\rho_p$\,$\sim$\,2~g\,cm$^{-3}$), which might correspond to the properties of TOI-396\,c. On a more general perspective, these simulations confirm that stellar activity destructively interferes with Keplerian signals having $P$\,$\sim$\,$\Prot$ \citep[e.g.][]{vanderburg2016}, and furthermore, they indicate that -- even in the case of firm detection -- values of $K_{\mathrm{out}}$ are significantly underestimated. 

Longer-baseline RV observations may help disentangle coherent signals originated by Keplerian motions from non-coherent signals due to stellar activity, even if degeneracy issues still hold when the planet orbital period is close to the stellar rotation period \citep{kossakowski2022}. Alternatively, a possible constraint on $M_c$ may come from TTVs, as planets b and c are close to an MMR of the second order.
Indeed, the TTV amplitudes of the two planets show a characteristic anti-correlation pattern, as expected; however, the phase coverage given by the available observations is too poor to perform a conclusive TTV dynamical analysis based on the observed transit timings of the planets. 
We also attempted to fit the TTV and RV simultaneously while integrating the orbits of the system.
We found that the masses and densities of planets b and d are consistent with the results from the joint LC and RV analysis.
TOI-396\,c shows a dynamical mass of $M_{c,\mathrm{dyn}}=2.24_{-0.67}^{+0.13}\, M_\oplus$, which is greater than that inferred from the joint LC and RV analysis, but it is consistent ($\text{Z-score} = 1.2\sigma$); the density is consistent at the $1.1\sigma$ level. However, we emphasise that, although formally precise, the $M_{c,\mathrm{dyn}}$ estimate might not be accurate, as the full coverage of the TTV phase is needed to reliably compute TTV-based masses. Therefore, to fully confirm the system architecture, a reliable estimate of the mass of TOI-396\,c is still missing.

We also checked the evolution of the system over 10\,000 years, and the critical resonance angles showed that planets b and c are close to but not in a 5:3 MMR.
We further
performed forward N-body simulations over a temporal baseline of $\sim$\,5.2 years in order to track the transit epochs and evaluate the expected TTV amplitudes during time. 
It turns out that TOI-396\,b and TOI-396\,c may exhibit TTVs 
with a super-period of about 5 years and semi-amplitudes of
$\sim$\,2 and $\sim$\,5 hours, respectively. This translates into a temporal drift of the transit timings that can rise up to $\sim$\,5 and $\sim$\,10 hours with respect to the linear ephemerides computed from \tess data.

Studying the planetary atmospheres with JWST would take advantage of the favourable spectroscopy metrics of the system \citep{KemptonEtal2018paspTransitSpectroscopicMetric}. Therefore, we set up 1D cloud-free atmospheric models, generated the synthetic emission spectra of the three planets, and simulated eclipse observations with JWST. It turns out that 2, 4, and 8 eclipses (for TOI-396\,b, c, and d, respectively) would be sufficient to distinguish between primary and secondary atmosphere scenarios at the 3$\sigma$ level.

Characterising the nature of the planetary atmosphere is also key to correctly assessing the planetary bulk densities (in particular for planet c). The potentially high TTVs inferred from our simulations should be duly taken into account when scheduling future observations of the target. This holds not only for JWST, but also for \cheops \citep{benz2021}, which appears especially suitable for collecting exquisite photometric data to enable the full characterisation of the system.

\begin{acknowledgements}
We thank the anonymous referee for all the valuable comments that significantly improved the quality of the manuscript.
This work has made use of data from the European Space Agency (ESA) mission {\it Gaia} (\url{https://www.cosmos.esa.int/gaia}), processed by the {\it Gaia} Data Processing and Analysis Consortium (DPAC, \url{https://www.cosmos.esa.int/web/gaia/dpac/consortium}).
This research made use of Lightkurve, a Python package for Kepler and TESS data analysis \citep{lightkurve2018}. We thank contributors to \texttt{NumPy} \citep{numpy2020}, \texttt{SciPy} \citep{sciPy2020}, \texttt{matplotlib} \citep{matplotlib2007}, \texttt{astropy} \citep{astropy2013,astropy2018,astropy2022}, \texttt{astroquery} \citep{astroquery2019}, and \texttt{tesscut} \citep{tesscut2019}.
Funding for the DPAC has been provided by national institutions, in particular the institutions participating in the {\it Gaia} Multilateral Agreement.
We acknowledge financial support from the Agencia Estatal de Investigaci\'on of the Ministerio de Ciencia e Innovaci\'on MCIN/AEI/10.13039/501100011033 and the ERDF “A way of making Europe” through project PID2021-125627OB-C32, and from the Centre of Excellence “Severo Ochoa” award to the Instituto de Astrofisica de Canarias.
This research was funded in part by the UKRI, (Grants ST/X001121/1, EP/X027562/1).
This work was supported by FCT - Fundação para a Ciência e a Tecnologia through national funds and by FEDER through COMPETE2020 - Programa Operacional Competitividade e Internacionalização by these grants: UIDB/04434/2020; UIDP/04434/2020.
D.G., A.B., L.F., and L.M.S. gratefully acknowledge the financial support from the grant for internationalization (GAND\_GFI\_23\_01) provided by the University of Turin (Italy).
S.G.S acknowledges the support from FCT through Investigador FCT contract nr. CEECIND/00826/2018 and  POPH/FSE (EC).
P.J.W. acknowledges support from the UK Science and Technology Facilities Council (STFC) through consolidated grants ST/P000495/1, ST/T000406/1 and ST/X001121/1.
N.C.S. is funded by the European Union (ERC, FIERCE, 101052347). Views and opinions expressed are however those of the author(s) only and do not necessarily reflect those of the European Union or the European Research Council. Neither the European Union nor the granting authority can be held responsible for them.
J.L.-B. is funded by the MICIU/AEI/10.13039/501100011033 and NextGenerationEU/PRTR grant PID2019-107061GB-C61 and CNS2023-144309.
X.D. acknowledges the support from the European Research Council (ERC) under the European Union’s Horizon 2020 research and innovation programme (grant agreement SCORE No 851555) and from the Swiss National Science Foundation under the grant SPECTRE (No 200021\_215200). This work has been carried out within the framework of the NCCR PlanetS supported by the Swiss National Science Foundation under grants 51NF40\_182901 and 51NF40\_205606.
G.N. thanks for the research funding from the Ministry of Science and Higher Education programme the "Excellence Initiative - Research University" conducted at the Centre of Excellence in Astrophysics and Astrochemistry of the Nicolaus Copernicus University in Toru\'n, Poland.
Research activities of the Board of Observational and Instrumental Astronomy at the Federal University of Rio Grande do Norte are supported by continuous grants from the Brazilian funding agencies CNPq. This study was financed in part by the Coordenação de Aperfeiçoamento de Pessoal de Nível Superior—Brasil (CAPES)—Finance Code 001 and CAPES-Print program. B.L.C.M., I.C.L., and J.R.M. acknowledge CNPq research fellowships.
K.W.F.L. was supported by Deutsche Forschungsgemeinschaft grants RA714/14-1, RA714/14-2 within the DFG Schwerpunkt SPP 1992, Exploring the Diversity of Extrasolar Planets.
L.B. acknowledges support from CHEOPS ASI-INAF agreement n. 2019-29-HH.0.
D.G. sincerely thanks Stefano Camera for the inspiring and valuable discussions on the properties of TOI-396.

\end{acknowledgements}

\bibliographystyle{aa}
\bibliography{biblio}

\begin{thebibliography}{159}
\expandafter\ifx\csname natexlab\endcsname\relax\def\natexlab#1{#1}\fi

\bibitem[{{Addison} {et~al.}(2019){Addison}, {Wright}, {Wittenmyer}, {Horner},
  {Mengel}, {Johns}, {Marti}, {Nicholson}, {Soutter}, {Bowler}, {Crossfield},
  {Kane}, {Kielkopf}, {Plavchan}, {Tinney}, {Zhang}, {Clark}, {Clerte},
  {Eastman}, {Swift}, {Bottom}, {Muirhead}, {McCrady}, {Herzig}, {Hogstrom},
  {Wilson}, {Sliski}, {Johnson}, {Wright}, {Johnson}, {Blake}, {Riddle}, {Lin},
  {Cornachione}, {Bedding}, {Stello}, {Huber}, {Marsden}, \&
  {Carter}}]{addison2019}
{Addison}, B., {Wright}, D.~J., {Wittenmyer}, R.~A., {et~al.} 2019, \pasp, 131,
  115003

\bibitem[{{Adibekyan} {et~al.}(2021){Adibekyan}, {Dorn}, {Sousa}, {Santos},
  {Bitsch}, {Israelian}, {Mordasini}, {Barros}, {Delgado Mena}, {Demangeon},
  {Faria}, {Figueira}, {Hakobyan}, {Oshagh}, {Soares}, {Kunitomo}, {Takeda},
  {Jofr{\'e}}, {Petrucci}, \& {Martioli}}]{Adibekyan2021}
{Adibekyan}, V., {Dorn}, C., {Sousa}, S.~G., {et~al.} 2021, Science, 374, 330

\bibitem[{{Adibekyan} {et~al.}(2015){Adibekyan}, {Figueira}, {Santos}, {Sousa},
  {Faria}, {Delgado-Mena}, {Oshagh}, {Tsantaki}, {Hakobyan}, {Gonz{\'a}lez
  Hern{\'a}ndez}, {Su{\'a}rez-Andr{\'e}s}, \& {Israelian}}]{Adibekyan-15}
{Adibekyan}, V., {Figueira}, P., {Santos}, N.~C., {et~al.} 2015, \aap, 583, A94

\bibitem[{{Adibekyan} {et~al.}(2012){Adibekyan}, {Sousa}, {Santos}, {Delgado
  Mena}, {Gonz{\'a}lez Hern{\'a}ndez}, {Israelian}, {Mayor}, \&
  {Khachatryan}}]{Adibekyan-12}
{Adibekyan}, V.~Z., {Sousa}, S.~G., {Santos}, N.~C., {et~al.} 2012, \aap, 545,
  A32

\bibitem[{{Agol} \& {Fabrycky}(2018)}]{AgolFabrycky2018}
{Agol}, E. \& {Fabrycky}, D.~C. 2018, in Handbook of Exoplanets, ed. H.~J.
  {Deeg} \& J.~A. {Belmonte} (Cambridge University Press), 7

\bibitem[{{Agol} {et~al.}(2005){Agol}, {Steffen}, {Sari}, \&
  {Clarkson}}]{agol2005}
{Agol}, E., {Steffen}, J., {Sari}, R., \& {Clarkson}, W. 2005, \mnras, 359, 567

\bibitem[{{Aguichine} {et~al.}(2021){Aguichine}, {Mousis}, {Deleuil}, \&
  {Marcq}}]{aguichine2021}
{Aguichine}, A., {Mousis}, O., {Deleuil}, M., \& {Marcq}, E. 2021, \apj, 914,
  84

\bibitem[{{Alexander} \& {Armitage}(2009)}]{alexander2009}
{Alexander}, R.~D. \& {Armitage}, P.~J. 2009, \apj, 704, 989

\bibitem[{{Astropy Collaboration} {et~al.}(2022){Astropy Collaboration},
  {Price-Whelan}, {Lim}, {Earl}, {Starkman}, {Bradley}, {Shupe}, {Patil},
  {Corrales}, {Brasseur}, {N{\"o}the}, {Donath}, {Tollerud}, {Morris},
  {Ginsburg}, {Vaher}, {Weaver}, {Tocknell}, {Jamieson}, {van Kerkwijk},
  {Robitaille}, {Merry}, {Bachetti}, {G{\"u}nther}, {Aldcroft},
  {Alvarado-Montes}, {Archibald}, {B{\'o}di}, {Bapat}, {Barentsen},
  {Baz{\'a}n}, {Biswas}, {Boquien}, {Burke}, {Cara}, {Cara}, {Conroy},
  {Conseil}, {Craig}, {Cross}, {Cruz}, {D'Eugenio}, {Dencheva}, {Devillepoix},
  {Dietrich}, {Eigenbrot}, {Erben}, {Ferreira}, {Foreman-Mackey}, {Fox},
  {Freij}, {Garg}, {Geda}, {Glattly}, {Gondhalekar}, {Gordon}, {Grant},
  {Greenfield}, {Groener}, {Guest}, {Gurovich}, {Handberg}, {Hart},
  {Hatfield-Dodds}, {Homeier}, {Hosseinzadeh}, {Jenness}, {Jones}, {Joseph},
  {Kalmbach}, {Karamehmetoglu}, {Ka{\l}uszy{\'n}ski}, {Kelley}, {Kern},
  {Kerzendorf}, {Koch}, {Kulumani}, {Lee}, {Ly}, {Ma}, {MacBride}, {Maljaars},
  {Muna}, {Murphy}, {Norman}, {O'Steen}, {Oman}, {Pacifici}, {Pascual},
  {Pascual-Granado}, {Patil}, {Perren}, {Pickering}, {Rastogi}, {Roulston},
  {Ryan}, {Rykoff}, {Sabater}, {Sakurikar}, {Salgado}, {Sanghi}, {Saunders},
  {Savchenko}, {Schwardt}, {Seifert-Eckert}, {Shih}, {Jain}, {Shukla}, {Sick},
  {Simpson}, {Singanamalla}, {Singer}, {Singhal}, {Sinha}, {Sip{\H{o}}cz},
  {Spitler}, {Stansby}, {Streicher}, {{\v{S}}umak}, {Swinbank}, {Taranu},
  {Tewary}, {Tremblay}, {de Val-Borro}, {Van Kooten}, {Vasovi{\'c}}, {Verma},
  {de Miranda Cardoso}, {Williams}, {Wilson}, {Winkel}, {Wood-Vasey}, {Xue},
  {Yoachim}, {Zhang}, {Zonca}, \& {Astropy Project Contributors}}]{astropy2022}
{Astropy Collaboration}, {Price-Whelan}, A.~M., {Lim}, P.~L., {et~al.} 2022,
  \apj, 935, 167

\bibitem[{{Astropy Collaboration} {et~al.}(2018){Astropy Collaboration},
  {Price-Whelan}, {Sip{\H{o}}cz}, {G{\"u}nther}, {Lim}, {Crawford}, {Conseil},
  {Shupe}, {Craig}, {Dencheva}, {Ginsburg}, {VanderPlas}, {Bradley},
  {P{\'e}rez-Su{\'a}rez}, {de Val-Borro}, {Aldcroft}, {Cruz}, {Robitaille},
  {Tollerud}, {Ardelean}, {Babej}, {Bach}, {Bachetti}, {Bakanov}, {Bamford},
  {Barentsen}, {Barmby}, {Baumbach}, {Berry}, {Biscani}, {Boquien}, {Bostroem},
  {Bouma}, {Brammer}, {Bray}, {Breytenbach}, {Buddelmeijer}, {Burke},
  {Calderone}, {Cano Rodr{\'\i}guez}, {Cara}, {Cardoso}, {Cheedella}, {Copin},
  {Corrales}, {Crichton}, {D'Avella}, {Deil}, {Depagne}, {Dietrich}, {Donath},
  {Droettboom}, {Earl}, {Erben}, {Fabbro}, {Ferreira}, {Finethy}, {Fox},
  {Garrison}, {Gibbons}, {Goldstein}, {Gommers}, {Greco}, {Greenfield},
  {Groener}, {Grollier}, {Hagen}, {Hirst}, {Homeier}, {Horton}, {Hosseinzadeh},
  {Hu}, {Hunkeler}, {Ivezi{\'c}}, {Jain}, {Jenness}, {Kanarek}, {Kendrew},
  {Kern}, {Kerzendorf}, {Khvalko}, {King}, {Kirkby}, {Kulkarni}, {Kumar},
  {Lee}, {Lenz}, {Littlefair}, {Ma}, {Macleod}, {Mastropietro}, {McCully},
  {Montagnac}, {Morris}, {Mueller}, {Mumford}, {Muna}, {Murphy}, {Nelson},
  {Nguyen}, {Ninan}, {N{\"o}the}, {Ogaz}, {Oh}, {Parejko}, {Parley}, {Pascual},
  {Patil}, {Patil}, {Plunkett}, {Prochaska}, {Rastogi}, {Reddy Janga},
  {Sabater}, {Sakurikar}, {Seifert}, {Sherbert}, {Sherwood-Taylor}, {Shih},
  {Sick}, {Silbiger}, {Singanamalla}, {Singer}, {Sladen}, {Sooley},
  {Sornarajah}, {Streicher}, {Teuben}, {Thomas}, {Tremblay}, {Turner},
  {Terr{\'o}n}, {van Kerkwijk}, {de la Vega}, {Watkins}, {Weaver}, {Whitmore},
  {Woillez}, {Zabalza}, \& {Astropy Contributors}}]{astropy2018}
{Astropy Collaboration}, {Price-Whelan}, A.~M., {Sip{\H{o}}cz}, B.~M., {et~al.}
  2018, \aj, 156, 123

\bibitem[{{Astropy Collaboration} {et~al.}(2013){Astropy Collaboration},
  {Robitaille}, {Tollerud}, {Greenfield}, {Droettboom}, {Bray}, {Aldcroft},
  {Davis}, {Ginsburg}, {Price-Whelan}, {Kerzendorf}, {Conley}, {Crighton},
  {Barbary}, {Muna}, {Ferguson}, {Grollier}, {Parikh}, {Nair}, {Unther},
  {Deil}, {Woillez}, {Conseil}, {Kramer}, {Turner}, {Singer}, {Fox}, {Weaver},
  {Zabalza}, {Edwards}, {Azalee Bostroem}, {Burke}, {Casey}, {Crawford},
  {Dencheva}, {Ely}, {Jenness}, {Labrie}, {Lim}, {Pierfederici}, {Pontzen},
  {Ptak}, {Refsdal}, {Servillat}, \& {Streicher}}]{astropy2013}
{Astropy Collaboration}, {Robitaille}, T.~P., {Tollerud}, E.~J., {et~al.} 2013,
  \aap, 558, A33

\bibitem[{Bai \& Perron(2003)}]{BaiPerron2003}
Bai, J. \& Perron, P. 2003, Journal of Applied Econometrics, 18, 1

\bibitem[{{Baranne} {et~al.}(1996){Baranne}, {Queloz}, {Mayor}, {Adrianzyk},
  {Knispel}, {Kohler}, {Lacroix}, {Meunier}, {Rimbaud}, \& {Vin}}]{Baranne1996}
{Baranne}, A., {Queloz}, D., {Mayor}, M., {et~al.} 1996, \aaps, 119, 373

\bibitem[{{Barnes}(2010)}]{barnes10}
{Barnes}, S.~A. 2010, \apj, 722, 222

\bibitem[{{Benz} {et~al.}(2021){Benz}, {Broeg}, {Fortier}, {Rando}, {Beck},
  {Beck}, {Queloz}, {Ehrenreich}, {Maxted}, {Isaak}, {Billot}, {Alibert},
  {Alonso}, {Ant{\'o}nio}, {Asquier}, {Bandy}, {B{\'a}rczy}, {Barrado},
  {Barros}, {Baumjohann}, {Bekkelien}, {Bergomi}, {Biondi}, {Bonfils},
  {Borsato}, {Brandeker}, {Busch}, {Cabrera}, {Cessa}, {Charnoz}, {Chazelas},
  {Collier Cameron}, {Corral Van Damme}, {Cortes}, {Davies}, {Deleuil},
  {Deline}, {Delrez}, {Demangeon}, {Demory}, {Erikson}, {Farinato}, {Fossati},
  {Fridlund}, {Futyan}, {Gandolfi}, {Garcia Munoz}, {Gillon}, {Guterman},
  {Gutierrez}, {Hasiba}, {Heng}, {Hernandez}, {Hoyer}, {Kiss}, {Kovacs},
  {Kuntzer}, {Laskar}, {Lecavelier des Etangs}, {Lendl}, {L{\'o}pez}, {Lora},
  {Lovis}, {L{\"u}ftinger}, {Magrin}, {Malvasio}, {Marafatto}, {Michaelis}, {de
  Miguel}, {Modrego}, {Munari}, {Nascimbeni}, {Olofsson}, {Ottacher},
  {Ottensamer}, {Pagano}, {Palacios}, {Pall{\'e}}, {Peter}, {Piazza}, {Piotto},
  {Pizarro}, {Pollaco}, {Ragazzoni}, {Ratti}, {Rauer}, {Ribas}, {Rieder},
  {Rohlfs}, {Safa}, {Salatti}, {Santos}, {Scandariato}, {S{\'e}gransan},
  {Simon}, {Smith}, {Sordet}, {Sousa}, {Steller}, {Szab{\'o}}, {Szoke},
  {Thomas}, {Tschentscher}, {Udry}, {Van Grootel}, {Viotto}, {Walter},
  {Walton}, {Wildi}, \& {Wolter}}]{benz2021}
{Benz}, W., {Broeg}, C., {Fortier}, A., {et~al.} 2021, Experimental Astronomy,
  51, 109

\bibitem[{{Bergner} {et~al.}(2020){Bergner}, {{\"O}berg}, {Bergin}, {Andrews},
  {Blake}, {Carpenter}, {Cleeves}, {Guzm{\'a}n}, {Huang}, {J{\o}rgensen}, {Qi},
  {Schwarz}, {Williams}, \& {Wilner}}]{bergner2020}
{Bergner}, J.~B., {{\"O}berg}, K.~I., {Bergin}, E.~A., {et~al.} 2020, \apj,
  898, 97

\bibitem[{{Bonfanti} {et~al.}(2021){Bonfanti}, {Delrez}, {Hooton}, {Wilson},
  {Fossati}, {Alibert}, {Hoyer}, {Mustill}, {Osborn}, {Adibekyan}, {Gandolfi},
  {Salmon}, {Sousa}, {Tuson}, {Van Grootel}, {Cabrera}, {Nascimbeni}, {Maxted},
  {Barros}, {Billot}, {Bonfils}, {Borsato}, {Broeg}, {Davies}, {Deleuil},
  {Demangeon}, {Fridlund}, {Lacedelli}, {Lendl}, {Persson}, {Santos},
  {Scandariato}, {Szab{\'o}}, {Collier Cameron}, {Udry}, {Benz}, {Beck},
  {Ehrenreich}, {Fortier}, {Isaak}, {Queloz}, {Alonso}, {Asquier}, {Bandy},
  {B{\'a}rczy}, {Barrado}, {Barrag{\'a}n}, {Baumjohann}, {Beck}, {Bekkelien},
  {Bergomi}, {Brandeker}, {Busch}, {Cessa}, {Charnoz}, {Chazelas}, {Corral Van
  Damme}, {Demory}, {Erikson}, {Farinato}, {Futyan}, {Garcia Mu{\~n}oz},
  {Gillon}, {Guedel}, {Guterman}, {Hasiba}, {Heng}, {Hernandez}, {Kiss},
  {Kuntzer}, {Laskar}, {Lecavelier des Etangs}, {Lovis}, {Magrin}, {Malvasio},
  {Marafatto}, {Michaelis}, {Munari}, {Olofsson}, {Ottacher}, {Ottensamer},
  {Pagano}, {Pall{\'e}}, {Peter}, {Piazza}, {Piotto}, {Pollacco}, {Ragazzoni},
  {Rando}, {Ratti}, {Rauer}, {Ribas}, {Rieder}, {Rohlfs}, {Safa}, {Salatti},
  {S{\'e}gransan}, {Simon}, {Smith}, {Sordet}, {Steller}, {Thomas},
  {Tschentscher}, {Van Eylen}, {Viotto}, {Walter}, {Walton}, {Wildi}, \&
  {Wolter}}]{bonfanti2021}
{Bonfanti}, A., {Delrez}, L., {Hooton}, M.~J., {et~al.} 2021, \aap, 646, A157

\bibitem[{{Bonfanti} {et~al.}(2023){Bonfanti}, {Gandolfi}, {Egger}, {Fossati},
  {Cabrera}, {Krenn}, {Alibert}, {Benz}, {Billot}, {Flor{\'e}n}, {Lendl},
  {Adibekyan}, {Salmon}, {Santos}, {Sousa}, {Wilson}, {Barrag{\'a}n}, {Collier
  Cameron}, {Delrez}, {Esposito}, {Goffo}, {Osborne}, {Osborn}, {Serrano}, {Van
  Eylen}, {Alarcon}, {Alonso}, {Anglada}, {B{\'a}rczy}, {Barrado Navascues},
  {Barros}, {Baumjohann}, {Beck}, {Beck}, {Bedell}, {Bonfils}, {Borsato},
  {Brandeker}, {Broeg}, {Charnoz}, {Corral Van Damme}, {Csizmadia}, {Cubillos},
  {Davies}, {Deleuil}, {Demangeon}, {Demory}, {Ehrenreich}, {Erikson},
  {Fortier}, {Fridlund}, {Gillon}, {G{\"u}del}, {Hoyer}, {Isaak}, {Kerschbaum},
  {Kiss}, {Laskar}, {Lecavelier des Etangs}, {Lorenzo-Oliveira}, {Lovis},
  {Magrin}, {Marafatto}, {Maxted}, {Mel{\'e}ndez}, {Mordasini}, {Nascimbeni},
  {Olofsson}, {Ottensamer}, {Pagano}, {Pall{\'e}}, {Peter}, {Piazza}, {Piotto},
  {Pollacco}, {Queloz}, {Ragazzoni}, {Rando}, {Rauer}, {Ribas}, {Scandariato},
  {S{\'e}gransan}, {Simon}, {Smith}, {Steller}, {Szab{\'o}}, {Thomas}, {Udry},
  {Ulmer}, {Van Grootel}, {Venturini}, \& {Walton}}]{bonfanti2023}
{Bonfanti}, A., {Gandolfi}, D., {Egger}, J.~A., {et~al.} 2023, \aap, 671, L8

\bibitem[{{Bonfanti} \& {Gillon}(2020)}]{bonfanti2020}
{Bonfanti}, A. \& {Gillon}, M. 2020, \aap, 635, A6

\bibitem[{{Bonfanti} {et~al.}(2016){Bonfanti}, {Ortolani}, \&
  {Nascimbeni}}]{bonfanti2016}
{Bonfanti}, A., {Ortolani}, S., \& {Nascimbeni}, V. 2016, \aap, 585, A5

\bibitem[{{Bonfanti} {et~al.}(2015){Bonfanti}, {Ortolani}, {Piotto}, \&
  {Nascimbeni}}]{bonfanti2015}
{Bonfanti}, A., {Ortolani}, S., {Piotto}, G., \& {Nascimbeni}, V. 2015, \aap,
  575, A18

\bibitem[{{Borsato} {et~al.}(2024){Borsato}, {Degen}, {Leleu}, {Hooton},
  {Egger}, {Bekkelien}, {Brandeker}, {Collier Cameron}, {G{\"u}nther},
  {Nascimbeni}, {Persson}, {Bonfanti}, {Wilson}, {Correia}, {Zingales},
  {Guillot}, {Triaud}, {Piotto}, {Gandolfi}, {Abe}, {Alibert}, {Alonso},
  {B{\'a}rczy}, {Navascues}, {Barros}, {Baumjohann}, {Beck}, {Bendjoya},
  {Benz}, {Billot}, {Broeg}, {Busch}, {Csizmadia}, {Cubillos}, {Davies},
  {Deleuil}, {Deline}, {Delrez}, {Demangeon}, {Demory}, {Derekas}, {Edwards},
  {Ehrenreich}, {Erikson}, {Fortier}, {Fossati}, {Fridlund}, {Gazeas},
  {Gillon}, {G{\"u}del}, {Heitzmann}, {Helling}, {Hoyer}, {Isaak}, {Kiss},
  {Korth}, {Lam}, {Laskar}, {Lecavelier des Etangs}, {Lendl}, {Magrin},
  {Marafatto}, {Maxted}, {Mecina}, {M{\'e}karnia}, {Mordasini}, {Mura},
  {Olofsson}, {Ottensamer}, {Pagano}, {Pall{\'e}}, {Peter}, {Pollacco},
  {Queloz}, {Ragazzoni}, {Rando}, {Ratti}, {Rauer}, {Ribas}, {Salmon},
  {Santos}, {Scandariato}, {S{\'e}gransan}, {Simon}, {Smith}, {Sousa},
  {Stalport}, {Suarez}, {Sulis}, {Szab{\'o}}, {Udry}, {Van Grootel},
  {Venturini}, {Villaver}, {Walton}, \& {Wolter}}]{Borsato2024A&A...689A..52B}
{Borsato}, L., {Degen}, D., {Leleu}, A., {et~al.} 2024, \aap, 689, A52

\bibitem[{{Borsato} {et~al.}(2019){Borsato}, {Malavolta}, {Piotto}, {Buchhave},
  {Mortier}, {Rice}, {Collier Cameron}, {Coffinet}, {Sozzetti}, {Charbonneau},
  {Cosentino}, {Dumusque}, {Figueira}, {Latham}, {Lopez-Morales}, {Mayor},
  {Micela}, {Molinari}, {Pepe}, {Phillips}, {Poretti}, {Udry}, \&
  {Watson}}]{Borsato2019MNRAS.484.3233B}
{Borsato}, L., {Malavolta}, L., {Piotto}, G., {et~al.} 2019, \mnras, 484, 3233

\bibitem[{{Borsato} {et~al.}(2014){Borsato}, {Marzari}, {Nascimbeni}, {Piotto},
  {Granata}, {Bedin}, \& {Malavolta}}]{Borsato2014A&A...571A..38B}
{Borsato}, L., {Marzari}, F., {Nascimbeni}, V., {et~al.} 2014, \aap, 571, A38

\bibitem[{{Borsato} {et~al.}(2021){Borsato}, {Piotto}, {Gandolfi},
  {Nascimbeni}, {Lacedelli}, {Marzari}, {Billot}, {Maxted}, {Sousa}, {Cameron},
  {Bonfanti}, {Wilson}, {Serrano}, {Garai}, {Alibert}, {Alonso}, {Asquier},
  {B{\'a}rczy}, {Bandy}, {Barrado}, {Barros}, {Baumjohann}, {Beck}, {Beck},
  {Benz}, {Bonfils}, {Brandeker}, {Broeg}, {Cabrera}, {Charnoz}, {Csizmadia},
  {Davies}, {Deleuil}, {Delrez}, {Demangeon}, {Demory}, {des Etangs},
  {Ehrenreich}, {Erikson}, {Escud{\'e}}, {Fortier}, {Fossati}, {Fridlund},
  {Gillon}, {Guedel}, {Hasiba}, {Heng}, {Hoyer}, {Isaak}, {Kiss}, {Kopp},
  {Laskar}, {Lendl}, {Lovis}, {Magrin}, {Munari}, {Olofsson}, {Ottensamer},
  {Pagano}, {Pall{\'e}}, {Peter}, {Pollacco}, {Queloz}, {Ragazzoni}, {Rando},
  {Rauer}, {Ribas}, {S{\'e}gransan}, {Santos}, {Scandariato}, {Simon}, {Smith},
  {Steller}, {Szab{\'o}}, {Thomas}, {Udry}, {Van Grootel}, \&
  {Walton}}]{Borsato2021MNRAS.506.3810B}
{Borsato}, L., {Piotto}, G., {Gandolfi}, D., {et~al.} 2021, \mnras, 506, 3810

\bibitem[{{Borysow}(2002)}]{Borysow2002jqsrtH2H2lowT}
{Borysow}, A. 2002, \aap, 390, 779

\bibitem[{{Borysow} {et~al.}(2001){Borysow}, {Jorgensen}, \&
  {Fu}}]{BorysowEtal2001jqsrtH2H2highT}
{Borysow}, A., {Jorgensen}, U.~G., \& {Fu}, Y. 2001, \jqsrt, 68, 235

\bibitem[{{Brasseur} {et~al.}(2019){Brasseur}, {Phillip}, {Fleming},
  {Mullally}, \& {White}}]{tesscut2019}
{Brasseur}, C.~E., {Phillip}, C., {Fleming}, S.~W., {Mullally}, S.~E., \&
  {White}, R.~L. 2019, {Astrocut: Tools for creating cutouts of TESS images},
  Astrophysics Source Code Library, record ascl:1905.007

\bibitem[{{Bruntt} {et~al.}(2010){Bruntt}, {Deleuil}, {Fridlund}, {Alonso},
  {Bouchy}, {Hatzes}, {Mayor}, {Moutou}, \& {Queloz}}]{bruntt2010}
{Bruntt}, H., {Deleuil}, M., {Fridlund}, M., {et~al.} 2010, \aap, 519, A51

\bibitem[{{Burrows} {et~al.}(2000){Burrows}, {Marley}, \&
  {Sharp}}]{BurrowsEtal2000apjBDspectra}
{Burrows}, A., {Marley}, M.~S., \& {Sharp}, C.~M. 2000, \apj, 531, 438

\bibitem[{{Ciardi} {et~al.}(2013){Ciardi}, {Fabrycky}, {Ford}, {Gautier},
  {Howell}, {Lissauer}, {Ragozzine}, \& {Rowe}}]{ciardi2013}
{Ciardi}, D.~R., {Fabrycky}, D.~C., {Ford}, E.~B., {et~al.} 2013, \apj, 763, 41

\bibitem[{{Correia} {et~al.}(2018){Correia}, {Delisle}, \&
  {Laskar}}]{correia2018}
{Correia}, A. C.~M., {Delisle}, J.-B., \& {Laskar}, J. 2018, in Handbook of
  Exoplanets, ed. H.~J. {Deeg} \& J.~A. {Belmonte} ({Springer International
  Publishing AG, part of Springer Nature}), 12

\bibitem[{{Cubillos}(2017)}]{Cubillos2017apjRepack}
{Cubillos}, P.~E. 2017, \apj, 850, 32

\bibitem[{{Cubillos} \& {Blecic}(2021)}]{CubillosBlecic2021mnrasPyratBay}
{Cubillos}, P.~E. \& {Blecic}, J. 2021, \mnras, 505, 2675

\bibitem[{{Cutri} {et~al.}(2003){Cutri}, {Skrutskie}, {van Dyk}, {Beichman},
  {Carpenter}, {Chester}, {Cambresy}, {Evans}, {Fowler}, {Gizis}, {Howard},
  {Huchra}, {Jarrett}, {Kopan}, {Kirkpatrick}, {Light}, {Marsh}, {McCallon},
  {Schneider}, {Stiening}, {Sykes}, {Weinberg}, {Wheaton}, {Wheelock}, \&
  {Zacarias}}]{cutri2003}
{Cutri}, R.~M., {Skrutskie}, M.~F., {van Dyk}, S., {et~al.} 2003, {2MASS All
  Sky Catalog of point sources.} (University of Massachusetts and Infrared
  Processing and Analysis Center (IPAC/California Institute of Technology))

\bibitem[{{D'Angelo} \& {Lubow}(2008)}]{dangelo2008}
{D'Angelo}, G. \& {Lubow}, S.~H. 2008, \apj, 685, 560

\bibitem[{{Delisle} {et~al.}(2012){Delisle}, {Laskar}, {Correia}, \&
  {Bou{\'e}}}]{delisle2012}
{Delisle}, J.~B., {Laskar}, J., {Correia}, A.~C.~M., \& {Bou{\'e}}, G. 2012,
  \aap, 546, A71

\bibitem[{{Delrez} {et~al.}(2021){Delrez}, {Ehrenreich}, {Alibert}, {Bonfanti},
  {Borsato}, {Fossati}, {Hooton}, {Hoyer}, {Pozuelos}, {Salmon}, {Sulis},
  {Wilson}, {Adibekyan}, {Bourrier}, {Brandeker}, {Charnoz}, {Deline},
  {Guterman}, {Haldemann}, {Hara}, {Oshagh}, {Sousa}, {Van Grootel}, {Alonso},
  {Anglada-Escud{\'e}}, {B{\'a}rczy}, {Barrado}, {Barros}, {Baumjohann},
  {Beck}, {Bekkelien}, {Benz}, {Billot}, {Bonfils}, {Broeg}, {Cabrera},
  {Collier Cameron}, {Davies}, {Deleuil}, {Delisle}, {Demangeon}, {Demory},
  {Erikson}, {Fortier}, {Fridlund}, {Futyan}, {Gandolfi}, {Garcia Mu{\~n}oz},
  {Gillon}, {Guedel}, {Heng}, {Kiss}, {Laskar}, {Lecavelier des Etangs},
  {Lendl}, {Lovis}, {Maxted}, {Nascimbeni}, {Olofsson}, {Osborn}, {Pagano},
  {Pall{\'e}}, {Piotto}, {Pollacco}, {Queloz}, {Rauer}, {Ragazzoni}, {Ribas},
  {Santos}, {Scandariato}, {S{\'e}gransan}, {Simon}, {Smith}, {Steller},
  {Szab{\'o}}, {Thomas}, {Udry}, \& {Walton}}]{delrez2021}
{Delrez}, L., {Ehrenreich}, D., {Alibert}, Y., {et~al.} 2021, Nature Astronomy,
  5, 775

\bibitem[{{Dorn} {et~al.}(2017){Dorn}, {Venturini}, {Khan}, {Heng}, {Alibert},
  {Helled}, {Rivoldini}, \& {Benz}}]{dorn2017}
{Dorn}, C., {Venturini}, J., {Khan}, A., {et~al.} 2017, \aap, 597, A37

\bibitem[{{Doyle} {et~al.}(2014){Doyle}, {Davies}, {Smalley}, {Chaplin}, \&
  {Elsworth}}]{Doyle2014}
{Doyle}, A.~P., {Davies}, G.~R., {Smalley}, B., {Chaplin}, W.~J., \&
  {Elsworth}, Y. 2014, \mnras, 444, 3592

\bibitem[{{Dragomir} {et~al.}(2012){Dragomir}, {Kane}, {Henry}, {Ciardi},
  {Fischer}, {Howard}, {Jensen}, {Laughlin}, {Mahadevan}, {Matthews},
  {Pilyavsky}, {von Braun}, {Wang}, \& {Wright}}]{dragomir2012}
{Dragomir}, D., {Kane}, S.~R., {Henry}, G.~W., {et~al.} 2012, \apj, 754, 37

\bibitem[{{Dumusque} {et~al.}(2011{\natexlab{a}}){Dumusque}, {Santos}, {Udry},
  {Lovis}, \& {Bonfils}}]{Dumusque2011a}
{Dumusque}, X., {Santos}, N.~C., {Udry}, S., {Lovis}, C., \& {Bonfils}, X.
  2011{\natexlab{a}}, \aap, 527, A82

\bibitem[{{Dumusque} {et~al.}(2011{\natexlab{b}}){Dumusque}, {Udry}, {Lovis},
  {Santos}, \& {Monteiro}}]{Dumusque2011b}
{Dumusque}, X., {Udry}, S., {Lovis}, C., {Santos}, N.~C., \& {Monteiro},
  M.~J.~P.~F.~G. 2011{\natexlab{b}}, \aap, 525, A140

\bibitem[{{Egger} {et~al.}(2024){Egger}, {Osborn}, {Kubyshkina}, {Mordasini},
  {Alibert}, {G{\"u}nther}, {Lendl}, {Brandeker}, {Heitzmann}, {Leleu},
  {Damasso}, {Bonfanti}, {Wilson}, {Sousa}, {Haldemann}, {Delrez}, {Hooton},
  {Zingales}, {Luque}, {Alonso}, {Asquier}, {B{\'a}rczy}, {Navascues},
  {Barros}, {Baumjohann}, {Benz}, {Billot}, {Borsato}, {Broeg}, {Buder},
  {Castro-Gonz{\'a}lez}, {Cameron}, {Correia}, {Cortes}, {Csizmadia},
  {Cubillos}, {Davies}, {Deleuil}, {Deline}, {Demangeon}, {Demory}, {Derekas},
  {Edwards}, {Ehrenreich}, {Erikson}, {Fortier}, {Fossati}, {Fridlund},
  {Gandolfi}, {Gazeas}, {Gillon}, {G{\"u}del}, {Helling}, {Isaak}, {Kiss},
  {Korth}, {Lam}, {Laskar}, {Lavie}, {des Etangs}, {Lovis}, {Luntzer},
  {Magrin}, {Maxted}, {Mer{\'\i}n}, {Munari}, {Nascimbeni}, {Olofsson},
  {Ottensamer}, {Pagano}, {Pall{\'e}}, {Peter}, {Piazza}, {Piotto}, {Pollacco},
  {Queloz}, {Ragazzoni}, {Rando}, {Rauer}, {Ribas}, {Rodrigues}, {Santos},
  {Scandariato}, {S{\'e}gransan}, {Simon}, {Smith}, {Stalport}, {Sulis},
  {Szab{\'o}}, {Udry}, {Van Grootel}, {Venturini}, {Villaver}, \&
  {Walton}}]{Egger+2024}
{Egger}, J.~A., {Osborn}, H.~P., {Kubyshkina}, D., {et~al.} 2024, \aap, 688,
  A223

\bibitem[{{Espinoza} \& {Jord{\'a}n}(2015)}]{espinoza2015}
{Espinoza}, N. \& {Jord{\'a}n}, A. 2015, \mnras, 450, 1879

\bibitem[{{Fabrycky} {et~al.}(2014){Fabrycky}, {Lissauer}, {Ragozzine}, {Rowe},
  {Steffen}, {Agol}, {Barclay}, {Batalha}, {Borucki}, {Ciardi}, {Ford},
  {Gautier}, {Geary}, {Holman}, {Jenkins}, {Li}, {Morehead}, {Morris},
  {Shporer}, {Smith}, {Still}, \& {Van Cleve}}]{fabrycky2014}
{Fabrycky}, D.~C., {Lissauer}, J.~J., {Ragozzine}, D., {et~al.} 2014, \apj,
  790, 146

\bibitem[{{Foreman-Mackey} {et~al.}(2019){Foreman-Mackey}, {Farr}, {Sinha},
  {Archibald}, {Hogg}, {Sanders}, {Zuntz}, {Williams}, {Nelson}, {de
  Val-Borro}, {Erhardt}, {Pashchenko}, \& {Pla}}]{DFM2019JOSS....4.1864F}
{Foreman-Mackey}, D., {Farr}, W., {Sinha}, M., {et~al.} 2019, The Journal of
  Open Source Software, 4, 1864

\bibitem[{{Foreman-Mackey} {et~al.}(2013){Foreman-Mackey}, {Hogg}, {Lang}, \&
  {Goodman}}]{Foreman2013}
{Foreman-Mackey}, D., {Hogg}, D.~W., {Lang}, D., \& {Goodman}, J. 2013, \pasp,
  125, 306

\bibitem[{{Fridlund} {et~al.}(2024){Fridlund}, {Georgieva}, {Bonfanti},
  {Alibert}, {Persson}, {Gandolfi}, {Beck}, {Deline}, {Hoyer}, {Olofsson},
  {Wilson}, {Barrag{\'a}n}, {Fossati}, {Mustill}, {Brandeker}, {Hatzes},
  {Flor{\'e}n}, {Simola}, {Hooton}, {Luque}, {Sousa}, {Egger},
  {Antoniadis-Karnavas}, {Salmon}, {Adibekyan}, {Alonso}, {Anglada},
  {B{\'a}rczy}, {Barrado Navascues}, {Barros}, {Baumjohann}, {Beck}, {Benz},
  {Bonfils}, {Broeg}, {Cabrera}, {Charnoz}, {Collier Cameron}, {Csizmadia},
  {Davies}, {Deeg}, {Deleuil}, {Delrez}, {Demangeon}, {Demory}, {Ehrenreich},
  {Erikson}, {Esposito}, {Fortier}, {Gillon}, {G{\"u}del}, {Heng}, {Isaak},
  {Kiss}, {Korth}, {Laskar}, {Lecavelier des Etangs}, {Lendl}, {Livingston},
  {Lovis}, {Magrin}, {Maxted}, {Muresan}, {Nascimbeni}, {Ottensamer}, {Pagano},
  {Pall{\'e}}, {Peter}, {Piotto}, {Pollacco}, {Queloz}, {Ragazzoni}, {Rando},
  {Rauer}, {Redfield}, {Ribas}, {Santos}, {Scandariato}, {S{\'e}gransan},
  {Serrano}, {Simon}, {Smith}, {Steller}, {Szab{\'o}}, {Thomas}, {Udry}, {Van
  Eylen}, {Van Grootel}, \& {Walton}}]{fridlund2024}
{Fridlund}, M., {Georgieva}, I.~Y., {Bonfanti}, A., {et~al.} 2024, \aap, 684,
  A12

\bibitem[{{Gaia Collaboration} {et~al.}(2023){Gaia Collaboration}, {Vallenari},
  {Brown}, {Prusti}, {de Bruijne}, {Arenou}, {Babusiaux}, {Biermann},
  {Creevey}, {Ducourant}, {Evans}, {Eyer}, {Guerra}, {Hutton}, {Jordi},
  {Klioner}, {Lammers}, {Lindegren}, {Luri}, {Mignard}, {Panem}, {Pourbaix},
  {Randich}, {Sartoretti}, {Soubiran}, {Tanga}, {Walton}, {Bailer-Jones},
  {Bastian}, {Drimmel}, {Jansen}, {Katz}, {Lattanzi}, {van Leeuwen}, {Bakker},
  {Cacciari}, {Casta{\~n}eda}, {De Angeli}, {Fabricius}, {Fouesneau},
  {Fr{\'e}mat}, {Galluccio}, {Guerrier}, {Heiter}, {Masana}, {Messineo},
  {Mowlavi}, {Nicolas}, {Nienartowicz}, {Pailler}, {Panuzzo}, {Riclet}, {Roux},
  {Seabroke}, {Sordo}, {Th{\'e}venin}, {Gracia-Abril}, {Portell}, {Teyssier},
  {Altmann}, {Andrae}, {Audard}, {Bellas-Velidis}, {Benson}, {Berthier},
  {Blomme}, {Burgess}, {Busonero}, {Busso}, {C{\'a}novas}, {Carry}, {Cellino},
  {Cheek}, {Clementini}, {Damerdji}, {Davidson}, {de Teodoro}, {Nu{\~n}ez
  Campos}, {Delchambre}, {Dell'Oro}, {Esquej}, {Fern{\'a}ndez-Hern{\'a}ndez},
  {Fraile}, {Garabato}, {Garc{\'\i}a-Lario}, {Gosset}, {Haigron}, {Halbwachs},
  {Hambly}, {Harrison}, {Hern{\'a}ndez}, {Hestroffer}, {Hodgkin}, {Holl},
  {Jan{\ss}en}, {Jevardat de Fombelle}, {Jordan}, {Krone-Martins}, {Lanzafame},
  {L{\"o}ffler}, {Marchal}, {Marrese}, {Moitinho}, {Muinonen}, {Osborne},
  {Pancino}, {Pauwels}, {Recio-Blanco}, {Reyl{\'e}}, {Riello}, {Rimoldini},
  {Roegiers}, {Rybizki}, {Sarro}, {Siopis}, {Smith}, {Sozzetti}, {Utrilla},
  {van Leeuwen}, {Abbas}, {{\'A}brah{\'a}m}, {Abreu Aramburu}, {Aerts},
  {Aguado}, {Ajaj}, {Aldea-Montero}, {Altavilla}, {{\'A}lvarez}, {Alves},
  {Anders}, {Anderson}, {Anglada Varela}, {Antoja}, {Baines}, {Baker},
  {Balaguer-N{\'u}{\~n}ez}, {Balbinot}, {Balog}, {Barache}, {Barbato},
  {Barros}, {Barstow}, {Bartolom{\'e}}, {Bassilana}, {Bauchet}, {Becciani},
  {Bellazzini}, {Berihuete}, {Bernet}, {Bertone}, {Bianchi}, {Binnenfeld},
  {Blanco-Cuaresma}, {Blazere}, {Boch}, {Bombrun}, {Bossini}, {Bouquillon},
  {Bragaglia}, {Bramante}, {Breedt}, {Bressan}, {Brouillet}, {Brugaletta},
  {Bucciarelli}, {Burlacu}, {Butkevich}, {Buzzi}, {Caffau}, {Cancelliere},
  {Cantat-Gaudin}, {Carballo}, {Carlucci}, {Carnerero}, {Carrasco},
  {Casamiquela}, {Castellani}, {Castro-Ginard}, {Chaoul}, {Charlot}, {Chemin},
  {Chiaramida}, {Chiavassa}, {Chornay}, {Comoretto}, {Contursi}, {Cooper},
  {Cornez}, {Cowell}, {Crifo}, {Cropper}, {Crosta}, {Crowley}, {Dafonte},
  {Dapergolas}, {David}, {David}, {de Laverny}, {De Luise}, {De March}, {De
  Ridder}, {de Souza}, {de Torres}, {del Peloso}, {del Pozo}, {Delbo},
  {Delgado}, {Delisle}, {Demouchy}, {Dharmawardena}, {Di Matteo}, {Diakite},
  {Diener}, {Distefano}, {Dolding}, {Edvardsson}, {Enke}, {Fabre}, {Fabrizio},
  {Faigler}, {Fedorets}, {Fernique}, {Fienga}, {Figueras}, {Fournier},
  {Fouron}, {Fragkoudi}, {Gai}, {Garcia-Gutierrez}, {Garcia-Reinaldos},
  {Garc{\'\i}a-Torres}, {Garofalo}, {Gavel}, {Gavras}, {Gerlach}, {Geyer},
  {Giacobbe}, {Gilmore}, {Girona}, {Giuffrida}, {Gomel}, {Gomez},
  {Gonz{\'a}lez-N{\'u}{\~n}ez}, {Gonz{\'a}lez-Santamar{\'\i}a},
  {Gonz{\'a}lez-Vidal}, {Granvik}, {Guillout}, {Guiraud},
  {Guti{\'e}rrez-S{\'a}nchez}, {Guy}, {Hatzidimitriou}, {Hauser}, {Haywood},
  {Helmer}, {Helmi}, {Sarmiento}, {Hidalgo}, {Hilger}, {H{\l}adczuk}, {Hobbs},
  {Holland}, {Huckle}, {Jardine}, {Jasniewicz}, {Jean-Antoine Piccolo},
  {Jim{\'e}nez-Arranz}, {Jorissen}, {Juaristi Campillo}, {Julbe}, {Karbevska},
  {Kervella}, {Khanna}, {Kontizas}, {Kordopatis}, {Korn}, {K{\'o}sp{\'a}l},
  {Kostrzewa-Rutkowska}, {Kruszy{\'n}ska}, {Kun}, {Laizeau}, {Lambert},
  {Lanza}, {Lasne}, {Le Campion}, {Lebreton}, {Lebzelter}, {Leccia}, {Leclerc},
  {Lecoeur-Taibi}, {Liao}, {Licata}, {Lindstr{\o}m}, {Lister}, {Livanou},
  {Lobel}, {Lorca}, {Loup}, {Madrero Pardo}, {Magdaleno Romeo}, {Managau},
  {Mann}, {Manteiga}, {Marchant}, {Marconi}, {Marcos}, {Marcos Santos},
  {Mar{\'\i}n Pina}, {Marinoni}, {Marocco}, {Marshall}, {Martin Polo},
  {Mart{\'\i}n-Fleitas}, {Marton}, {Mary}, {Masip}, {Massari},
  {Mastrobuono-Battisti}, {Mazeh}, {McMillan}, {Messina}, {Michalik}, {Millar},
  {Mints}, {Molina}, {Molinaro}, {Moln{\'a}r}, {Monari}, {Mongui{\'o}},
  {Montegriffo}, {Montero}, {Mor}, {Mora}, {Morbidelli}, {Morel}, {Morris},
  {Muraveva}, {Murphy}, {Musella}, {Nagy}, {Noval}, {Oca{\~n}a}, {Ogden},
  {Ordenovic}, {Osinde}, {Pagani}, {Pagano}, {Palaversa}, {Palicio},
  {Pallas-Quintela}, {Panahi}, {Payne-Wardenaar}, {Pe{\~n}alosa Esteller},
  {Penttil{\"a}}, {Pichon}, {Piersimoni}, {Pineau}, {Plachy}, {Plum}, {Poggio},
  {Pr{\v{s}}a}, {Pulone}, {Racero}, {Ragaini}, {Rainer}, {Raiteri}, {Rambaux},
  {Ramos}, {Ramos-Lerate}, {Re Fiorentin}, {Regibo}, {Richards}, {Rios Diaz},
  {Ripepi}, {Riva}, {Rix}, {Rixon}, {Robichon}, {Robin}, {Robin}, {Roelens},
  {Rogues}, {Rohrbasser}, {Romero-G{\'o}mez}, {Rowell}, {Royer}, {Ruz Mieres},
  {Rybicki}, {Sadowski}, {S{\'a}ez N{\'u}{\~n}ez}, {Sagrist{\`a} Sell{\'e}s},
  {Sahlmann}, {Salguero}, {Samaras}, {Sanchez Gimenez}, {Sanna},
  {Santove{\~n}a}, {Sarasso}, {Schultheis}, {Sciacca}, {Segol}, {Segovia},
  {S{\'e}gransan}, {Semeux}, {Shahaf}, {Siddiqui}, {Siebert}, {Siltala},
  {Silvelo}, {Slezak}, {Slezak}, {Smart}, {Snaith}, {Solano}, {Solitro},
  {Souami}, {Souchay}, {Spagna}, {Spina}, {Spoto}, {Steele},
  {Steidelm{\"u}ller}, {Stephenson}, {S{\"u}veges}, {Surdej}, {Szabados},
  {Szegedi-Elek}, {Taris}, {Taylor}, {Teixeira}, {Tolomei}, {Tonello}, {Torra},
  {Torra}, {Torralba Elipe}, {Trabucchi}, {Tsounis}, {Turon}, {Ulla}, {Unger},
  {Vaillant}, {van Dillen}, {van Reeven}, {Vanel}, {Vecchiato}, {Viala},
  {Vicente}, {Voutsinas}, {Weiler}, {Wevers}, {Wyrzykowski}, {Yoldas}, {Yvard},
  {Zhao}, {Zorec}, {Zucker}, \& {Zwitter}}]{GaiaDR3-2023}
{Gaia Collaboration}, {Vallenari}, A., {Brown}, A.~G.~A., {et~al.} 2023, \aap,
  674, A1

\bibitem[{{Gandolfi} {et~al.}(2017){Gandolfi}, {Barrag{\'a}n}, {Hatzes},
  {Fridlund}, {Fossati}, {Donati}, {Johnson}, {Nowak}, {Prieto-Arranz},
  {Albrecht}, {Dai}, {Deeg}, {Endl}, {Grziwa}, {Hjorth}, {Korth}, {Nespral},
  {Saario}, {Smith}, {Antoniciello}, {Alarcon}, {Bedell}, {Blay}, {Brems},
  {Cabrera}, {Csizmadia}, {Cusano}, {Cochran}, {Eigm{\"u}ller}, {Erikson},
  {Gonz{\'a}lez Hern{\'a}ndez}, {Guenther}, {Hirano}, {Su{\'a}rez
  Mascare{\~n}o}, {Narita}, {Palle}, {Parviainen}, {P{\"a}tzold}, {Persson},
  {Rauer}, {Saviane}, {Schmidtobreick}, {Van Eylen}, {Winn}, \&
  {Zakhozhay}}]{Gandolfi2017}
{Gandolfi}, D., {Barrag{\'a}n}, O., {Hatzes}, A.~P., {et~al.} 2017, \aj, 154,
  123

\bibitem[{{Gardner} {et~al.}(2006){Gardner}, {Mather}, {Clampin}, {Doyon},
  {Greenhouse}, {Hammel}, {Hutchings}, {Jakobsen}, {Lilly}, {Long}, {Lunine},
  {McCaughrean}, {Mountain}, {Nella}, {Rieke}, {Rieke}, {Rix}, {Smith},
  {Sonneborn}, {Stiavelli}, {Stockman}, {Windhorst}, \& {Wright}}]{gardner2006}
{Gardner}, J.~P., {Mather}, J.~C., {Clampin}, M., {et~al.} 2006, \ssr, 123, 485

\bibitem[{{Gelman} \& {Rubin}(1992)}]{gelman1992}
{Gelman}, A. \& {Rubin}, D.~B. 1992, Statistical Science, 7, 457

\bibitem[{Geweke(1991)}]{geweke1991}
Geweke, J.~F. 1991, {Evaluating the accuracy of sampling-based approaches to
  the calculation of posterior moments}, Staff Report 148, Federal Reserve Bank
  of Minneapolis

\bibitem[{{Ginsburg} {et~al.}(2019){Ginsburg}, {Sip{\H{o}}cz}, {Brasseur},
  {Cowperthwaite}, {Craig}, {Deil}, {Guillochon}, {Guzman}, {Liedtke}, {Lian
  Lim}, {Lockhart}, {Mommert}, {Morris}, {Norman}, {Parikh}, {Persson},
  {Robitaille}, {Segovia}, {Singer}, {Tollerud}, {de Val-Borro}, {Valtchanov},
  {Woillez}, {Astroquery Collaboration}, \& {a subset of astropy
  Collaboration}}]{astroquery2019}
{Ginsburg}, A., {Sip{\H{o}}cz}, B.~M., {Brasseur}, C.~E., {et~al.} 2019, \aj,
  157, 98

\bibitem[{{Goldreich} \& {Tremaine}(1980)}]{goldreich1980}
{Goldreich}, P. \& {Tremaine}, S. 1980, \apj, 241, 425

\bibitem[{Goodman \& Weare(2010)}]{AffineInvariantGoodmanWeare2010}
Goodman, J. \& Weare, J. 2010, Communications in Applied Mathematics and
  Computational Science, 5, 65

\bibitem[{{Gray} {et~al.}(2006){Gray}, {Corbally}, {Garrison}, {McFadden},
  {Bubar}, {McGahee}, {O'Donoghue}, \& {Knox}}]{gray2006}
{Gray}, R.~O., {Corbally}, C.~J., {Garrison}, R.~F., {et~al.} 2006, \aj, 132,
  161

\bibitem[{{Haldemann} {et~al.}(2024){Haldemann}, {Dorn}, {Venturini},
  {Alibert}, \& {Benz}}]{haldemann2024}
{Haldemann}, J., {Dorn}, C., {Venturini}, J., {Alibert}, Y., \& {Benz}, W.
  2024, \aap, 681, A96

\bibitem[{{Handberg} {et~al.}(2021){Handberg}, {Lund}, {White}, {Hall},
  {Buzasi}, {Pope}, {Hansen}, {von Essen}, {Carboneau}, {Huber}, {Vanderspek},
  {Fausnaugh}, {Tenenbaum}, {Jenkins}, \& {T'DA Collaboration}}]{handberg2021}
{Handberg}, R., {Lund}, M.~N., {White}, T.~R., {et~al.} 2021, \aj, 162, 170

\bibitem[{Harris {et~al.}(2020)Harris, Millman, van~der Walt, Gommers,
  Virtanen, Cournapeau, Wieser, Taylor, Berg, Smith, Kern, Picus, Hoyer, van
  Kerkwijk, Brett, Haldane, del R{\'{i}}o, Wiebe, Peterson,
  G{\'{e}}rard-Marchant, Sheppard, Reddy, Weckesser, Abbasi, Gohlke, \&
  Oliphant}]{numpy2020}
Harris, C.~R., Millman, K.~J., van~der Walt, S.~J., {et~al.} 2020, Nature, 585,
  357

\bibitem[{{Hatzes}(2016)}]{hatzes2016}
{Hatzes}, A.~P. 2016, in Astrophysics and Space Science Library, Vol. 428,
  Methods of Detecting Exoplanets: 1st Advanced School on Exoplanetary Science,
  ed. V.~{Bozza}, L.~{Mancini}, \& A.~{Sozzetti}, 3

\bibitem[{{Hatzes}(2019)}]{Hatzes2019}
{Hatzes}, A.~P. 2019, {The Doppler Method for the Detection of Exoplanets}
  (Institute of Physics Publishing)

\bibitem[{{Hatzes} {et~al.}(2010){Hatzes}, {Dvorak}, {Wuchterl}, {Guterman},
  {Hartmann}, {Fridlund}, {Gandolfi}, {Guenther}, \&
  {P{\"a}tzold}}]{Hatzes2010}
{Hatzes}, A.~P., {Dvorak}, R., {Wuchterl}, G., {et~al.} 2010, \aap, 520, A93

\bibitem[{{Haywood} {et~al.}(2014){Haywood}, {Collier Cameron}, {Queloz},
  {Barros}, {Deleuil}, {Fares}, {Gillon}, {Lanza}, {Lovis}, {Moutou}, {Pepe},
  {Pollacco}, {Santerne}, {S{\'e}gransan}, \& {Unruh}}]{Haywood2014}
{Haywood}, R.~D., {Collier Cameron}, A., {Queloz}, D., {et~al.} 2014, \mnras,
  443, 2517

\bibitem[{{H{\o}g} {et~al.}(2000){H{\o}g}, {Fabricius}, {Makarov}, {Urban},
  {Corbin}, {Wycoff}, {Bastian}, {Schwekendiek}, \& {Wicenec}}]{hog2000}
{H{\o}g}, E., {Fabricius}, C., {Makarov}, V.~V., {et~al.} 2000, \aap, 355, L27

\bibitem[{{Holman} {et~al.}(2010){Holman}, {Fabrycky}, {Ragozzine}, {Ford},
  {Steffen}, {Welsh}, {Lissauer}, {Latham}, {Marcy}, {Walkowicz}, {Batalha},
  {Jenkins}, {Rowe}, {Cochran}, {Fressin}, {Torres}, {Buchhave}, {Sasselov},
  {Borucki}, {Koch}, {Basri}, {Brown}, {Caldwell}, {Charbonneau}, {Dunham},
  {Gautier}, {Geary}, {Gilliland}, {Haas}, {Howell}, {Ciardi}, {Endl},
  {Fischer}, {F{\"u}r{\'e}sz}, {Hartman}, {Isaacson}, {Johnson}, {MacQueen},
  {Moorhead}, {Morehead}, \& {Orosz}}]{holman2010}
{Holman}, M.~J., {Fabrycky}, D.~C., {Ragozzine}, D., {et~al.} 2010, Science,
  330, 51

\bibitem[{{Holman} {et~al.}(2006){Holman}, {Winn}, {Latham}, {O'Donovan},
  {Charbonneau}, {Bakos}, {Esquerdo}, {Hergenrother}, {Everett}, \&
  {P{\'a}l}}]{holman2006}
{Holman}, M.~J., {Winn}, J.~N., {Latham}, D.~W., {et~al.} 2006, \apj, 652, 1715

\bibitem[{Hunter(2007)}]{matplotlib2007}
Hunter, J.~D. 2007, Computing in Science \& Engineering, 9, 90

\bibitem[{{Izidoro} {et~al.}(2017){Izidoro}, {Ogihara}, {Raymond},
  {Morbidelli}, {Pierens}, {Bitsch}, {Cossou}, \& {Hersant}}]{izidoro2017}
{Izidoro}, A., {Ogihara}, M., {Raymond}, S.~N., {et~al.} 2017, \mnras, 470,
  1750

\bibitem[{{Jenkins} {et~al.}(2016){Jenkins}, {Twicken}, {McCauliff},
  {Campbell}, {Sanderfer}, {Lung}, {Mansouri-Samani}, {Girouard}, {Tenenbaum},
  {Klaus}, {Smith}, {Caldwell}, {Chacon}, {Henze}, {Heiges}, {Latham},
  {Morgan}, {Swade}, {Rinehart}, \& {Vanderspek}}]{Jenkins2016}
{Jenkins}, J.~M., {Twicken}, J.~D., {McCauliff}, S., {et~al.} 2016, in Society
  of Photo-Optical Instrumentation Engineers (SPIE) Conference Series, Vol.
  9913, Software and Cyberinfrastructure for Astronomy IV, ed. G.~{Chiozzi} \&
  J.~C. {Guzman}, 99133E

\bibitem[{{Kass} \& {Raftery}(1995)}]{kass95}
{Kass}, R.~E. \& {Raftery}, A.~E. 1995, Journal of the American Statistical
  Association, 90, 773

\bibitem[{{Kempton} {et~al.}(2018){Kempton}, {Bean}, {Louie}, {Deming}, {Koll},
  {Mansfield}, {Christiansen}, {L{\'o}pez-Morales}, {Swain}, {Zellem},
  {Ballard}, {Barclay}, {Barstow}, {Batalha}, {Beatty}, {Berta-Thompson},
  {Birkby}, {Buchhave}, {Charbonneau}, {Cowan}, {Crossfield}, {de Val-Borro},
  {Doyon}, {Dragomir}, {Gaidos}, {Heng}, {Hu}, {Kane}, {Kreidberg}, {Mallonn},
  {Morley}, {Narita}, {Nascimbeni}, {Pall{\'e}}, {Quintana}, {Rauscher},
  {Seager}, {Shkolnik}, {Sing}, {Sozzetti}, {Stassun}, {Valenti}, \& {von
  Essen}}]{KemptonEtal2018paspTransitSpectroscopicMetric}
{Kempton}, E. M.~R., {Bean}, J.~L., {Louie}, D.~R., {et~al.} 2018, \pasp, 130,
  114401

\bibitem[{{Kipping}(2010)}]{kipping2010}
{Kipping}, D.~M. 2010, \mnras, 408, 1758

\bibitem[{{Kossakowski} {et~al.}(2022){Kossakowski}, {K{\"u}rster}, {Henning},
  {Trifonov}, {Caballero}, {Lafarga}, {Bauer}, {Stock}, {Kemmer}, {Jeffers},
  {Amado}, {P{\'e}rez-Torres}, {B{\'e}jar}, {Cort{\'e}s-Contreras}, {Ribas},
  {Reiners}, {Quirrenbach}, {Aceituno}, {Baroch}, {Cifuentes}, {Dreizler},
  {Hatzes}, {Kaminski}, {Montes}, {Morales}, {Pavlov}, {Pena}, {Perdelwitz},
  {Reffert}, {Revilla}, {Rodr{\'\i}guez Lopez}, {Rosich}, {Sadegi},
  {Sanz-Forcada}, {Sch{\"o}fer}, {Schweitzer}, \&
  {Zechmeister}}]{kossakowski2022}
{Kossakowski}, D., {K{\"u}rster}, M., {Henning}, T., {et~al.} 2022, \aap, 666,
  A143

\bibitem[{{Kuerster} {et~al.}(1997){Kuerster}, {Schmitt}, {Cutispoto}, \&
  {Dennerl}}]{Kuerster1997}
{Kuerster}, M., {Schmitt}, J.~H.~M.~M., {Cutispoto}, G., \& {Dennerl}, K. 1997,
  \aap, 320, 831

\bibitem[{{Kurucz}(1970)}]{Kurucz1970saorsAtlas}
{Kurucz}, R.~L. 1970, SAO Special Report, 309

\bibitem[{{Kurucz}(1993)}]{Kurucz-93}
{Kurucz}, R.~L. 1993, {SYNTHE spectrum synthesis programs and line data}
  (Astrophysics Source Code Library)

\bibitem[{{Kurucz}(2013)}]{Kurucz2013}
{Kurucz}, R.~L. 2013, {ATLAS12: Opacity sampling model atmosphere program}

\bibitem[{{Laskar}(1997)}]{Laskar1997A&A...317L..75L}
{Laskar}, J. 1997, \aap, 317, L75

\bibitem[{{Laskar}(2000)}]{Laskar2000PhRvL..84.3240L}
{Laskar}, J. 2000, \prl, 84, 3240

\bibitem[{{Laskar} \& {Petit}(2017)}]{LaskarPetit2017A&A...605A..72L}
{Laskar}, J. \& {Petit}, A.~C. 2017, \aap, 605, A72

\bibitem[{{Lee} \& {Peale}(2002)}]{lee2002}
{Lee}, M.~H. \& {Peale}, S.~J. 2002, arXiv e-prints, astro

\bibitem[{{Leleu} {et~al.}(2021){Leleu}, {Alibert}, {Hara}, {Hooton}, {Wilson},
  {Robutel}, {Delisle}, {Laskar}, {Hoyer}, {Lovis}, {Bryant}, {Ducrot},
  {Cabrera}, {Delrez}, {Acton}, {Adibekyan}, {Allart}, {Allende Prieto},
  {Alonso}, {Alves}, {Anderson}, {Angerhausen}, {Anglada Escud{\'e}},
  {Asquier}, {Barrado}, {Barros}, {Baumjohann}, {Bayliss}, {Beck}, {Beck},
  {Bekkelien}, {Benz}, {Billot}, {Bonfanti}, {Bonfils}, {Bouchy}, {Bourrier},
  {Bou{\'e}}, {Brandeker}, {Broeg}, {Buder}, {Burdanov}, {Burleigh},
  {B{\'a}rczy}, {Cameron}, {Chamberlain}, {Charnoz}, {Cooke}, {Corral Van
  Damme}, {Correia}, {Cristiani}, {Damasso}, {Davies}, {Deleuil}, {Demangeon},
  {Demory}, {Di Marcantonio}, {Di Persio}, {Dumusque}, {Ehrenreich}, {Erikson},
  {Figueira}, {Fortier}, {Fossati}, {Fridlund}, {Futyan}, {Gandolfi},
  {Garc{\'\i}a Mu{\~n}oz}, {Garcia}, {Gill}, {Gillen}, {Gillon}, {Goad},
  {Gonz{\'a}lez Hern{\'a}ndez}, {Guedel}, {G{\"u}nther}, {Haldemann},
  {Henderson}, {Heng}, {Hogan}, {Isaak}, {Jehin}, {Jenkins}, {Jord{\'a}n},
  {Kiss}, {Kristiansen}, {Lam}, {Lavie}, {Lecavelier des Etangs}, {Lendl},
  {Lillo-Box}, {Lo Curto}, {Magrin}, {Martins}, {Maxted}, {McCormac}, {Mehner},
  {Micela}, {Molaro}, {Moyano}, {Murray}, {Nascimbeni}, {Nunes}, {Olofsson},
  {Osborn}, {Oshagh}, {Ottensamer}, {Pagano}, {Pall{\'e}}, {Pedersen}, {Pepe},
  {Persson}, {Peter}, {Piotto}, {Polenta}, {Pollacco}, {Poretti}, {Pozuelos},
  {Queloz}, {Ragazzoni}, {Rando}, {Ratti}, {Rauer}, {Raynard}, {Rebolo},
  {Reimers}, {Ribas}, {Santos}, {Scandariato}, {Schneider}, {Sebastian},
  {Sestovic}, {Simon}, {Smith}, {Sousa}, {Sozzetti}, {Steller}, {Su{\'a}rez
  Mascare{\~n}o}, {Szab{\'o}}, {S{\'e}gransan}, {Thomas}, {Thompson},
  {Tilbrook}, {Triaud}, {Turner}, {Udry}, {Van Grootel}, {Venus}, {Verrecchia},
  {Vines}, {Walton}, {West}, {Wheatley}, {Wolter}, \& {Zapatero
  Osorio}}]{leleu2021}
{Leleu}, A., {Alibert}, Y., {Hara}, N.~C., {et~al.} 2021, \aap, 649, A26

\bibitem[{{Li} {et~al.}(2020){Li}, {Huang}, {Petaev}, {Zhu}, \&
  {Steffen}}]{li2020}
{Li}, M., {Huang}, S., {Petaev}, M.~I., {Zhu}, Z., \& {Steffen}, J.~H. 2020,
  \mnras, 495, 2543

\bibitem[{{Lightkurve Collaboration} {et~al.}(2018){Lightkurve Collaboration},
  {Cardoso}, {Hedges}, {Gully-Santiago}, {Saunders}, {Cody}, {Barclay}, {Hall},
  {Sagear}, {Turtelboom}, {Zhang}, {Tzanidakis}, {Mighell}, {Coughlin}, {Bell},
  {Berta-Thompson}, {Williams}, {Dotson}, \& {Barentsen}}]{lightkurve2018}
{Lightkurve Collaboration}, {Cardoso}, J.~V.~d.~M., {Hedges}, C., {et~al.}
  2018, {Lightkurve: Kepler and TESS time series analysis in Python},
  Astrophysics Source Code Library

\bibitem[{{Lin} \& {Papaloizou}(1979)}]{lin1979}
{Lin}, D.~N.~C. \& {Papaloizou}, J. 1979, \mnras, 186, 799

\bibitem[{{Lindegren} {et~al.}(2021){Lindegren}, {Bastian}, {Biermann},
  {Bombrun}, {de Torres}, {Gerlach}, {Geyer}, {Hern{\'a}ndez}, {Hilger},
  {Hobbs}, {Klioner}, {Lammers}, {McMillan}, {Ramos-Lerate},
  {Steidelm{\"u}ller}, {Stephenson}, \& {van Leeuwen}}]{lindegren2021}
{Lindegren}, L., {Bastian}, U., {Biermann}, M., {et~al.} 2021, \aap, 649, A4

\bibitem[{{Lissauer} {et~al.}(2011){Lissauer}, {Ragozzine}, {Fabrycky},
  {Steffen}, {Ford}, {Jenkins}, {Shporer}, {Holman}, {Rowe}, {Quintana},
  {Batalha}, {Borucki}, {Bryson}, {Caldwell}, {Carter}, {Ciardi}, {Dunham},
  {Fortney}, {Gautier}, {Howell}, {Koch}, {Latham}, {Marcy}, {Morehead}, \&
  {Sasselov}}]{lissauer2011}
{Lissauer}, J.~J., {Ragozzine}, D., {Fabrycky}, D.~C., {et~al.} 2011, \apjs,
  197, 8

\bibitem[{{Lovis, C.} \& {Pepe, F.}(2007)}]{Lovis2007}
{Lovis, C.} \& {Pepe, F.} 2007, A\&A, 468, 1115

\bibitem[{{Lund} {et~al.}(2021){Lund}, {Handberg}, {Buzasi}, {Carboneau},
  {Hall}, {Pereira}, {Huber}, {Hey}, {Van Reeth}, \& {T'DA
  Collaboration}}]{lund2021}
{Lund}, M.~N., {Handberg}, R., {Buzasi}, D.~L., {et~al.} 2021, \apjs, 257, 53

\bibitem[{{Luque} {et~al.}(2023){Luque}, {Osborn}, {Leleu}, {Pall{\'e}},
  {Bonfanti}, {Barrag{\'a}n}, {Wilson}, {Broeg}, {Cameron}, {Lendl}, {Maxted},
  {Alibert}, {Gandolfi}, {Delisle}, {Hooton}, {Egger}, {Nowak}, {Lafarga},
  {Rapetti}, {Twicken}, {Morales}, {Carleo}, {Orell-Miquel}, {Adibekyan},
  {Alonso}, {Alqasim}, {Amado}, {Anderson}, {Anglada-Escud{\'e}}, {Bandy},
  {B{\'a}rczy}, {Barrado Navascues}, {Barros}, {Baumjohann}, {Bayliss}, {Bean},
  {Beck}, {Beck}, {Benz}, {Billot}, {Bonfils}, {Borsato}, {Boyle}, {Brandeker},
  {Bryant}, {Cabrera}, {Carrazco-Gaxiola}, {Charbonneau}, {Charnoz}, {Ciardi},
  {Cochran}, {Collins}, {Crossfield}, {Csizmadia}, {Cubillos}, {Dai}, {Davies},
  {Deeg}, {Deleuil}, {Deline}, {Delrez}, {Demangeon}, {Demory}, {Ehrenreich},
  {Erikson}, {Esparza-Borges}, {Falk}, {Fortier}, {Fossati}, {Fridlund},
  {Fukui}, {Garcia-Mejia}, {Gill}, {Gillon}, {Goffo}, {G{\'o}mez Maqueo Chew},
  {G{\"u}del}, {Guenther}, {G{\"u}nther}, {Hatzes}, {Helling}, {Hesse},
  {Howell}, {Hoyer}, {Ikuta}, {Isaak}, {Jenkins}, {Kagetani}, {Kiss}, {Kodama},
  {Korth}, {Lam}, {Laskar}, {Latham}, {Lecavelier des Etangs}, {Leon},
  {Livingston}, {Magrin}, {Matson}, {Matthews}, {Mordasini}, {Mori}, {Moyano},
  {Munari}, {Murgas}, {Narita}, {Nascimbeni}, {Olofsson}, {Osborne},
  {Ottensamer}, {Pagano}, {Parviainen}, {Peter}, {Piotto}, {Pollacco},
  {Queloz}, {Quinn}, {Quirrenbach}, {Ragazzoni}, {Rando}, {Ratti}, {Rauer},
  {Redfield}, {Ribas}, {Ricker}, {Rudat}, {Sabin}, {Salmon}, {Santos},
  {Scandariato}, {Schanche}, {Schlieder}, {Seager}, {S{\'e}gransan}, {Shporer},
  {Simon}, {Smith}, {Sousa}, {Stalport}, {Szab{\'o}}, {Thomas}, {Tuson},
  {Udry}, {Vanderburg}, {Van Eylen}, {Van Grootel}, {Venturini}, {Walter},
  {Walton}, {Watanabe}, {Winn}, \& {Zingales}}]{luque2023}
{Luque}, R., {Osborn}, H.~P., {Leleu}, A., {et~al.} 2023, \nat, 623, 932

\bibitem[{{Mamajek} \& {Hillenbrand}(2008)}]{Mamajek2008}
{Mamajek}, E.~E. \& {Hillenbrand}, L.~A. 2008, \apj, 687, 1264

\bibitem[{{Marigo} {et~al.}(2017){Marigo}, {Girardi}, {Bressan}, {Rosenfield},
  {Aringer}, {Chen}, {Dussin}, {Nanni}, {Pastorelli}, {Rodrigues}, {Trabucchi},
  {Bladh}, {Dalcanton}, {Groenewegen}, {Montalb{\'a}n}, \& {Wood}}]{marigo2017}
{Marigo}, P., {Girardi}, L., {Bressan}, A., {et~al.} 2017, \apj, 835, 77

\bibitem[{{Mayor} {et~al.}(2003){Mayor}, {Pepe}, {Queloz}, {Bouchy},
  {Rupprecht}, {Lo Curto}, {Avila}, {Benz}, {Bertaux}, {Bonfils}, {Dall},
  {Dekker}, {Delabre}, {Eckert}, {Fleury}, {Gilliotte}, {Gojak}, {Guzman},
  {Kohler}, {Lizon}, {Longinotti}, {Lovis}, {Megevand}, {Pasquini}, {Reyes},
  {Sivan}, {Sosnowska}, {Soto}, {Udry}, {van Kesteren}, {Weber}, \&
  {Weilenmann}}]{mayor2003}
{Mayor}, M., {Pepe}, F., {Queloz}, D., {et~al.} 2003, The Messenger, 114, 20

\bibitem[{{McKay} {et~al.}(2019){McKay}, {DiSanti}, {Kelley}, {Knight},
  {Womack}, {Wierzchos}, {Harrington Pinto}, {Bonev}, {Villanueva}, {Dello
  Russo}, {Cochran}, {Biver}, {Bauer}, {Vervack}, {Gibb}, {Roth}, \&
  {Kawakita}}]{McKayEtal2019ajPanSTARRScometCO}
{McKay}, A.~J., {DiSanti}, M.~A., {Kelley}, M. S.~P., {et~al.} 2019, \aj, 158,
  128

\bibitem[{{Mishra} {et~al.}(2023){Mishra}, {Alibert}, {Udry}, \&
  {Mordasini}}]{mishra2023}
{Mishra}, L., {Alibert}, Y., {Udry}, S., \& {Mordasini}, C. 2023, \aap, 670,
  A68

\bibitem[{{Morton}(2012)}]{morton2012}
{Morton}, T.~D. 2012, \apj, 761, 6

\bibitem[{{Morton}(2015)}]{morton2015}
{Morton}, T.~D. 2015, {VESPA: False positive probabilities calculator},
  Astrophysics Source Code Library, record ascl:1503.011

\bibitem[{{Mumma} \&
  {Charnley}(2011)}]{MummaCharnley2011araaCometsChemicalComposition}
{Mumma}, M.~J. \& {Charnley}, S.~B. 2011, \araa, 49, 471

\bibitem[{{Murdoch} {et~al.}(1993){Murdoch}, {Hearnshaw}, \&
  {Clark}}]{Murdoch1993}
{Murdoch}, K.~A., {Hearnshaw}, J.~B., \& {Clark}, M. 1993, \apj, 413, 349

\bibitem[{{Nascimbeni} {et~al.}(2024){Nascimbeni}, {Borsato}, {Leonardi},
  {Sousa}, {Wilson}, {Fortier}, {Heitzmann}, {Mantovan}, {Luque}, {Zingales},
  {Piotto}, {Alibert}, {Alonso}, {B{\'a}rczy}, {Barrado Navascues}, {Barros},
  {Baumjohann}, {Beck}, {Benz}, {Billot}, {Biondi}, {Brandeker}, {Broeg},
  {Busch}, {Collier Cameron}, {Correia}, {Csizmadia}, {Cubillos}, {Davies},
  {Deleuil}, {Deline}, {Delrez}, {Demangeon}, {Demory}, {Derekas}, {Edwards},
  {Ehrenreich}, {Erikson}, {Fossati}, {Fridlund}, {Gandolfi}, {Gazeas},
  {Gillon}, {G{\"u}del}, {G{\"u}nther}, {Helling}, {Isaak}, {Kerschbaum},
  {Kiss}, {Korth}, {Lam}, {Laskar}, {Lecavelier des Etangs}, {Leleu}, {Lendl},
  {Magrin}, {Maxted}, {Mer{\'\i}n}, {Mordasini}, {Olofsson}, {Ottensamer},
  {Pagano}, {Pall{\'e}}, {Peter}, {Pollacco}, {Queloz}, {Ragazzoni}, {Rando},
  {Rauer}, {Ribas}, {Santos}, {Scandariato}, {S{\'e}gransan}, {Simon}, {Smith},
  {Southworth}, {Stalport}, {Sulis}, {Szab{\'o}}, {Udry}, {Ulmer}, {Van
  Grootel}, {Venturini}, {Villaver}, \& {Walton}}]{nascimbeni2024}
{Nascimbeni}, V., {Borsato}, L., {Leonardi}, P., {et~al.} 2024, \aap, 690, A349

\bibitem[{{Nascimbeni} {et~al.}(2023){Nascimbeni}, {Borsato}, {Zingales},
  {Piotto}, {Pagano}, {Beck}, {Broeg}, {Ehrenreich}, {Hoyer}, {Majidi},
  {Granata}, {Sousa}, {Wilson}, {Van Grootel}, {Bonfanti}, {Salmon}, {Mustill},
  {Delrez}, {Alibert}, {Alonso}, {Anglada}, {B{\'a}rczy}, {Barrado}, {Barros},
  {Baumjohann}, {Beck}, {Benz}, {Bergomi}, {Billot}, {Bonfils}, {Brandeker},
  {Cabrera}, {Charnoz}, {Collier Cameron}, {Csizmadia}, {Cubillos}, {Davies},
  {Deleuil}, {Deline}, {Demangeon}, {Demory}, {Erikson}, {Fortier}, {Fossati},
  {Fridlund}, {Gandolfi}, {Gillon}, {G{\"u}del}, {Isaak}, {Kiss}, {Laskar},
  {Lecavelier des Etangs}, {Lendl}, {Lovis}, {Luque}, {Magrin}, {Maxted},
  {Mordasini}, {Olofsson}, {Ottensamer}, {Pall{\'e}}, {Peter}, {Piazza},
  {Pollacco}, {Queloz}, {Ragazzoni}, {Rando}, {Ratti}, {Rauer}, {Ribas},
  {Santos}, {Scandariato}, {S{\'e}gransan}, {Simon}, {Smith}, {Steinberger},
  {Steller}, {Szab{\'o}}, {Thomas}, {Udry}, {Venturini}, {Walton}, \&
  {Wolter}}]{Nascimbeni2023A&A...673A..42N}
{Nascimbeni}, V., {Borsato}, L., {Zingales}, T., {et~al.} 2023, \aap, 673, A42

\bibitem[{{Nelson} {et~al.}(2014){Nelson}, {Ford}, \&
  {Payne}}]{DENelson2014ApJS..210...11N}
{Nelson}, B., {Ford}, E.~B., \& {Payne}, M.~J. 2014, \apjs, 210, 11

\bibitem[{{Noyes} {et~al.}(1984){Noyes}, {Hartmann}, {Baliunas}, {Duncan}, \&
  {Vaughan}}]{Noyes1984}
{Noyes}, R.~W., {Hartmann}, L.~W., {Baliunas}, S.~L., {Duncan}, D.~K., \&
  {Vaughan}, A.~H. 1984, \apj, 279, 763

\bibitem[{{Otegi} {et~al.}(2020{\natexlab{a}}){Otegi}, {Bouchy}, \&
  {Helled}}]{otegi2020MR}
{Otegi}, J.~F., {Bouchy}, F., \& {Helled}, R. 2020{\natexlab{a}}, \aap, 634,
  A43

\bibitem[{{Otegi} {et~al.}(2020{\natexlab{b}}){Otegi}, {Dorn}, {Helled},
  {Bouchy}, {Haldemann}, \& {Alibert}}]{otegi2020internalStructure}
{Otegi}, J.~F., {Dorn}, C., {Helled}, R., {et~al.} 2020{\natexlab{b}}, \aap,
  640, A135

\bibitem[{{Otegi} {et~al.}(2022){Otegi}, {Helled}, \& {Bouchy}}]{otegi2022}
{Otegi}, J.~F., {Helled}, R., \& {Bouchy}, F. 2022, \aap, 658, A107

\bibitem[{{Parviainen} {et~al.}(2016){Parviainen}, {Pall{\'e}}, {Nortmann},
  {Nowak}, {Iro}, {Murgas}, \& {Aigrain}}]{Parviainen2016}
{Parviainen}, H., {Pall{\'e}}, E., {Nortmann}, L., {et~al.} 2016, \aap, 585,
  A114

\bibitem[{{Pepe} {et~al.}(2002){Pepe}, {Mayor}, {Galland}, {Naef}, {Queloz},
  {Santos}, {Udry}, \& {Burnet}}]{Pepe2002}
{Pepe}, F., {Mayor}, M., {Galland}, F., {et~al.} 2002, \aap, 388, 632

\bibitem[{{Perryman} {et~al.}(1997){Perryman}, {Lindegren}, {Kovalevsky},
  {Hoeg}, {Bastian}, {Bernacca}, {Cr{\'e}z{\'e}}, {Donati}, {Grenon},
  {Grewing}, {van Leeuwen}, {van der Marel}, {Mignard}, {Murray}, {Le Poole},
  {Schrijver}, {Turon}, {Arenou}, {Froeschl{\'e}}, \&
  {Petersen}}]{perryman1997}
{Perryman}, M.~A.~C., {Lindegren}, L., {Kovalevsky}, J., {et~al.} 1997, \aap,
  323, L49

\bibitem[{{Persson} {et~al.}(2018){Persson}, {Fridlund}, {Barrag{\'a}n}, {Dai},
  {Gandolfi}, {Hatzes}, {Hirano}, {Grziwa}, {Korth}, {Prieto-Arranz},
  {Fossati}, {Van Eylen}, {Justesen}, {Livingston}, {Kubyshkina}, {Deeg},
  {Guenther}, {Nowak}, {Cabrera}, {Eigm{\"u}ller}, {Csizmadia}, {Smith},
  {Erikson}, {Albrecht}, {Sobrino}, {Cochran}, {Endl}, {Esposito}, {Fukui},
  {Heeren}, {Hidalgo}, {Hjorth}, {Kuzuhara}, {Narita}, {Nespral}, {Palle},
  {P{\"a}tzold}, {Rauer}, {Rodler}, \& {Winn}}]{persson2018}
{Persson}, C.~M., {Fridlund}, M., {Barrag{\'a}n}, O., {et~al.} 2018, \aap, 618,
  A33

\bibitem[{{Petigura} {et~al.}(2018){Petigura}, {Benneke}, {Batygin}, {Fulton},
  {Werner}, {Krick}, {Gorjian}, {Sinukoff}, {Deck}, {Mills}, \&
  {Deming}}]{petigura2018}
{Petigura}, E.~A., {Benneke}, B., {Batygin}, K., {et~al.} 2018, \aj, 156, 89

\bibitem[{{Petigura} {et~al.}(2016){Petigura}, {Howard}, {Lopez}, {Deck},
  {Fulton}, {Crossfield}, {Ciardi}, {Chiang}, {Lee}, {Isaacson}, {Beichman},
  {Hansen}, {Schlieder}, \& {Sinukoff}}]{petigura2016}
{Petigura}, E.~A., {Howard}, A.~W., {Lopez}, E.~D., {et~al.} 2016, \apj, 818,
  36

\bibitem[{{Petit} {et~al.}(2018){Petit}, {Laskar}, \&
  {Bou{\'e}}}]{Petit2018A&A...617A..93P}
{Petit}, A.~C., {Laskar}, J., \& {Bou{\'e}}, G. 2018, \aap, 617, A93

\bibitem[{{Piskunov} \& {Valenti}(2017)}]{Piskunov2017}
{Piskunov}, N. \& {Valenti}, J.~A. 2017, \aap, 597, A16

\bibitem[{{Piskunov} {et~al.}(1995){Piskunov}, {Kupka}, {Ryabchikova}, {Weiss},
  \& {Jeffery}}]{Piskunov95}
{Piskunov}, N.~E., {Kupka}, F., {Ryabchikova}, T.~A., {Weiss}, W.~W., \&
  {Jeffery}, C.~S. 1995, \aaps, 112, 525

\bibitem[{{Pont} {et~al.}(2006){Pont}, {Zucker}, \& {Queloz}}]{pont2006}
{Pont}, F., {Zucker}, S., \& {Queloz}, D. 2006, \mnras, 373, 231

\bibitem[{{Pontoppidan} {et~al.}(2016){Pontoppidan}, {Pickering}, {Laidler},
  {Gilbert}, {Sontag}, {Slocum}, {Sienkiewicz}, {Hanley}, {Earl}, {Pueyo},
  {Ravindranath}, {Karakla}, {Robberto}, {Noriega-Crespo}, \&
  {Barker}}]{PontoppidanEtal2016spiePandeia}
{Pontoppidan}, K.~M., {Pickering}, T.~E., {Laidler}, V.~G., {et~al.} 2016, in
  Society of Photo-Optical Instrumentation Engineers (SPIE) Conference Series,
  Vol. 9910, Observatory Operations: Strategies, Processes, and Systems VI, ed.
  A.~B. {Peck}, R.~L. {Seaman}, \& C.~R. {Benn}, 991016

\bibitem[{{Queloz} {et~al.}(2001){Queloz}, {Henry}, {Sivan}, {Baliunas},
  {Beuzit}, {Donahue}, {Mayor}, {Naef}, {Perrier}, \& {Udry}}]{Queloz2001}
{Queloz}, D., {Henry}, G.~W., {Sivan}, J.~P., {et~al.} 2001, \aap, 379, 279

\bibitem[{{Rein} \& {Liu}(2012)}]{rebound}
{Rein}, H. \& {Liu}, S.~F. 2012, \aap, 537, A128

\bibitem[{{Rein} \& {Tamayo}(2015)}]{reboundwhfast}
{Rein}, H. \& {Tamayo}, D. 2015, \mnras, 452, 376

\bibitem[{{Richard} {et~al.}(2012){Richard}, {Gordon}, {Rothman}, {Abel},
  {Frommhold}, {Gustafsson}, {Hartmann}, {Hermans}, {Lafferty}, {Orton},
  {Smith}, \& {Tran}}]{RichardEtal2012jqsrtCIA}
{Richard}, C., {Gordon}, I.~E., {Rothman}, L.~S., {et~al.} 2012, \jqsrt, 113,
  1276

\bibitem[{{Ricker} {et~al.}(2015){Ricker}, {Winn}, {Vanderspek}, {Latham},
  {Bakos}, {Bean}, {Berta-Thompson}, {Brown}, {Buchhave}, {Butler}, {Butler},
  {Chaplin}, {Charbonneau}, {Christensen-Dalsgaard}, {Clampin}, {Deming},
  {Doty}, {De Lee}, {Dressing}, {Dunham}, {Endl}, {Fressin}, {Ge}, {Henning},
  {Holman}, {Howard}, {Ida}, {Jenkins}, {Jernigan}, {Johnson}, {Kaltenegger},
  {Kawai}, {Kjeldsen}, {Laughlin}, {Levine}, {Lin}, {Lissauer}, {MacQueen},
  {Marcy}, {McCullough}, {Morton}, {Narita}, {Paegert}, {Palle}, {Pepe},
  {Pepper}, {Quirrenbach}, {Rinehart}, {Sasselov}, {Sato}, {Seager},
  {Sozzetti}, {Stassun}, {Sullivan}, {Szentgyorgyi}, {Torres}, {Udry}, \&
  {Villasenor}}]{ricker2015}
{Ricker}, G.~R., {Winn}, J.~N., {Vanderspek}, R., {et~al.} 2015, Journal of
  Astronomical Telescopes, Instruments, and Systems, 1, 014003

\bibitem[{{Rothman} {et~al.}(2010){Rothman}, {Gordon}, {Barber}, {Dothe},
  {Gamache}, {Goldman}, {Perevalov}, {Tashkun}, \&
  {Tennyson}}]{RothmanEtal2010jqsrtHITEMP}
{Rothman}, L.~S., {Gordon}, I.~E., {Barber}, R.~J., {et~al.} 2010, \jqsrt, 111,
  2139

\bibitem[{{Santos} {et~al.}(2013){Santos}, {Sousa}, {Mortier}, {Neves},
  {Adibekyan}, {Tsantaki}, {Delgado Mena}, {Bonfils}, {Israelian}, {Mayor}, \&
  {Udry}}]{Santos-13}
{Santos}, N.~C., {Sousa}, S.~G., {Mortier}, A., {et~al.} 2013, \aap, 556, A150

\bibitem[{{Schwarz}(1978)}]{schwarz1978}
{Schwarz}, G. 1978, Annals of Statistics, 6, 461

\bibitem[{{Simola} {et~al.}(2022){Simola}, {Bonfanti}, {Dumusque},
  {Cisewski-Kehe}, {Kaski}, \& {Corander}}]{Simola2022}
{Simola}, U., {Bonfanti}, A., {Dumusque}, X., {et~al.} 2022, \aap, 664, A127

\bibitem[{{Simola} {et~al.}(2019){Simola}, {Dumusque}, \&
  {Cisewski-Kehe}}]{simola2019}
{Simola}, U., {Dumusque}, X., \& {Cisewski-Kehe}, J. 2019, \aap, 622, A131

\bibitem[{{Smith} {et~al.}(2012){Smith}, {Stumpe}, {Van Cleve}, {Jenkins},
  {Barclay}, {Fanelli}, {Girouard}, {Kolodziejczak}, {McCauliff}, {Morris}, \&
  {Twicken}}]{Smith2012}
{Smith}, J.~C., {Stumpe}, M.~C., {Van Cleve}, J.~E., {et~al.} 2012, \pasp, 124,
  1000

\bibitem[{{Sneden}(1973)}]{Sneden-73}
{Sneden}, C.~A. 1973, PhD thesis, THE UNIVERSITY OF TEXAS AT AUSTIN.

\bibitem[{{Sousa}(2014)}]{Sousa-14}
{Sousa}, S.~G. 2014, in Determination of Atmospheric Parameters of B-, A-, F-
  and G-Type Stars., ed. E.~{Niemczura}, B.~{Smalley}, \& W.~{Pych} (Springer
  International Publishing (Cham)), 297--310

\bibitem[{{Sousa} {et~al.}(2021){Sousa}, {Adibekyan}, {Delgado-Mena}, {Santos},
  {Rojas-Ayala}, {Soares}, {Legoinha}, {Ulmer-Moll}, {Camacho}, {Barros},
  {Demangeon}, {Hoyer}, {Israelian}, {Mortier}, {Tsantaki}, \&
  {Monteiro}}]{Sousa-21}
{Sousa}, S.~G., {Adibekyan}, V., {Delgado-Mena}, E., {et~al.} 2021, \aap, 656,
  A53

\bibitem[{{Sousa} {et~al.}(2015){Sousa}, {Santos}, {Adibekyan}, {Delgado-Mena},
  \& {Israelian}}]{Sousa-15}
{Sousa}, S.~G., {Santos}, N.~C., {Adibekyan}, V., {Delgado-Mena}, E., \&
  {Israelian}, G. 2015, \aap, 577, A67

\bibitem[{{Sousa} {et~al.}(2007){Sousa}, {Santos}, {Israelian}, {Mayor}, \&
  {Monteiro}}]{Sousa-07}
{Sousa}, S.~G., {Santos}, N.~C., {Israelian}, G., {Mayor}, M., \& {Monteiro},
  M.~J.~P.~F.~G. 2007, A\&A, 469, 783

\bibitem[{{Sousa} {et~al.}(2008){Sousa}, {Santos}, {Mayor}, {Udry},
  {Casagrande}, {Israelian}, {Pepe}, {Queloz}, \& {Monteiro}}]{Sousa-08}
{Sousa}, S.~G., {Santos}, N.~C., {Mayor}, M., {et~al.} 2008, \aap, 487, 373

\bibitem[{Storn \& Price(1997)}]{Storn1997}
Storn, R. \& Price, K. 1997, Journal of Global Optimization, 11, 341

\bibitem[{{Stumpe} {et~al.}(2014){Stumpe}, {Smith}, {Catanzarite}, {Van Cleve},
  {Jenkins}, {Twicken}, \& {Girouard}}]{Stumpe2014}
{Stumpe}, M.~C., {Smith}, J.~C., {Catanzarite}, J.~H., {et~al.} 2014, \pasp,
  126, 100

\bibitem[{{Stumpe} {et~al.}(2012){Stumpe}, {Smith}, {Van Cleve}, {Twicken},
  {Barclay}, {Fanelli}, {Girouard}, {Jenkins}, {Kolodziejczak}, {McCauliff}, \&
  {Morris}}]{Stumpe2012}
{Stumpe}, M.~C., {Smith}, J.~C., {Van Cleve}, J.~E., {et~al.} 2012, \pasp, 124,
  985

\bibitem[{{Su{\'a}rez Mascare{\~n}o} {et~al.}(2017){Su{\'a}rez Mascare{\~n}o},
  {Rebolo}, {Gonz{\'a}lez Hern{\'a}ndez}, \& {Esposito}}]{suarezMascareno2017}
{Su{\'a}rez Mascare{\~n}o}, A., {Rebolo}, R., {Gonz{\'a}lez Hern{\'a}ndez},
  J.~I., \& {Esposito}, M. 2017, \mnras, 468, 4772

\bibitem[{Sundman(1913)}]{sundman1913}
Sundman, K.~F. 1913, Acta Mathematica, 36, 105

\bibitem[{{Tanaka} {et~al.}(2002){Tanaka}, {Takeuchi}, \& {Ward}}]{tanaka2002}
{Tanaka}, H., {Takeuchi}, T., \& {Ward}, W.~R. 2002, \apj, 565, 1257

\bibitem[{{Tennyson} {et~al.}(2016){Tennyson}, {Yurchenko}, {Al-Refaie},
  {Barton}, {Chubb}, {Coles}, {Diamantopoulou}, {Gorman}, {Hill}, {Lam},
  {Lodi}, {McKemmish}, {Na}, {Owens}, {Polyansky}, {Rivlin}, {Sousa-Silva},
  {Underwood}, {Yachmenev}, \& {Zak}}]{TennysonEtal2016jmsExomol}
{Tennyson}, J., {Yurchenko}, S.~N., {Al-Refaie}, A.~F., {et~al.} 2016, Journal
  of Molecular Spectroscopy, 327, 73

\bibitem[{ter Braak \& Vrugt(2008)}]{terBraak2008}
ter Braak, C. J.~F. \& Vrugt, J.~A. 2008, Statistics and Computing, 18, 435

\bibitem[{{Thiabaud} {et~al.}(2014){Thiabaud}, {Marboeuf}, {Alibert}, {Cabral},
  {Leya}, \& {Mezger}}]{thiabaud2014}
{Thiabaud}, A., {Marboeuf}, U., {Alibert}, Y., {et~al.} 2014, \aap, 562, A27

\bibitem[{{Thiabaud} {et~al.}(2015){Thiabaud}, {Marboeuf}, {Alibert}, {Leya},
  \& {Mezger}}]{thiabaud2015}
{Thiabaud}, A., {Marboeuf}, U., {Alibert}, Y., {Leya}, I., \& {Mezger}, K.
  2015, \aap, 574, A138

\bibitem[{{Trotta}(2007)}]{trotta2007}
{Trotta}, R. 2007, \mnras, 378, 72

\bibitem[{{Van Eylen} {et~al.}(2019){Van Eylen}, {Albrecht}, {Huang},
  {MacDonald}, {Dawson}, {Cai}, {Foreman-Mackey}, {Lundkvist}, {Silva Aguirre},
  {Snellen}, \& {Winn}}]{VanEylen2019AJ....157...61V}
{Van Eylen}, V., {Albrecht}, S., {Huang}, X., {et~al.} 2019, \aj, 157, 61

\bibitem[{{Vanderburg} {et~al.}(2019){Vanderburg}, {Huang}, {Rodriguez},
  {Becker}, {Ricker}, {Vanderspek}, {Latham}, {Seager}, {Winn}, {Jenkins},
  {Addison}, {Bieryla}, {Brice{\~n}o}, {Bowler}, {Brown}, {Burke}, {Burt},
  {Caldwell}, {Clark}, {Crossfield}, {Dittmann}, {Dynes}, {Fulton}, {Guerrero},
  {Harbeck}, {Horner}, {Kane}, {Kielkopf}, {Kraus}, {Kreidberg}, {Law}, {Mann},
  {Mengel}, {Morton}, {Okumura}, {Pearce}, {Plavchan}, {Quinn}, {Rabus},
  {Rose}, {Rowden}, {Shporer}, {Siverd}, {Smith}, {Stassun}, {Tinney},
  {Wittenmyer}, {Wright}, {Zhang}, {Zhou}, \& {Ziegler}}]{Vanderburg2019}
{Vanderburg}, A., {Huang}, C.~X., {Rodriguez}, J.~E., {et~al.} 2019, \apjl,
  881, L19

\bibitem[{{Vanderburg} {et~al.}(2016){Vanderburg}, {Plavchan}, {Johnson},
  {Ciardi}, {Swift}, \& {Kane}}]{vanderburg2016}
{Vanderburg}, A., {Plavchan}, P., {Johnson}, J.~A., {et~al.} 2016, \mnras, 459,
  3565

\bibitem[{Virtanen {et~al.}(2020)Virtanen, Gommers, Oliphant, Haberland, Reddy,
  Cournapeau, Burovski, Peterson, Weckesser, Bright, {van der Walt}, Brett,
  Wilson, Millman, Mayorov, Nelson, Jones, Kern, Larson, Carey, Polat, Feng,
  Moore, {VanderPlas}, Laxalde, Perktold, Cimrman, Henriksen, Quintero, Harris,
  Archibald, Ribeiro, Pedregosa, {van Mulbregt}, \& {SciPy 1.0
  Contributors}}]{sciPy2020}
Virtanen, P., Gommers, R., Oliphant, T.~E., {et~al.} 2020, Nature Methods, 17,
  261

\bibitem[{{Walsh} {et~al.}(2015){Walsh}, {Nomura}, \& {van
  Dishoeck}}]{walsh2015}
{Walsh}, C., {Nomura}, H., \& {van Dishoeck}, E. 2015, \aap, 582, A88

\bibitem[{{Weiss} \& {Marcy}(2014)}]{weiss2014}
{Weiss}, L.~M. \& {Marcy}, G.~W. 2014, \apjl, 783, L6

\bibitem[{{Weiss} {et~al.}(2018){Weiss}, {Marcy}, {Petigura}, {Fulton},
  {Howard}, {Winn}, {Isaacson}, {Morton}, {Hirsch}, {Sinukoff}, {Cumming},
  {Hebb}, \& {Cargile}}]{weiss2018}
{Weiss}, L.~M., {Marcy}, G.~W., {Petigura}, E.~A., {et~al.} 2018, \aj, 155, 48

\bibitem[{{Wildi} {et~al.}(2010){Wildi}, {Pepe}, {Chazelas}, {Lo Curto}, \&
  {Lovis}}]{wildi2010}
{Wildi}, F., {Pepe}, F., {Chazelas}, B., {Lo Curto}, G., \& {Lovis}, C. 2010,
  in Society of Photo-Optical Instrumentation Engineers (SPIE) Conference
  Series, Vol. 7735, Ground-based and Airborne Instrumentation for Astronomy
  III, ed. I.~S. {McLean}, S.~K. {Ramsay}, \& H.~{Takami}, 77354X

\bibitem[{{Winn}(2010)}]{Winn2010}
{Winn}, J.~N. 2010, in Exoplanets, ed. S.~{Seager} (University of Arizona
  Press, Tucson, AZ), 55--77

\bibitem[{{Winn} \& {Fabrycky}(2015)}]{winn2015}
{Winn}, J.~N. \& {Fabrycky}, D.~C. 2015, \araa, 53, 409

\bibitem[{{Wisdom} \& {Holman}(1991)}]{wh}
{Wisdom}, J. \& {Holman}, M. 1991, \aj, 102, 1528

\bibitem[{{Zechmeister} \& {K{\"u}rster}(2009)}]{Zechmeister2009}
{Zechmeister}, M. \& {K{\"u}rster}, M. 2009, \aap, 496, 577

\end{thebibliography}

\begin{appendix}

\section{Supplementary material}

\begin{figure}
    \centering
    \includegraphics[width=\columnwidth]{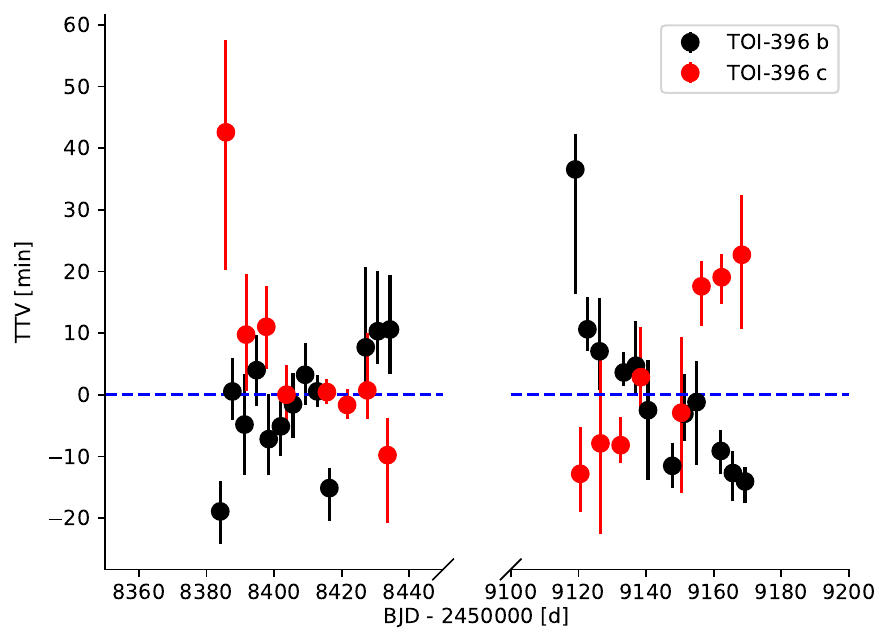}
    \caption{TTV amplitudes obtained for TOI-396\,b (black markers) and TOI-396\,c (red markers). This is a superposition of the first two panels of Figure~\ref{fig:TTVall} emphasising the anti-correlation pattern.}
    \label{fig:TTVbc}
\end{figure}

\newpage
\vfill\null

% \begin{figure*}
\begin{minipage}{\textwidth}
\centering
\includegraphics[height=0.231\textheight]{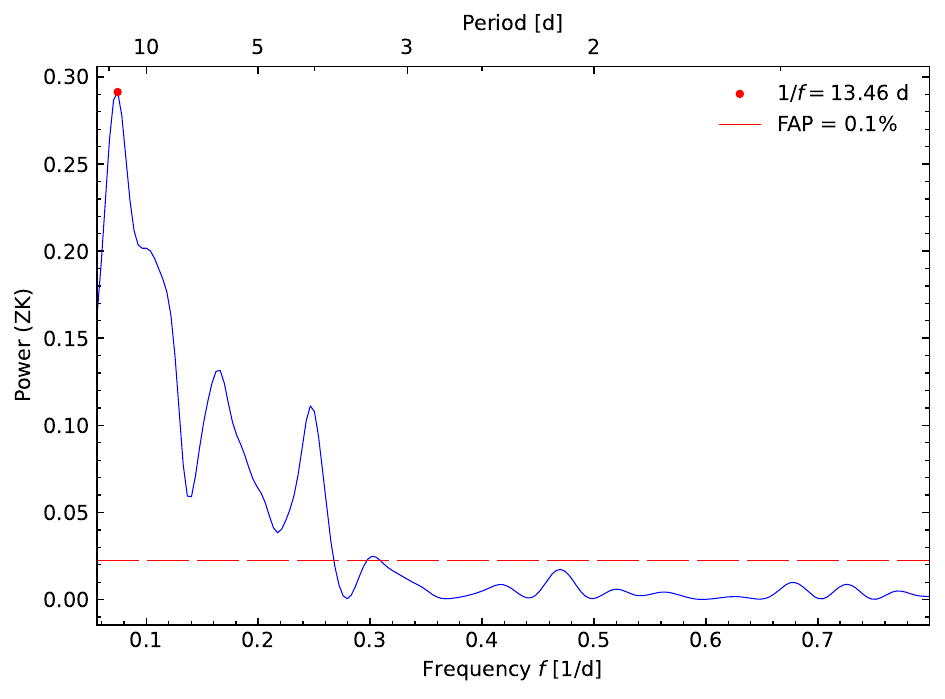}
\includegraphics[height=0.231\textheight]{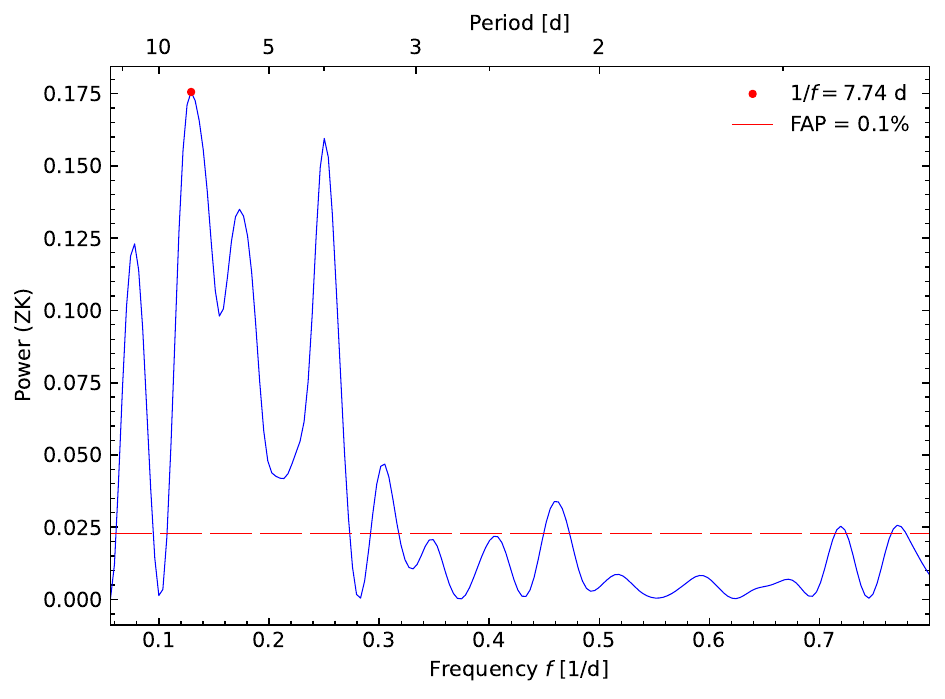} \\
\includegraphics[height=0.231\textheight]{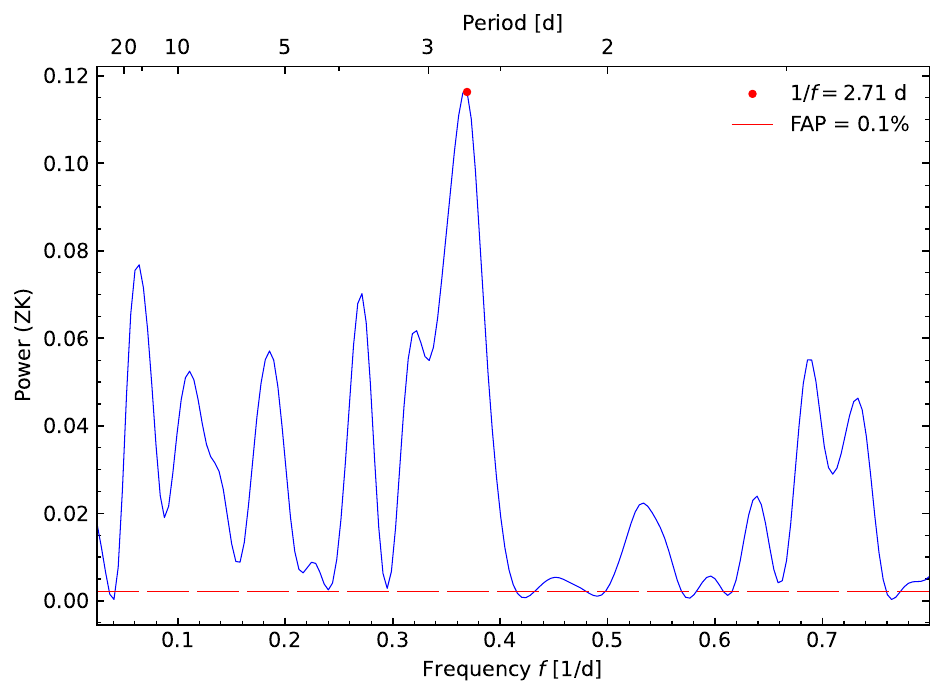}
\includegraphics[height=0.231\textheight]{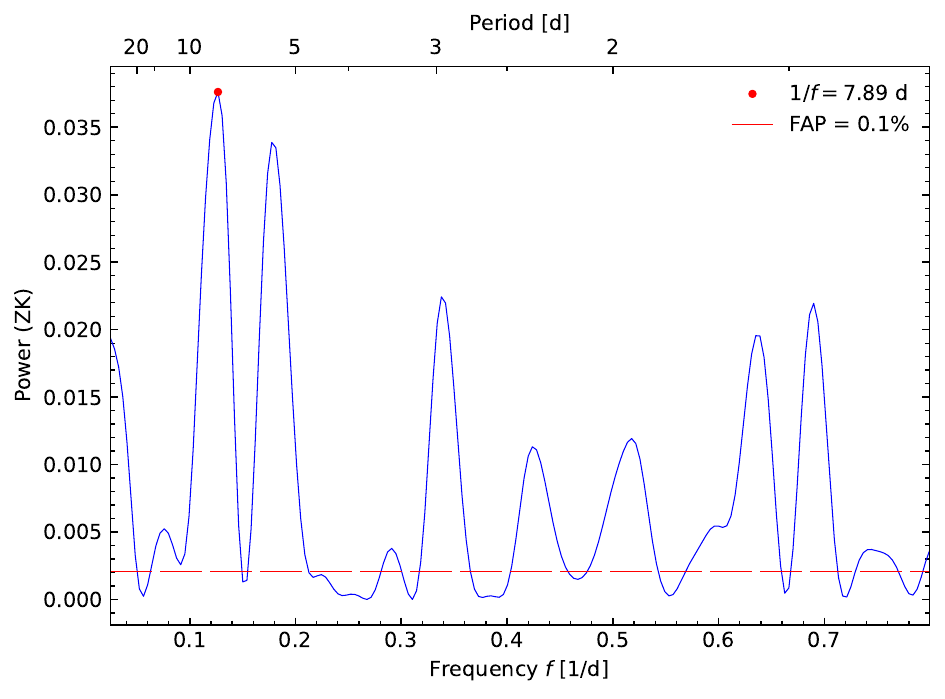} \\
\includegraphics[height=0.231\textheight]{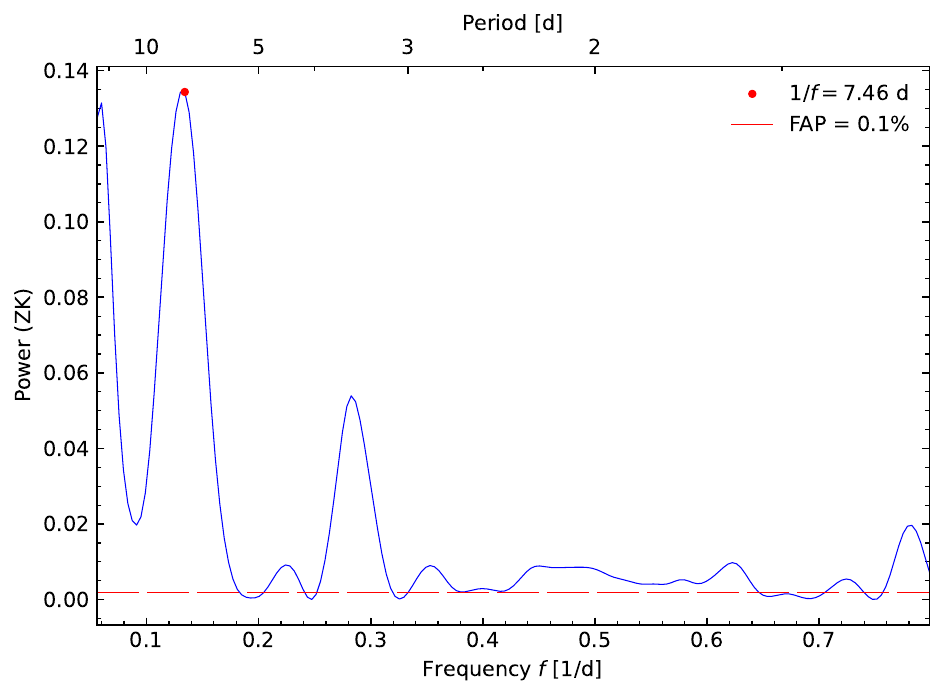}
\includegraphics[height=0.231\textheight]{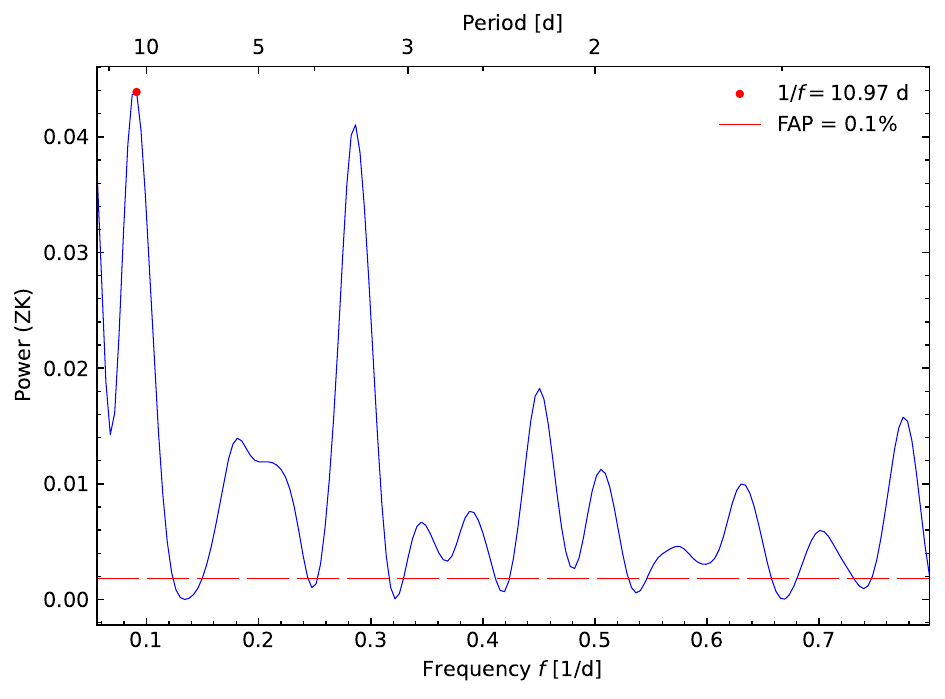} \\
\includegraphics[height=0.231\textheight]{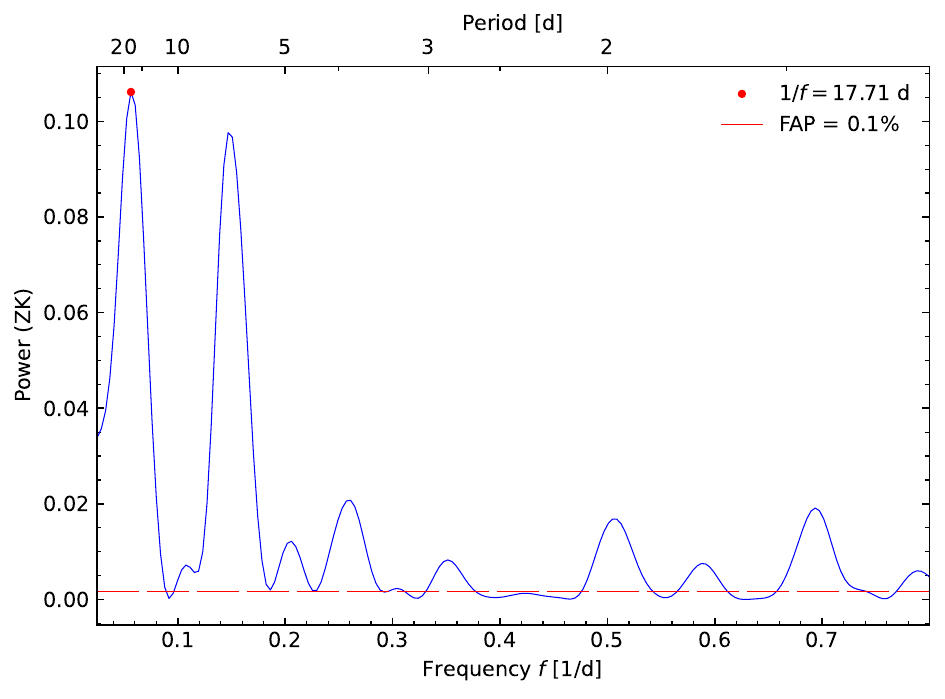}
\includegraphics[height=0.231\textheight]{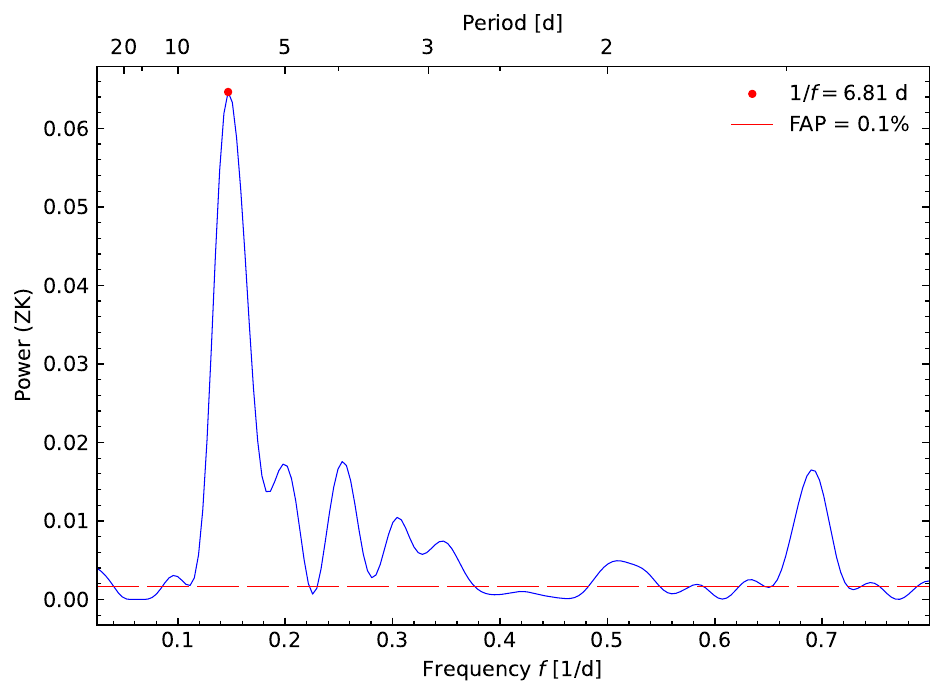}
\captionof{figure}{\emph{Left column}: Generalised Lomb-Scargle periodograms of \tess Sector 3 (\textit{first row}), 4 (\textit{second row}), 30 (\textit{third row}), and 31 (\textit{fourth row}) raw photometry. \textit{Right column}: Generalised Lomb-Scargle periodograms of the residuals after removing the best fitting sinusoidal signals at the most significant frequency identified in the periodograms of the corresponding \tess time series (left column). The long dashed red line marks the FAP at 0.1\%. Significant peaks are detected in the 6---8 day range, which we attribute to the presence of active regions co-rotating with the star.}
  \label{fig:gls-photometry}
% \end{figure*}
\end{minipage}

%optional number of longtab[] needed to be specified manually within the Appendix environment
%https://tex.stackexchange.com/questions/547412/long-table-appendix-numbering-a-b-c
% \longtab[1]{ 
% \input{TOI396_RVlongtable}
% }
\begin{table*}
\centering
\caption{Radial velocities $\overline{RV}$ as extracted from the centred CCFs with their errors $\sigma_{\mathrm{RV}}$. They are followed by the hyperparameters inferred from the SN fit onto the CCFs (i.e. $\mathrm{FWHM_{SN}}$, $A$, and $\gamma$) and by the \emph{bp}-detrended RV values ($RV_{\mathrm{bp}}$) with their errors which also account for the jitter ($\sigma_{\mathrm{RV(bp+jitter)}}$).}
\label{tab:RVdata}
\begin{tabular}{rrrrrrrr}
\hline\hline
\multicolumn{1}{c}{$\mathrm{BJD_{TDB}}$} & \multicolumn{1}{c}{$\overline{RV}$} & \multicolumn{1}{c}{$\sigma_{\mathrm{RV}}$} & \multicolumn{1}{c}{$\mathrm{FWHM_{SN}}$} & \multicolumn{1}{c}{$A$} & \multicolumn{1}{c}{$\gamma$} & \multicolumn{1}{c}{$RV_{\mathrm{bp}}$} & \multicolumn{1}{c}{$\sigma_{\mathrm{RV(bp+jitter)}}$} \\
\multicolumn{1}{c}{[$\mathrm{JD-2\,450\,000}$]} & \multicolumn{1}{c}{[m\,s$^{-1}$]} & \multicolumn{1}{c}{[m\,s$^{-1}$]} & \multicolumn{1}{c}{[km\,s$^{-1}$]} & \multicolumn{1}{c}{[\%]} & \multicolumn{1}{c}{} & \multicolumn{1}{c}{[m\,s$^{-1}$]} & \multicolumn{1}{c}{[m\,s$^{-1}$]} \\
\hline
  8514.522079 & $ -3.951$ & 1.406 & 12.065 & 27.510 & $ 0.0095$ & $-5.358$ & 2.027 \\
  8515.519139 & $  3.873$ & 1.058 & 12.071 & 27.524 & $ 0.0118$ & $ 2.414$ & 1.803 \\
  8516.523282 & $  0.300$ & 0.855 & 12.064 & 27.538 & $ 0.0109$ & $ 1.204$ & 1.692 \\
  8516.600186 & $ -0.492$ & 0.863 & 12.064 & 27.541 & $ 0.0117$ & $ 0.372$ & 1.696 \\
  8517.521720 & $ -1.049$ & 1.010 & 12.055 & 27.532 & $ 0.0105$ & $ 0.890$ & 1.776 \\
  8517.601089 & $  0.374$ & 1.059 & 12.063 & 27.540 & $ 0.0117$ & $ 1.412$ & 1.804 \\
  8518.527888 & $  1.859$ & 1.021 & 12.063 & 27.523 & $ 0.0087$ & $ 1.827$ & 1.782 \\
  8518.576207 & $  4.622$ & 1.062 & 12.072 & 27.533 & $ 0.0084$ & $ 3.890$ & 1.805 \\
  8519.514742 & $ -0.349$ & 0.884 & 12.066 & 27.541 & $ 0.0103$ & $ 0.765$ & 1.707 \\
  8519.526570 & $ -0.606$ & 0.995 & 12.066 & 27.550 & $ 0.0080$ & $ 0.544$ & 1.767 \\
  8519.592896 & $  0.359$ & 1.102 & 12.070 & 27.549 & $ 0.0099$ & $ 1.506$ & 1.829 \\
  8520.511989 & $  2.471$ & 0.962 & 12.064 & 27.534 & $ 0.0079$ & $ 2.884$ & 1.748 \\
  8520.559451 & $ -0.344$ & 1.093 & 12.065 & 27.568 & $ 0.0063$ & $ 1.891$ & 1.824 \\
  8521.517823 & $ -2.621$ & 0.854 & 12.064 & 27.544 & $ 0.0070$ & $-1.799$ & 1.691 \\
  8521.563294 & $ -2.900$ & 1.094 & 12.066 & 27.541 & $ 0.0072$ & $-2.440$ & 1.825 \\
  8522.515348 & $ -1.535$ & 0.840 & 12.067 & 27.555 & $ 0.0085$ & $ 0.283$ & 1.684 \\
  8522.541134 & $  0.229$ & 0.835 & 12.068 & 27.551 & $ 0.0066$ & $ 1.106$ & 1.682 \\
  8524.539218 & $ -4.858$ & 1.008 & 12.059 & 27.543 & $ 0.0135$ & $-3.407$ & 1.774 \\
  8528.508947 & $ -2.042$ & 1.224 & 12.062 & 27.520 & $ 0.0087$ & $-1.652$ & 1.905 \\
  8530.513044 & $  4.381$ & 0.903 & 12.063 & 27.535 & $ 0.0090$ & $ 5.814$ & 1.716 \\
  8532.521390 & $  2.482$ & 0.909 & 12.081 & 27.544 & $ 0.0128$ & $ 1.143$ & 1.720 \\
  8536.503152 & $ -3.627$ & 0.913 & 12.064 & 27.529 & $ 0.0101$ & $-2.800$ & 1.722 \\
  8537.508151 & $ -0.331$ & 1.032 & 12.050 & 27.534 & $ 0.0068$ & $ 1.840$ & 1.788 \\
  8538.506461 & $ -1.146$ & 1.260 & 12.068 & 27.531 & $ 0.0044$ & $-0.293$ & 1.929 \\
  8539.518544 & $  0.945$ & 0.941 & 12.064 & 27.520 & $ 0.0101$ & $ 0.914$ & 1.737 \\
  8540.504854 & $ -0.613$ & 1.169 & 12.057 & 27.523 & $ 0.0065$ & $-0.512$ & 1.870 \\
  8541.516591 & $  1.669$ & 1.039 & 12.072 & 27.558 & $ 0.0079$ & $ 2.180$ & 1.792 \\
  8542.503979 & $ -2.972$ & 0.905 & 12.067 & 27.551 & $ 0.0098$ & $-1.802$ & 1.718 \\
  8543.502941 & $ -4.827$ & 0.781 & 12.052 & 27.548 & $ 0.0110$ & $-1.815$ & 1.656 \\
  8545.522846 & $ -0.012$ & 0.999 & 12.060 & 27.550 & $ 0.0068$ & $ 0.913$ & 1.769 \\
  8546.513570 & $ -3.846$ & 1.053 & 12.069 & 27.547 & $ 0.0074$ & $-4.497$ & 1.800 \\
  8547.514665 & $ -4.125$ & 0.972 & 12.064 & 27.560 & $ 0.0081$ & $-2.977$ & 1.754 \\
  8549.505689 & $ -1.232$ & 0.824 & 12.062 & 27.577 & $ 0.0073$ & $ 0.758$ & 1.676 \\
  8550.519298 & $ -0.642$ & 1.050 & 12.067 & 27.556 & $ 0.0065$ & $-1.147$ & 1.798 \\
  8551.501312 & $ -0.164$ & 1.044 & 12.062 & 27.571 & $ 0.0104$ & $ 1.877$ & 1.795 \\
  8553.499287 & $  3.275$ & 1.525 & 12.061 & 27.536 & $ 0.0080$ & $ 2.157$ & 2.111 \\
  8554.511465 & $ -0.040$ & 1.684 & 12.067 & 27.530 & $ 0.0129$ & $-2.804$ & 2.229 \\
  8555.493401 & $  2.487$ & 0.982 & 12.062 & 27.548 & $ 0.0069$ & $ 1.501$ & 1.760 \\
  8556.494019 & $ -5.920$ & 0.876 & 12.076 & 27.578 & $ 0.0030$ & $-1.109$ & 1.703 \\
  8557.489695 & $  0.012$ & 1.313 & 12.057 & 27.541 & $ 0.0104$ & $-0.227$ & 1.964 \\
  8558.510256 & $  3.576$ & 1.523 & 12.062 & 27.554 & $ 0.0058$ & $ 2.370$ & 2.110 \\
  8656.922558 & $  5.571$ & 1.287 & 12.065 & 27.521 & $ 0.0096$ & $-0.169$ & 1.946 \\
  8657.917377 & $  6.901$ & 1.288 & 12.087 & 27.531 & $ 0.0074$ & $-1.501$ & 1.947 \\
  8659.926600 & $ -1.368$ & 2.267 & 12.068 & 27.527 & $ 0.0046$ & $-5.376$ & 2.697 \\
  8660.912021 & $ -1.737$ & 1.372 & 12.068 & 27.544 & $ 0.0094$ & $-4.360$ & 2.003 \\
\hline
\end{tabular}
\end{table*}

\setcounter{table}{0}
\begin{table*}[h!]
\centering
\caption{continued.}
\begin{tabular}{rrrrrrrr}
\hline\hline
\multicolumn{1}{c}{$\mathrm{BJD_{TDB}}$} & \multicolumn{1}{c}{$\overline{RV}$} & \multicolumn{1}{c}{$\sigma_{\mathrm{RV}}$} & \multicolumn{1}{c}{$\mathrm{FWHM_{SN}}$} & \multicolumn{1}{c}{$A$} & \multicolumn{1}{c}{$\gamma$} & \multicolumn{1}{c}{$RV_{\mathrm{bp}}$} & \multicolumn{1}{c}{$\sigma_{\mathrm{RV(bp+jitter)}}$} \\
\multicolumn{1}{c}{[$\mathrm{JD-2\,450\,000}$]} & \multicolumn{1}{c}{[m\,s$^{-1}$]} & \multicolumn{1}{c}{[m\,s$^{-1}$]} & \multicolumn{1}{c}{[km\,s$^{-1}$]} & \multicolumn{1}{c}{[\%]} & \multicolumn{1}{c}{} & \multicolumn{1}{c}{[m\,s$^{-1}$]} & \multicolumn{1}{c}{[m\,s$^{-1}$]} \\
\hline  
  8660.936119 & $ -0.427$ & 2.037 & 12.076 & 27.560 & $ 0.0121$ & $-3.455$ & 2.506 \\
  8666.907826 & $  6.470$ & 1.514 & 12.078 & 27.534 & $ 0.0104$ & $ 5.351$ & 2.104 \\
  8667.940719 & $ -3.649$ & 0.960 & 12.075 & 27.554 & $ 0.0093$ & $-2.530$ & 1.747 \\
  8668.949085 & $ -8.977$ & 1.057 & 12.076 & 27.552 & $ 0.0028$ & $-0.924$ & 1.802 \\
  8669.932275 & $ -1.509$ & 1.000 & 12.076 & 27.572 & $ 0.0055$ & $ 0.474$ & 1.770 \\
  8670.877547 & $ -3.853$ & 1.101 & 12.066 & 27.551 & $ 0.0048$ & $-2.679$ & 1.829 \\
  8670.943987 & $ -2.070$ & 0.934 & 12.074 & 27.566 & $ 0.0071$ & $-1.958$ & 1.733 \\
  8672.892348 & $  2.329$ & 1.552 & 12.073 & 27.495 & $ 0.0071$ & $-3.765$ & 2.131 \\
  8672.937224 & $  6.039$ & 2.184 & 12.066 & 27.493 & $ 0.0111$ & $ 3.382$ & 2.627 \\
  8673.885378 & $  3.408$ & 1.281 & 12.077 & 27.544 & $ 0.0092$ & $ 2.900$ & 1.942 \\
  8673.953311 & $  9.563$ & 1.450 & 12.080 & 27.534 & $ 0.0132$ & $ 1.387$ & 2.058 \\
  8674.893561 & $  4.520$ & 1.229 & 12.075 & 27.549 & $ 0.0080$ & $ 2.229$ & 1.908 \\
  8674.952407 & $  6.129$ & 1.587 & 12.074 & 27.516 & $ 0.0080$ & $ 1.317$ & 2.156 \\
  8675.896534 & $ -0.472$ & 2.590 & 12.069 & 27.541 & $ 0.0086$ & $-3.271$ & 2.973 \\
  8676.890399 & $  7.142$ & 1.127 & 12.055 & 27.536 & $ 0.0086$ & $ 3.314$ & 1.844 \\
  8676.931698 & $  4.179$ & 1.359 & 12.063 & 27.545 & $ 0.0084$ & $ 0.939$ & 1.994 \\
  8677.940205 & $  4.018$ & 1.075 & 12.076 & 27.551 & $ 0.0093$ & $ 2.179$ & 1.813 \\
  8678.854134 & $  4.830$ & 0.960 & 12.069 & 27.544 & $ 0.0078$ & $ 0.004$ & 1.747 \\
  8678.914902 & $  2.784$ & 0.925 & 12.067 & 27.547 & $ 0.0081$ & $-1.534$ & 1.729 \\
  8679.872861 & $  3.456$ & 1.000 & 12.067 & 27.523 & $ 0.0089$ & $-1.987$ & 1.770 \\
  8679.934961 & $  4.891$ & 0.831 & 12.071 & 27.532 & $ 0.0093$ & $ 0.768$ & 1.680 \\
  8680.893592 & $ -0.225$ & 0.902 & 12.077 & 27.551 & $ 0.0098$ & $-2.422$ & 1.716 \\
  8681.884748 & $  5.537$ & 1.281 & 12.064 & 27.528 & $ 0.0082$ & $-1.250$ & 1.943 \\
  8682.859272 & $ -7.039$ & 0.944 & 12.075 & 27.550 & $ 0.0020$ & $-2.313$ & 1.738 \\
  8682.939961 & $  2.124$ & 1.076 & 12.073 & 27.540 & $ 0.0067$ & $-4.614$ & 1.814 \\
  8683.875650 & $  4.838$ & 1.197 & 12.083 & 27.564 & $ 0.0094$ & $ 2.553$ & 1.888 \\
  8683.908141 & $  6.550$ & 1.150 & 12.082 & 27.560 & $ 0.0090$ & $ 3.403$ & 1.858 \\
  8684.816283 & $ -0.252$ & 0.967 & 12.076 & 27.559 & $ 0.0106$ & $-1.949$ & 1.751 \\
  8689.838032 & $  3.280$ & 1.025 & 12.072 & 27.556 & $ 0.0075$ & $-1.000$ & 1.784 \\
  8689.929566 & $ -6.178$ & 1.172 & 12.066 & 27.527 & $ 0.0018$ & $-1.467$ & 1.872 \\
  8690.851088 & $  6.751$ & 0.946 & 12.064 & 27.542 & $ 0.0062$ & $ 1.820$ & 1.740 \\
  8690.940875 & $  6.573$ & 2.035 & 12.059 & 27.530 & $ 0.0051$ & $ 2.266$ & 2.504 \\
  8691.920653 & $ -0.563$ & 0.781 & 12.093 & 27.551 & $ 0.0088$ & $-1.822$ & 1.656 \\
\hline
\end{tabular}
\end{table*}

%%%
%%%%
\begin{table}
\caption{Polynomial detrending baseline applied to the \tess LCs within the MCMC scheme.}
\label{tab:polyDetrending}
\centering     
\begin{tabular}{l c c}
\hline\hline            
Light curve & Planet & Detrending \\
\hline
TE 1  & b   &  \ttt$^2$ + \ttdy$^1$ \\
TE 2  & c   &  \ttt$^3$ + \ttdx$^2$ + \ttxy$^1$  \\
TE 3  & b d &  $c$    \\
TE 4  & b c &  $c$   \\
TE 5  & b   &  \ttt$^3$ +\ttdx$^1$   \\
TE 6  & c   &  $c$   \\
TE 7  & b d &  \ttt$^1$   \\
TE 8  & b   &  \ttt$^1$ + \ttdy$^1$    \\
TE 9  & c   &  \ttt$^1$   \\
TE 10 & b   &  \ttt$^1$   \\
TE 11 & b   &  \ttt$^4$ + \ttdy$^1$ \\
TE 12 & b   &  \ttt$^3$ + \ttdy$^2$  \\
TE 13 & c   &  \ttt$^2$ + \ttdy$^1$    \\
TE 14 & b   &  $c$   \\
TE 15 & c   &  \ttt$^2$ + \ttdy$^1$ + \ttxy$^1$   \\
TE 16 & b   &  $c$   \\
TE 17 & c   &  \ttt$^1$   \\
TE 18 & b   &  \ttt$^4$ + \ttdy$^1$   \\
TE 19 & d   &  \ttt$^2$   \\
TE 20 & c   &  \ttt$^4$ + \ttdy$^1$  \\
TE 21 & b   &  $c$ \\
TE 22 & d   &  $c$   \\
TE 23 & b   &  $c$    \\
TE 24 & c   &  $c$   \\
TE 25 & b   &  $c$   \\
TE 26 & b c &  $c$   \\
TE 27 & c   &  $c$   \\
TE 28 & b   &  \ttdy$^1$    \\
TE 29 & b   &  $c$   \\
TE 30 & c   &  \ttt$^1$ + \ttdx$^2$   \\
TE 31 & d   &  \ttdx$^2$ + \ttxy$^1$ \\
TE 32 & b   &  \ttdy$^1$   \\
TE 33 & b   &  $c$    \\
TE 34 & c   &  \ttt$^1$   \\
TE 35 & b d &  \ttt$^1$   \\
TE 36 & b   &  \ttdx$^1$   \\
TE 37 & c   &  \ttt$^2$   \\
TE 38 & b c d  & \ttdx$^1$     \\
TE 39 & b   &  \ttdx$^1$   \\
TE 40 & c   &  \ttt$^1$   \\
TE 41 & b   &  \ttt$^1$   \\
\hline
\end{tabular}
\tablefoot{\tess (TE) LCs are identified by a counter based on the chronological order of observation. In particular, LCs from 1 to 11, from 12 to 21, from 22 to 32, and from 33 to 41 were extracted from Sector 3, 4, 30, and 31, respectively. $c$ indicates a normalisation scalar; see text for further details.}
\end{table}

%%%
\begin{table*}
\caption{Timing of each transit event $T_{\mathrm{tr}}$ and the corresponding TTV amplitude as computed with respect to linear ephemerides. The last column specifies the \tess sector where the transit occurs.}
\label{tab:TTV}
\centering
\begin{minipage}{0.49\textwidth}
\begin{tabular}{l c c}
\hline\hline
\multicolumn{3}{c}{TOI-396\,b} \\
$T_{\mathrm{tr}}$ [BJD] & TTV [min] & Sector \\
\hline
$8384.0799_{-0.0037}^{+0.0035}$ & $-18.9_{-5.3}^{+5.0}$ & 3 \\
$8387.6787_{-0.0033}^{+0.0038}$ & $+0.5_{-4.7}^{+5.4}$ &  3 \\
$8391.2602_{-0.0056}^{+0.0057}$ & $-4.8_{-8.1}^{+8.2}$ &  3 \\
$8394.8516\pm0.0040$ & $+4.0\pm5.8$ &                     3 \\
$8398.4292_{-0.0041}^{+0.0051}$ & $-7.2_{-5.9}^{+7.3}$ &  3 \\
$8402.0159_{-0.0033}^{+0.0029}$ & $-5.1_{-4.8}^{+4.1}$ &  3 \\
$8405.6036_{-0.0038}^{+0.0036}$ & $-1.6_{-5.5}^{+5.1}$ &  3 \\
$8409.1923_{-0.0034}^{+0.0036}$ & $+3.2_{-4.9}^{+5.2}$ &  3 \\
$8412.7757_{-0.0017}^{+0.0019}$ & $+0.5_{-2.5}^{+2.7}$ &  4 \\
$8416.3501_{-0.0037}^{+0.0023}$ & $-15.1_{-5.3}^{+3.2}$ & 4 \\
$8427.1218_{-0.0041}^{+0.0091}$ & $+8_{-6}^{+13}$ &       4 \\
$8430.7089_{-0.0037}^{+0.0068}$ & $+10.3_{-5.3}^{+9.8}$ & 4 \\
$8434.2944_{-0.0050}^{+0.0062}$ & $+10.6_{-3.5}^{+5.2}$ & 4 \\
$9119.102_{-0.014}^{+0.004}$ & $+37_{-20}^{+6}$ &        30 \\
$9122.6695_{-0.0024}^{+0.0036}$ & $+10.6_{-3.5}^{+5.2}$ &30 \\
$9126.2523_{-0.0044}^{+0.0060}$ & $+7.0_{-6.3}^{+8.6}$ & 30 \\
$9133.4205_{-0.0015}^{+0.0023}$ & $+3.6_{-2.2}^{+3.3}$ & 30 \\
$9137.0065_{-0.0031}^{+0.0051}$ & $+4.7_{-4.4}^{+7.3}$ & 30 \\
$9140.5868_{-0.0079}^{+0.0056}$ & $-3_{-11}^{+8}$ &      30 \\
$9147.7511_{-0.0025}^{+0.0026}$ & $-11.5_{-3.6}^{+3.7}$ &31 \\
$9151.3423_{-0.0031}^{+0.0045}$ & $-3.1_{-4.4}^{+6.4}$ & 31 \\
$9154.9289_{-0.0071}^{+0.0046}$ & $-1_{-10}^{+7}$ &      31 \\
$9162.0940_{-0.0027}^{+0.0024}$ & $-9.1_{-3.8}^{+3.4}$ & 31 \\
$9165.6768_{-0.0032}^{+0.0025}$ & $-12.7_{-4.6}^{+3.6}$ &31 \\
$9169.2611_{-0.0024}^{+0.0016}$ & $-14.1_{-3.4}^{+2.4}$ &31 \\
\hline
\end{tabular}
\end{minipage}
\begin{minipage}{0.49\textwidth}
\begin{tabular}{l c c}
\hline\hline
\multicolumn{3}{c}{TOI-396\,c} \\
$T_{\mathrm{tr}}$ [BJD] & TTV [min] & Sector \\
\hline
$8385.794_{-0.016}^{+0.010}$ &  $+43_{-22}^{+15}$ &       3 \\
$8391.7447_{-0.0063}^{+0.0068}$ & $+9.7_{-9.1}^{+9.8}$ &  3 \\
$8397.7195_{-0.0047}^{+0.0046}$ & $+11.0_{-6.8}^{+6.6}$ & 3 \\
$8403.6857\pm0.0033$            & $0.0_{-4.8}^{+4.7}$ &   3 \\
$8415.6337_{-0.0013}^{+0.0015}$ & $+0.4_{-1.9}^{+2.2}$ &  4 \\
$8421.6062_{-0.0016}^{+0.0017}$ & $-1.7_{-2.3}^{+2.5}$ &  4 \\
$8427.5817_{-0.0033}^{+0.0065}$ & $+0.7_{-4.7}^{+9.4}$ &  4 \\
$8433.5482_{-0.0077}^{+0.0042}$ & $-10_{-11}^{+6}$ &      4 \\
$9120.5407_{-0.0043}^{+0.0052}$ & $-12.8_{-6.3}^{+7.5}$ &30 \\
$9126.518\pm0.010$              & $-8_{-15}^{+14}$ &     30 \\
$9132.4916_{-0.0020}^{+0.0032}$ & $-8.2_{-2.9}^{+4.5}$ & 30 \\
$9138.4731_{-0.0040}^{+0.0057}$ & $+2.8_{-5.7}^{+8.2}$ & 30 \\
$9150.4169_{-0.0090}^{+0.0085}$ & $-3_{-13}^{+12}$ &     31 \\
$9156.4050_{-0.0044}^{+0.0029}$ & $+17.6_{-6.4}^{+4.2}$ &31 \\
$9162.3798_{-0.0030}^{+0.0027}$ & $+19.0_{-4.4}^{+3.8}$ &31 \\
$9168.3562_{-0.0084}^{+0.0068}$ & $+23_{-12}^{+10}$ &    31 \\
\noalign{\smallskip}
\hline
\noalign{\smallskip}
\multicolumn{3}{c}{TOI-396\,d} \\
$8387.2711_{-0.0034}^{+0.0036}$ & $-0.4_{-4.9}^{+5.2}$ &  3 \\
$8398.5047_{-0.0051}^{+0.0056}$ & $4.0_{-7.4}^{+8.1}$ &   3 \\
$8432.1919_{-0.0034}^{+0.0050}$ & $-2.2_{-4.9}^{+7.2}$ &  4 \\
$9117.2589_{-0.0068}^{+0.0076}$ & $+6_{-10}^{+11}$ &     30 \\
$9139.7171_{-0.0025}^{+0.0047}$ & $+2.1_{-3.6}^{+6.8}$ & 30 \\
$9150.9371_{-0.0037}^{+0.0072}$ & $-13_{-5}^{+10}$ &     31 \\
$9162.1769_{-0.0031}^{+0.0038}$ & $+0.3_{-4.5}^{+5.5}$ & 31 \\
\hline

\end{tabular}
\end{minipage}
\end{table*}

\begin{table*}
\renewcommand{\arraystretch}{1.5}
\caption{Results of the internal structure modelling for TOI-396 b. The `w$_\cdot$' symbol represents the mass fraction with respect to the total planet mass, Z$_{\mathrm{envelope}}$ is the water mass fraction in the planet envelope, while the `x$_\cdot$' symbol represents the molar fraction of a given chemical element either in the planet core (x$_{\cdot,\mathrm{core}}$) or in the mantle (x$_{\cdot,\mathrm{mantle}}$).}
\centering
\begin{tabular}{r|ccc|ccc}
\hline \hline
Water prior &              \multicolumn{3}{c|}{Formation outside iceline (water-rich)} & \multicolumn{3}{c}{Formation inside iceline (water-poor)} \\
Si/Mg/Fe prior &           Stellar (A1) &       Iron-enriched (A2) &      Free (A3) &
                           Stellar (B1) &       Iron-enriched (B2) &      Free (B3) \\
\hline
w$_\textrm{core}$ [\%] &        $12.8_{-8.7}^{+9.1}$ &    $17.4_{-12.0}^{+14.3}$ &    $13.8_{-9.9}^{+15.1}$ &
                           $17.9_{-12.2}^{+12.4}$ &    $26.3_{-18.1}^{+21.4}$ &    $19.8_{-14.3}^{+22.4}$ \\
w$_\textrm{mantle}$ [\%] &      $58.5_{-10.8}^{+12.4}$ &    $49.7_{-15.3}^{+15.3}$ &    $56.4_{-17.5}^{+16.5}$ &
                           $82.1_{-12.4}^{+12.2}$ &    $73.7_{-21.4}^{+18.1}$ &    $80.2_{-22.4}^{+14.3}$ \\
w$_\textrm{envelope}$ [\%] &    $27.8_{-10.1}^{+10.1}$ &    $32.4_{-11.0}^{+9.4}$ &    $28.5_{-12.5}^{+10.8}$ &
                           $\left(1.0_{-0.5}^{+0.7}\right)$ $10^{-3}$ &    $\left(3.3_{-2.3}^{+6.7}\right)$ $10^{-3}$ &    $\left(1.7_{-1.2}^{+6.0}\right)$ $10^{-3}$ \\
\hline
Z$_\textrm{envelope}$ [\%] &        $99.9_{-1.5}^{+0.0}$ &    $99.9_{-1.8}^{+0.1}$ &    $99.9_{-1.5}^{+0.1}$ &
                           $0.5_{-0.2}^{+0.2}$ &    $0.5_{-0.2}^{+0.2}$ &    $0.5_{-0.2}^{+0.2}$ \\
\hline
x$_\textrm{Fe,core}$ [\%] &     $90.3_{-6.4}^{+6.5}$ &    $90.4_{-6.4}^{+6.5}$ &    $90.4_{-6.4}^{+6.5}$ &
                           $90.3_{-6.4}^{+6.5}$ &    $90.4_{-6.4}^{+6.5}$ &    $90.3_{-6.4}^{+6.5}$ \\
x$_\textrm{S,core}$ [\%] &      $9.7_{-6.5}^{+6.4}$ &    $9.6_{-6.5}^{+6.4}$ &    $9.6_{-6.5}^{+6.4}$ &
                           $9.7_{-6.5}^{+6.4}$ &    $9.6_{-6.5}^{+6.4}$ &    $9.7_{-6.5}^{+6.4}$ \\
\hline
x$_\textrm{Si,mantle}$ [\%] &   $38.7_{-6.2}^{+7.3}$ &    $32.5_{-9.2}^{+10.2}$ &    $34.3_{-23.2}^{+29.2}$ &
                           $38.6_{-6.1}^{+7.3}$ &    $32.1_{-9.2}^{+10.4}$ &    $32.6_{-22.5}^{+29.2}$ \\
x$_\textrm{Mg,mantle}$ [\%] &   $42.5_{-7.1}^{+7.8}$ &    $35.4_{-10.5}^{+11.4}$ &    $35.6_{-24.2}^{+29.1}$ &
                           $42.5_{-7.1}^{+7.9}$ &    $35.1_{-10.5}^{+11.5}$ &    $35.9_{-24.1}^{+29.4}$ \\
x$_\textrm{Fe,mantle}$ [\%] &   $18.6_{-12.0}^{+9.7}$ &    $31.2_{-20.0}^{+19.1}$ &    $22.2_{-15.7}^{+23.4}$ &
                           $18.8_{-12.1}^{+9.6}$ &    $31.9_{-20.2}^{+19.0}$ &    $23.2_{-16.6}^{+23.9}$ \\
\hline
\end{tabular}
\label{tab:internal_structure_results_b}
\end{table*}
\renewcommand{\arraystretch}{1.0}

\begin{table*}
\renewcommand{\arraystretch}{1.5}
\caption{Same as Tab.~\ref{tab:internal_structure_results_b}, but for TOI-396 d.}
\centering
\begin{tabular}{r|ccc|ccc}
\hline \hline
Water prior &              \multicolumn{3}{c|}{Formation outside iceline (water-rich)} & \multicolumn{3}{c}{Formation inside iceline (water-poor)} \\
Si/Mg/Fe prior &           Stellar (A1) &       Iron-enriched (A2) &      Free (A3) &
                           Stellar (B1) &       Iron-enriched (B2) &      Free (B3) \\
\hline
w$_\textrm{core}$ [\%] &        $13.5_{-9.2}^{+9.6}$ &    $18.4_{-12.7}^{+15.2}$ &    $14.4_{-10.3}^{+15.8}$ &
                           $18.2_{-12.4}^{+12.3}$ &    $24.3_{-16.8}^{+21.1}$ &    $17.4_{-12.5}^{+20.2}$ \\
w$_\textrm{mantle}$ [\%] &      $61.9_{-11.9}^{+13.2}$ &    $52.3_{-16.2}^{+16.5}$ &    $58.5_{-18.2}^{+17.2}$ &
                           $81.8_{-12.3}^{+12.4}$ &    $75.7_{-21.1}^{+16.8}$ &    $82.6_{-20.2}^{+12.5}$ \\
w$_\textrm{envelope}$ [\%] &    $23.2_{-10.3}^{+11.9}$ &    $28.0_{-11.3}^{+11.0}$ &    $25.1_{-12.2}^{+12.2}$ &
                           $\left(1.0_{-0.7}^{+1.4}\right)$ $10^{-2}$ &    $\left(4.3_{-3.3}^{+6.6}\right)$ $10^{-2}$ &    $\left(1.6_{-1.4}^{+6.0}\right)$ $10^{-2}$ \\
\hline
Z$_\textrm{envelope}$ [\%] &        $99.9_{-1.8}^{+0.1}$ &    $99.9_{-2.2}^{+0.1}$ &    $99.9_{-1.8}^{+0.1}$ &
                           $0.5_{-0.2}^{+0.2}$ &    $0.5_{-0.2}^{+0.2}$ &    $0.5_{-0.2}^{+0.2}$ \\
\hline
x$_\textrm{Fe,core}$ [\%] &     $90.3_{-6.4}^{+6.5}$ &    $90.4_{-6.4}^{+6.5}$ &    $90.3_{-6.4}^{+6.5}$ &
                           $90.3_{-6.4}^{+6.6}$ &    $90.4_{-6.4}^{+6.5}$ &    $90.3_{-6.4}^{+6.5}$ \\
x$_\textrm{S,core}$ [\%] &      $9.7_{-6.5}^{+6.4}$ &    $9.6_{-6.5}^{+6.4}$ &    $9.7_{-6.5}^{+6.4}$ &
                           $9.7_{-6.6}^{+6.4}$ &    $9.6_{-6.5}^{+6.4}$ &    $9.7_{-6.5}^{+6.4}$ \\
\hline
x$_\textrm{Si,mantle}$ [\%] &   $38.7_{-6.2}^{+7.3}$ &    $32.4_{-9.2}^{+10.2}$ &    $33.9_{-22.9}^{+29.1}$ &
                           $38.7_{-6.2}^{+7.3}$ &    $32.9_{-9.3}^{+10.2}$ &    $35.7_{-24.3}^{+28.8}$ \\
x$_\textrm{Mg,mantle}$ [\%] &   $42.5_{-7.0}^{+7.8}$ &    $35.4_{-10.5}^{+11.5}$ &    $35.6_{-24.2}^{+29.1}$ &
                           $42.6_{-7.1}^{+7.8}$ &    $36.0_{-10.7}^{+11.3}$ &    $36.5_{-25.2}^{+30.3}$ \\
x$_\textrm{Fe,mantle}$ [\%] &   $18.7_{-12.0}^{+9.7}$ &    $31.4_{-20.0}^{+19.0}$ &    $22.5_{-15.9}^{+23.4}$ &
                           $18.6_{-11.9}^{+9.8}$ &    $30.2_{-19.6}^{+19.4}$ &    $19.9_{-14.3}^{+22.6}$ \\
\hline
\end{tabular}
\label{tab:internal_structure_results_d}
\end{table*}
\renewcommand{\arraystretch}{1.0}

\end{appendix}

\end{document}